\documentclass[sigconf,nonacm]{acmart}

\usepackage{amsmath,amsfonts,amsthm}
\usepackage{booktabs}
\usepackage{multirow}
\usepackage{subfigure}
\usepackage{url}
\usepackage{graphicx}
\usepackage{textcomp}
\usepackage{xcolor}
\usepackage{float}
\usepackage{caption}
\usepackage{bbding}
\usepackage{enumitem}

\usepackage{circledsteps}
\pgfkeys{/csteps/fill color=black}
\pgfkeys{/csteps/inner color=white}

\usepackage{algorithmic}
\usepackage[linesnumbered,ruled, vlined]{algorithm2e}

\newcommand\scalemath[2]{\scalebox{#1}{\mbox{\ensuremath{\displaystyle #2}}}}

\begin{document}
\sloppy

\title{PriSampler: Mitigating Property Inference of Diffusion Models}

\author{Hailong Hu}
\affiliation{\institution{SnT, University of Luxembourg}
\city{~}
\country{~}
}

\author{Jun Pang}
\affiliation{\institution {FSTM \& SnT, University of Luxembourg}\city{~}
\country{~}
}

\begin{abstract}
Diffusion models have been remarkably successful in data synthesis.
However, when these models are applied to sensitive datasets, such as banking and human face data, they might bring up severe privacy concerns.
This work systematically presents the first privacy study about property inference attacks against diffusion models, where adversaries aim to extract sensitive global properties of its training set from a diffusion model.
Specifically, we focus on the most practical attack scenario: adversaries are restricted to accessing only synthetic data.
Under this realistic scenario, we conduct a comprehensive evaluation of property inference attacks on various diffusion models trained on diverse data types, including tabular and image datasets.
A broad range of evaluations reveals that diffusion models and their samplers are universally vulnerable to property inference attacks.
In response, we propose a new model-agnostic plug-in method {\sf PriSampler} to mitigate the risks of the property inference of diffusion models.
{\sf PriSampler} can be directly applied to well-trained diffusion models and support both stochastic and deterministic sampling.
Extensive experiments illustrate the effectiveness of our defense, and it can lead adversaries to infer the proportion of properties as close as predefined values that model owners wish.
Notably, {\sf PriSampler} also shows its significantly superior performance to diffusion models trained with differential privacy on both model utility and defense performance. 
This work will elevate the awareness of preventing property inference attacks and encourage privacy-preserving synthetic data release.

\end{abstract}

\maketitle
\thispagestyle{plain}
\pagestyle{plain}

%%%%%%%%% BODY TEXT
%------------------------------------------
\section{Introduction}
\label{sec:intro}
%------------------------------------------

Diffusion models~\cite{sohl2015deep}, as an emerging class of generative models, have gained widespread adoption in a large number of application areas, such as image synthesis~\cite{song2019generative,ho2020denoising,song2020score,karras2022elucidating}, tabular data generation~\cite{kotelnikov2023tabddpm}, and even text synthesis~\cite{gong2023sequence}.
However, when sensitive datasets, such as banking, medical, and human face data, are applied to train diffusion models, it might cause potential privacy breaches.

Property inference attacks~\cite{ateniese2015hacking} constitute a significant privacy risk by aiming to infer global properties of the whole training set used in a machine learning model, i.e. the proportion of the training data for a certain property.
These attacks can compromise sensitive statistical information that model owners intend to keep confidential. 
On the one hand, such disclosures can be exploited for competitive advantages, such as a company using inferred data to tailor advertising strategies to the detriment of a rival~\cite{mahloujifar2022property,chaudhari2022snap}.
On the other hand, the real statistics inferred by property inference attacks can assist in other types of privacy attacks, such as membership inference and data extraction~\cite{shokri2017membership,carlini2023extracting}.
In more severe scenarios, the leakage of demographic information might unintentionally expose the unfairness in a model's certain properties, such as gender and race, which potentially lead to accusations against companies for discrimination.
This is because these discriminations existing in a model will contravene principles of data regulations, such as the General Data Protection Regulation (GDPR)~\cite{GDPR}.

Although a considerable body of works has studied property inference attacks in classification models, such as support vector machines~\cite{ateniese2015hacking}, fully-connected neural networks~\cite{ganju2018property}, convolution neural networks~\cite{suri2022formalizing,chaudhari2022snap,mahloujifar2022property},
and generative adversarial networks~\cite{zhou2021property}, even graph neural networks~\cite{zhang2022inference},
property inference of diffusion models has not been explored to date.
In the era of generative artificial intelligence~(AI), brand-new diffusion models have become a dominant paradigm in deep generative modeling~\cite{dhariwal2021diffusion,kazerouni2022diffusion,yang2022diffusion}. 
This underscores the urgency for evaluating privacy vulnerabilities associated with these novel diffusion models.

\noindent
\textbf{Attacks.} 
In this work, we investigate the privacy risks of diffusion models through the lens of property inference attacks.
Our threat model assumes that adversaries can only have access to generated samples from a diffusion model.
Based on generated samples, our property inference attack aims to estimate the proportion of various properties of a diffusion model.
Our work considers two types of data generation: tabular and image data generation.
We explore the pioneering and state-of-the-art diffusion model in tabular data generation --- TabDDPM~\cite{kotelnikov2023tabddpm} and 
four different types of diffusion models in image generation, including the discrete variance preserving~(VP) model --- DDPM~\cite{ho2020denoising},  the discrete variance exploding preserving~(VE) model --- SMLD~\cite{song2019generative}, the continuous VP model --- VPSDE~\cite{song2020score} and the continuous VE model --- VESDE~\cite{song2020score}.
Unlike other generative models, such as generative adversarial networks~\cite{goodfellow2014generative} or variational autoencoder~\cite{kingma2013auto}, for a trained diffusion model in image generation, there are different sampling methods to generate samples, in which these methods aim to improve the quality of generated images or sampling speed during the sampling process.
Thus, we study three samplers over two different types of sampling mechanisms in image generation, including stochastic sampling --- PC sampler~\cite{song2020score} and deterministic sampling --- the black-box ODE sampler~\cite{dormand1980family} and the DPM sampler~\cite{lu2022dpm}.

We conduct our experiments on four datasets from different privacy-sensitive application domains, census data Adult~\cite{misc_adult_2}, banking data Churn~\cite{misc_Churn}, disease data Cardio~\cite{misc_Cardiovascular} and human face data CelebA~\cite{liu2015faceattributes}.
Our comprehensive evaluations demonstrate that state-of-the-art diffusion models and their samplers are vulnerable to property inference attacks~(see Section~\ref{ssec:Results_tabular} and Section~\ref{ssec:Results_image}).
For example, in tabular data generation, adversaries can precisely infer the proportion of sensitive properties with 0.11\% absolute difference in the best case and below 2\% absolute difference in the worst case, where the absolute difference refers to the difference between the inferred proportion and the real proportion.
In image generation, adversaries can accurately infer the proportion of properties with the absolute difference ranging from 0\% to 7\%.
We also explore the performance of property inference in terms of different properties, the number of generated samples, and utility performance.

\smallskip\noindent
\textbf{Defenses.}  
To defend against property inference attacks, we propose a property aware sampling method --- {\sf PriSampler}, which manipulates diffusion models in the sampling process to conceal the real proportion of sensitive properties.
More specifically, our defense method first finds the hyperplanes of properties in the diffusion models, and the learned hyperplanes are utilized to guide one sampler to synthesize samples in this property space~(see Figure~\ref{fig:defense_flow}).

We show the effectiveness of our defense method {\sf PriSampler} on tabular and image generation scenarios, and analyze defense performance in terms of different types of samplers, diffusion models, more properties, and the number of generated samples and different diffusion steps~(see Section~\ref{ssec:Defenses_res_tabular} and Section~\ref{ssec:Defenses_res_image}).
We also compare {\sf PriSampler} with differentially private diffusion models~(DPDMs)~\cite{dockhorn2022differentially}, and 
evaluations show that {\sf PriSampler} is superior to DPDMs on model utility and defense performance~(see Section~\ref{ssec:cmp_dp}).
More importantly, {\sf PriSampler} does not require re-training a diffusion model because it operates in the sampling process.

\smallskip\noindent
\textbf{Contributions.} Our contributions lie in revealing and highlighting property inference risks of diffusion models in the emerging field of generative AI, and providing a foundation defense for securing diffusion models against such attacks. Specifically, our contributions in this work are twofold.
\begin{itemize}[leftmargin=5mm]
	\item[(1)] We perform the first study of property inference attacks against diffusion models on various sensitive datasets under the most practical attack scenario, showing that diffusion models and their samplers are vulnerable to property inference attacks.
	\item[(2)] 
	We propose the first model-agnostic and plug-in defense --- {\sf PriSampler} to mitigate property inference risks of diffusion models, illustrating our method achieves state-of-the-art performance in both model utility and defense performance across various scenarios, including tabular and image generation.
\end{itemize}

%------------------------------------------
\section{Background}
\label{sec:Background}
%------------------------------------------

\subsection{Diffusion Models on Image Data}
\label{ssec:diff_models}
A diffusion model is a generative model, and it aims to learn the distribution~$p_{\it data}$ of a training set and generate new unseen data samples.
In general, it consists of two processes: a forward process and a reverse process.
In the forward process, it adds different levels of noise~$0=\sigma_{0}<\sigma_{1}<,...,<\sigma_{T}=\sigma_{\it max}$ into training data, in order to transform a training data's distribution into a Gaussian distribution within $T$ time steps. 
In the reverse process, it randomly samples a noise image from the Gaussian distribution and gradually denoises it into an image.
In the following, we introduce three fundamental types of diffusion models in image generation.

\smallskip\noindent
\textbf{DDPM.}  
The denoising diffusion probabilistic model~(DDPM) is proposed by Ho et al.~\cite{ho2020denoising}.
In the forward process, a sample at the $t$ time step is perturbed by: $x_t\gets \sqrt{{\alpha_t}}x_0+\sqrt{1-{\alpha_t}}\varepsilon$, where $\varepsilon\sim \mathcal{N}(0,\mathrm{I})$, $x_0 \sim p_{\it data}$, and ${\alpha_t} \in [0,1]$ is a variance schedule to control the magnitude of noise in each time step.
${\alpha_0}=1$ means that an image at $t=0$ time step is not perturbed  and ${\alpha_T}=0$ indicates that the perturbed image at $t=T$ time step becomes pure Gaussian noise.
In the reverse process, a noise image from~$\mathcal{N}(0,\mathrm{I})$ is step by step denoised and eventually recovers a noise-free image, and during the process
a neural network~$\epsilon_\theta(x_t,t)$ is trained to predict noise by minimizing the following loss:
\begin{equation}
\scalemath{0.85}{
	L(\theta) = \mathbb{E}_{t \sim [1,T],x \sim p_{\it data},\varepsilon\sim \mathcal{N}(0,\mathrm{I})}[||\varepsilon - \varepsilon_\theta(\sqrt{{\alpha_t}}x + \sqrt{1-{\alpha_t}}\varepsilon ,t)||^2]. 
 }
	\label{eq:ddpm}
\end{equation} 

\smallskip\noindent
\textbf{SMLD.} 
The score matching with Langevin dynamics~(SMLD) is proposed by Song et. al~\cite{song2019generative}. 
In the forward process, a perturbed sample at the $t$ time step is obtained by: $x_t\gets x_0+\sigma_t\epsilon$, where $\sigma_t$ is the noise schedule to control the magnitude of noise.
In the reverse process, a neural network~$s_\theta(x_t,\sigma_i)$ is trained to predict \textit{score}.
The \textit{score} refers to the gradient of the log probability density to data, i.e. $\nabla_x\text{\it log}\: p(x)$. 
SMLD minimizes the following loss:
\begin{equation}
\scalemath{0.83}{
	L_\theta=\mathbb{E}_{t\sim[1,T],x \sim p_{\it data},x_t \sim q(x_t|x)}[\lambda(\sigma_t)||s_\theta(x_t,\sigma_t) -\nabla_{x_t} \text{log}\:q(x_t|x)||^2],
 }
	\label{eq:SMLD}
\end{equation}
where $\lambda(\sigma_t)$ is a coefficient function and $\nabla_{x_t}\text{log}\:q(x_t|x)=-\frac{x_t-x}{\sigma_t^2}$.

\smallskip\noindent
\textbf{SSDE.} 
The score-based stochastic differential equation~(SSDE) proposed by Song et. al~\cite{song2020score} presents a general and unified framework for generative modeling.
The process of a diffusion model is described as a stochastic differential equation.
Specifically, SSDE defines the forward process as:~$\text{d}x=\text{f}(x,t)\text{d}t+g(t)\text{dw}$, where $\text{f}(x,t)$, $g(t)$ and $\text{dw}$ are the drift coefficient, the diffusion coefficient and a standard Wiener process, respectively.
In the reverse process, it can be expressed by a reverse-time SDE:~$\text{d}x=[\text{f}(x,t)-g(t)^2\nabla _x\text{log}~q_t(x)]\text{d}t+g(t)\text{d}\bar{\text w}$, where $\bar{\text w}$ is a standard Wiener process in the reverse time.
A neural network is used for predicting \textit{score} by minimizing the loss:
\begin{equation}
\scalemath{0.85}{
	L_\theta=\mathbb{E}_{t\in \mathcal{U}(0,T),x \sim p_{\it data},x_t \sim q(x_t|x)}[\lambda(t)||s_\theta(x_t,t)-\nabla_{x_t}\text{log}~q(x_t|x)||^2].
 }
	\label{eq:SSDE}
\end{equation}
Different coefficients, i.e. $\text{f}(x,t)$ and $g(t)$,  correspond to different types of SSDE, and in their work, two types of SSDE are proposed: variance preserving~(VP) and variance exploding (VE). We call their corresponding models as VPSDE and VESDE and they are continuous diffusion models.
Furthermore, under this framework, DDPM and SMLD can be considered discrete VP and VE, respectively.
In this work, we will systematically study these four types of diffusion models: DDPM, SMLD, VPSDE, and VESDE.

\subsection{Diffusion Models on Tabular Data}
\label{ssec:tabddpm}

Driven by the success of image generation, diffusion models have also been studied for tabular data generation~\cite{kotelnikov2023tabddpm,sattarov2023findiff,kim2023stasy}.
Different from image data, tabular data exhibits heterogeneity: a tabular sample~$x$ consists of numeric properties~$x_{\it num}$ and categorical properties~$x_{\it cat}$, i.e. $x=[x_{\it num}, x_{\it cat}]$. 
This mixed type of samples means that diffusion models in image generation cannot be directly used in tabular generation. 
To address this problem, Kotelnikov et al.~\cite{kotelnikov2023tabddpm} propose TabDDPM based on the DDPM framework to generate tabular data.
Specifically, TabDDPM learns numeric properties by the Gaussian diffusion model which is the same as the model in DDPM, while for categorical properties, TabDDPM firstly transforms them into one-hot properties and then utilizes the multinomial diffusion model~\cite{hoogeboom2021argmax} to learn.
Finally, TabDDPM minimizes a sum of loss from numerical properties and categorical properties:
\begin{equation}
	\scalemath{0.85}{
L_{\theta}=L_{\it num} + \frac{\sum_{i\le C}{L_{\it cat}^{i}}}{C},
}
	\label{eq:tabddpm}
\end{equation}
where $L_{\it num}$ is inherently the same with Equation~\ref{eq:ddpm} in DDPM and $L_{\it cat}^{i}$ is the KL divergences for each categorical property and $C$ is the number of categorical properties.
In this work, we will focus on TabDDPM considering their excellent performance.

\subsection{Samplers}
\label{ssec:diff_sampler}
\noindent
\textbf{Image data.} 
After the training of a diffusion model finishes, we can use different samplers to synthesize new data.
Based on the unified framework of SSDE, there are two types of sampling: stochastic sampling and deterministic sampling.

\smallskip\noindent
\textbf{$\bullet$~Stochastic sampling.} 
Because a diffusion model can be described as a stochastic differential equation~(SDE), we can generate a new sample by solving the corresponding reverse-time SDE~\cite{anderson1982reverse}.
Existing general-purpose numerical solvers, such as Euler-Maruyama and stochastic Runge-Kutta methods~\cite{platen2010numerical}, can be used for solving the SDE.
Song et. al~\cite{song2020score} propose Predictor-Corrector methods to further improve the sampling quality by utilizing the score-based model.
In this work, we call it as PC sampler.

\smallskip\noindent
\textbf{$\bullet$~Deterministic sampling.} 
In addition to solving a reverse-time SDE, Song et. al~\cite{song2020score} find that a reverse-time SDE also corresponds to a probability flow ordinary differential equation~(ODE) in which they have the same marginal probability densities.
It indicates that we can generate a new sample by solving a probability flow ODE.
Existing black-box ODE solver~\cite{dormand1980family} can be used to generate samples.
In this work, we call it as ODE sampler.

In addition to directly using a black-box ODE solver, there are many works about designing efficient samplers based on solving the probability flow ODE~\cite{liupseudo,lu2022dpm,zhang2022fast}.
For example, 
DPM~\cite{lu2022dpm} analyzes the ODE consisting of a linear function of the data variable and a nonlinear function parametrized by neural networks.
By deriving an exact formulation for the linear part, DPM can improve the quality of generated samples and speed up the sampling process.
In this work, we call it as DPM sampler.
Considering their excellent sampling performance, we will systematically study three samplers from two different sampling mechanisms: PC sampler, ODE sampler, and DPM sampler.

\smallskip\noindent
\textbf{Tabular data.}
Sampling methods on tabular data are much less explored than those on image data because tabular data generation does not significantly suffer from slow sampling speed. In addition, this is also due to the fact that the dimension of tabular data is significantly lower than that of image data.
Therefore, in this work, we directly use the stochastic sampling method in TabDDPM. 

%\smallskip
In this work, we will show that our proposed {\sf PriSampler} can be effectively integrated into these different types of samplers regardless of image and tabular data generation.

%------------------------------------------
\section{Property Inference Attacks}
\label{sec:method}
%------------------------------------------
The objective of a property inference attack is to predict the proportion of a property in the training set of a trained diffusion model.
This makes adversaries reveal some sensitive information that is not shared by model owners.
For instance, in addition to directly utilizing generated samples from a diffusion model, adversaries could also attempt to infer sensitive information disclosed by these generated samples, such as the proportion of the property gender and race.  
It is thus important to investigate the feasibility of property inference attacks against diffusion models.
This section starts with problem formulation. We then introduce the threat model and attack method and experimental setups. Finally, we present attack results and novel insights. 

%------------------------------------------
\subsection{Problem Formulation}
%------------------------------------------

A training dataset~$\mathcal{D}$ has different properties including sensitive ones.
Each property is either categorical or numeric.
Categorical property has limited discrete values, such as gender = $\{${male, female$\}$.
Numeric property has continuous real numbers, such as credit scores of customers in banking, creditScore $\in [0,1000]$.
Each property has a real proportion~$p_{s_i}$ in the dataset~$\mathcal{D}$.
A diffusion model~$\mathcal G$ is trained on the dataset~$\mathcal{D}$.
Now, given $m$ generated samples~$X$ from the diffusion model~$\mathcal G$, and a  property $s_i$, adversaries aim to infer the proportion of the property~$\hat{p}_{s_i}$, in order to make~$\hat{p}_{s_i}$ as close~$p_{s_i}$ as possible. More specifically, the adversaries need to design an attack algorithm~$\mathcal{A}$ to estimate the proportion~$\hat{p}_{s_i} = \mathcal{A}(X)$.

Note that, for a categorical property, the adversaries aim to infer the proportion of one property, such as the proportion of gender=male, while for a numeric property, the adversaries target to infer the proportion within the range of one property, such as the proportion of creditScore$<$600.

\subsection{Threat Model}
\label{ssec:Threatmodel}
Our threat model considers that adversaries only obtain generated samples from a diffusion model.
The adversaries do not know the type of diffusion models and the type of their samplers, which is usually the strictest and most practical scenario.
For image generation, we also assume that the adversaries have a shadow dataset that contains properties that adversaries intend to infer.
The shadow dataset is utilized by adversaries to build property classifiers. This assumption using shadow datasets is widely used in the privacy research~\cite{chaudhari2022snap,shokri2017membership,carlini2021membership,salemml,ganju2018property}.
However, we can relax this assumption by directly using a pre-trained classifier, and we demonstrate this by a case study in Appendix~\ref{ssec:case}.

\begin{figure}[!t]
	\centering
	\includegraphics[width=0.95\linewidth]{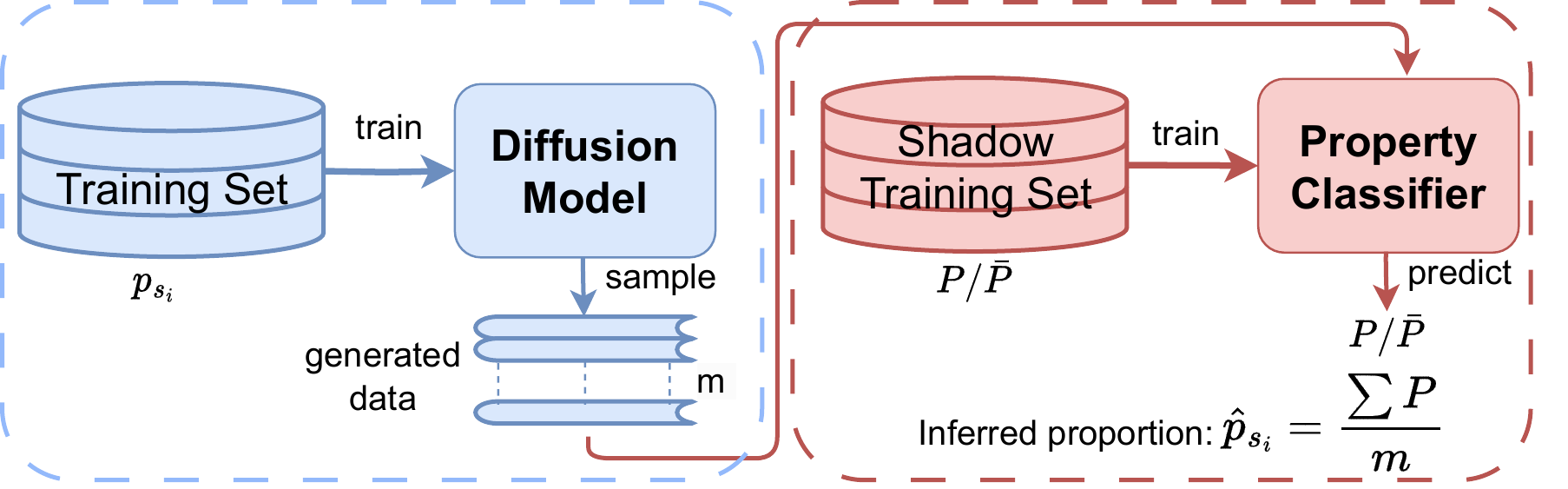}
	\caption{The attack process of the property inference attack.}
	\label{fig:attack_flow}
\end{figure}  

%------------------------------------------
\subsection{Attack Method}
%------------------------------------------
\label{ssec:Method}
The intuition of our attack is that generated samples have a similar distribution to the training set because a diffusion model learns the distribution of a training set and these generated samples are produced by the diffusion model.
Thus, adversaries might infer the proportion of the properties of the training set from these samples.

\smallskip\noindent
\textbf{Image data.} 
For image generation, we will first deploy a property classifier to predict the property of generated samples, and then use the average statistics of these generated samples containing the property as the inferred proportions. 
Figure~\ref{fig:attack_flow} illustrates the attack process of the property inference.
Firstly, a property classifier is trained on a shadow training set.
The shadow training set is labeled by the property that adversaries are interested in. 
$P$ refers to one sample containing this property while $\bar{P}$ refers to one sample not containing this property. 
After finishing the training on the shadow data set, the property classifier takes as input generated data from the diffusion model and outputs predictions.
Finally, the inferred proportion of one property is estimated by $ \hat p_{s_i}=\frac{\sum P}m$, where  $m$ is the number of generated samples.
For $k$  properties, we will train $k$ property classifiers.
Note that, unlike property inference attacks that consider them as a classification problem, i.e. inferring whether a machine learning model contains a property~\cite{ganju2018property,mahloujifar2022property,zhang2022inference}, here we directly estimate the proportion of a property for diffusion models, which is more precise.

\smallskip\noindent
\textbf{Tabular data.} 
For tabular data generation, the values of all properties are explicitly shown in generated samples.
For instance, one sample in synthetic tabular data shows it is female in the column gender.
Thus, we can directly infer the proportion of one property~${s_i}$ by calculating the ratio of generated samples containing this property ${s_i}$, i.e. $\hat p_{s_i}=\frac{\sum P}m$, where $P$ refers to one sample containing the property~${s_i}$.

%------------------------------------------
\subsection{Experimental Setup}
%------------------------------------------
\label{ssec:Experimental}
\noindent
\textbf{Datasets.} 
We conduct our experiments on four datasets from various application domains, such as census, banking, healthcare, and computer vision. The Adult and CelebA datasets are widely used in prior works~\cite{ganju2018property,chaudhari2022snap, mahloujifar2022property,zhou2021property}.

\smallskip\noindent
\textbf{$\bullet$~Adult.} The Adult dataset~\cite{misc_adult_2} contains 48,842 samples which are extracted from the 1994 U.S. Census database. Each sample has 14 properties and in this work we choose five properties, such as gender, race, age, marital status, and workclass (related to occupations).

\noindent
\textbf{$\bullet$~Churn.} The Churn dataset~\cite{misc_Churn} contains 10,000 records about a bank's customers. Each record has 11 properties and in this work we consider four sensitive properties: gender, age, geography, and CreditScore (related to default risks).

\noindent
\textbf{$\bullet$~Cardio.} The Cardio dataset~\cite{misc_Cardiovascular} includes 70,000 records of patient data. Each record has 11 properties and we choose three properties: gender, age and smoking (related to personal habits).

\noindent
\textbf{$\bullet$~CelebA.} The CelebA dataset includes 202,599 images of celebrity faces~\cite{liu2015faceattributes}.
Each image annotates 40 binary properties and we choose four representative properties including gender, age, smiling (related to sentiment), and eyeglasses (related to personal style).

\smallskip\noindent
\textbf{Target models.}
We utilize the state-of-the-art TabDDPM~\cite{kotelnikov2023tabddpm} as target models for tabular data generation on the Adult, Churn, and Cardio datasets.
We use the default dataset split of TabDDPM for each dataset, and
train the model through official codes\footnote{\url{https://github.com/yandex-research/tab-ddpm}} with optimal hyperparameters.

We choose four types of diffusion models: DDPM~\cite{ho2020denoising}, SMLD~\cite{song2019generative}, VPSDE~\cite{song2020score}, and VESDE~\cite{song2020score}, as target models for image generation on the CelebA dataset. 
We use open source codes in this library\footnote{\url{https://github.com/yang-song/score_sde_pytorch}\label{footnote_1}} with their suggested training hyperparameters to train each diffusion model.

\smallskip\noindent
\textbf{Samplers.}
For TabDDPM on tabular data generation, we use its default stochastic sampling method.
For image generation, we choose three typical samplers: one stochastic sampling --- PC sampler~\cite{song2020score}, and two deterministic samplings --- ODE sampler~\cite{dormand1980family} and DPM sampler~\cite{lu2022dpm}.
We adopt this library$^{\ref{footnote_1}}$ for PC and ODE samplers and this library\footnote{\url{https://github.com/LuChengTHU/dpm-solver}} for the DPM sampler.
The recommended sampling hyperparameters in each implementation are adopted.

\smallskip\noindent
\textbf{Attack models.}
For image data, we use the ResNet-50\footnote{\url{https://download.pytorch.org/models/resnet50-19c8e357.pth}} pre-trained on ImageNet~\cite{russakovsky2015imagenet} to train a property inference classifier.
The shadow training set used for training the classifier is from the remaining samples of the CelebA dataset.
In other words, one part of the CelebA dataset is used for training diffusion models while the other part, i.e. the shadow training set, is used for training property classifiers.  
These two parts are disjoint.
This is a common practice in the community of privacy in machine learning~\cite{ganju2018property,chaudhari2022snap,zhou2021property}.
In Appendix~\ref{ssec:case}, we show that the classifier works well for diffusion models trained on different humane face datasets.
More specifically, we train classifiers with the stochastic gradient descent optimizer.
The learning rate and weight decay of all property classifiers are 0.01, except for the classifier of the property age where both values are set as 0.005.
The number of training epochs is set as 5.

\smallskip\noindent
\textbf{Metrics.}
In this work, we comprehensively report utility and attack performance.

\smallskip\noindent
\textbf{$\bullet$~Utility.}
Utility refers to the performance of target models.
For tabular data generation, we adopt the widely-used  F1 score. 
We follow the `\textit{train in synthetic, test in real}' framework to compute F1 scores.
Specifically, a classifier is trained on generated tabular data from a diffusion model and tests the classier performance on a real dataset.
A higher F1 score indicates a higher quality of generated samples. In this work, we use the current state-of-the-art model on tabular data CatBoost as a classifier ~\cite{prokhorenkova2018catboost} and compute an F1 score with 50k generated samples.
For image generation, we use the widely-adopted Fr{\'e}chet Inception Distance~(FID) metric~\cite{FID2017gans}.
A lower FID means that a sampler of a diffusion model can generate more realistic and diverse samples.
In this work, by default, we compute an FID with all training samples and 50k generated samples for the ODE and DPM samplers, and 500 generated samples for the PC sampler.
This is because the PC sampler in image generation requires a much longer time to synthesize data, compared with deterministic ODE and DPM samplers.

\smallskip\noindent
\textbf{$\bullet$~Attack performance.}
Attack performance refers to the performance of property inference attacks.
Our property inference attacks predict a real number, i.e. the proportion of a property.
Thus, we show the attack performance by directly presenting the predicted value.
In addition, we also report the absolute difference~$\Delta s_i$ between the predicted value and the real value, i.e. $\Delta s_i = |\hat{p}_{s_i}-p_{s_i}|$.
A smaller absolute difference value means a more precise inference.

\subsection{Attack Results on Tabular Data}
\label{ssec:Results_tabular}

In this subsection, we present the attack results against diffusion models trained on tabular data.
We choose the state-of-the-art model TabDDPM as our target model.
Three different privacy-sensitive datasets including Adult, Churn, and Cardio are used to train TabDDPM, respectively.
Target models with their best utility performance (F1 score) are chosen to be attacked, and the F1 scores range from 73\% to 80\%, which is also reported in Table~\ref{tab:att_tabddpm}.

\smallskip\noindent
\textbf{Attack performance on different properties.} 
Table~\ref{tab:att_tabddpm} shows attack performance on TabDDPM with regard to different sensitive properties.
We choose 12 different properties whose real proportions range from 6\% to 70\%.
Overall, our attacks can precisely infer the proportion of each property with at most 2\% absolute difference errors.
The best attack inference can be seen on the properties Martial-status=Divorced on Adult and Age$<$30 on Churn, where the absolute difference is as low as 0.11\%.
In addition, the attack performance on the common privacy-sensitive properties Gender=Male and  Race=Black on Adult can achieve 0.38\% and 0.76\% absolute difference respectively.
For the location-related property Geography=Germany on Churn, we can infer its proportion as 25.71\%, which is quite close to the real proportion of 25.20\%. 
For the default risks-related property CreditScore$<$600, the absolute difference of our inference is 0.77\%.
The credit score range CreditScore$<$600, below average, signifies a customer's potential default risk and partially reflects a bank's risk profile.

\smallskip\noindent
\textbf{Attack performance on different datasets.} 
Table~\ref{tab:att_tabddpm} shows property inference performance on datasets from three different application domains.
No matter the census-related dataset Adult, the banking-related dataset Churn, and the disease-related dataset Cardio, our attacks consistently perform well. In particular, the absolute difference in all properties on Adult is less than 0.80\%.

\begin{table}
	\centering
	\caption{Attack performance on different sensitive properties across different datasets. The target model is TabDDPM. Prop.: proportion. Abs. Diff. : absolute difference. $\downarrow$ means smaller is better, while $\uparrow$ means larger is better.}
	\label{tab:att_tabddpm}
	\renewcommand{\arraystretch}{1.0}
	\scalebox{0.8}{	
		\begin{tabular}{ccl|rrr} 
			\toprule
			Dataset                 & \begin{tabular}[c]{@{}c@{}}Utility\\F1 (\%) $\uparrow$ \end{tabular} & Property                & \begin{tabular}[c]{@{}r@{}}~Real\\ Prop. \\ (\%)\end{tabular} & \begin{tabular}[c]{@{}r@{}}Inferred\\ Prop. \\ (\%)\end{tabular} & \begin{tabular}[c]{@{}r@{}}Abs.\\ Diff. \\ (\%) $\downarrow$\end{tabular}  \\
			\hline
			\multirow{5}{*}{Adult}  & \multirow{5}{*}{79.58}                                & Gender=Male             & 67.05                                                         & 67.43                                                            & 0.38                                                          \\
			&                                                       & Age$<$30                   & 29.81                                                         & 29.46                                                            & 0.35                                                          \\
			&                                                       & Race=Black              & 9.64                                                          & 8.88                                                             & 0.76                                                          \\
			&                                                       & Martial-status=Divorced & 13.61                                                         & 13.72                                                            & 0.11                                                          \\
			&                                                       & Workclass=Local-gov     & 6.42                                                          & 6.09                                                             & 0.33                                    \\ \hline
			\multirow{4}{*}{Churn}  & \multirow{4}{*}{75.75}                                & Gender=Male             & 54.47                                                         & 56.16                                                            & 1.69                                                          \\
			&                                                       & Age$<$30                   & 16.75                                                         & 16.64                                                            & 0.11                                                          \\
			&                                                       & Geography=Germany       & 25.20                                                         & 25.71                                                            & 0.51                                                          \\
			&                                                       & CreditScore$<$600           & 30.19                                                         & 29.42                                                            & 0.77                                                          \\ \hline
			\multirow{3}{*}{Cardio} & \multirow{3}{*}{73.54}                                & Gender=Male             & 34.87                                                         & 34.50                                                            & 0.37                                                          \\
			&                                                       & Age$>=$50                   & 69.34                                                         & 69.18                                                            & 0.16                                                          \\
			&                                                       & Smoking=Yes             & 8.84                                                          & 7.84                                                             & 1.00                                                          \\
			\bottomrule
		\end{tabular}
		
	}
\end{table}

\smallskip\noindent
\textbf{Attack performance on different numbers of generated samples.} 
Figure~\ref{fig:num_gen_samples_tabddpm} shows attack performance about the number of generated samples.
Here, we choose TabDDPM trained on Adult as the target model and explore different proportions of sensitive properties which represent low, medium, and high. 
The grey dashed line is the real proportion of each property.
Overall, for properties on different proportions, we can observe that the attack performance stabilizes and is close to the real proportions after the number of generated samples increases to 500.

%------------------------------------------
\subsection{Attack Results on Image Data}
\label{ssec:Results_image}
%------------------------------------------

In this subsection, we present the attack results against diffusion models trained on image data.
We choose the dataset CelebA as our main dataset. This is because this dataset provides extensive property-related information. We can systematically study property inference risks by considering different proportions of properties.

Specifically, on the basis of CelebA, we design ten datasets with five different proportions of sensitive properties and two different sizes of the training set (i.e. $10=5 \times 2$), to investigate the privacy risks of diffusion models.
Two sizes of training sets include 1k samples and 50k samples, respectively.
Five different proportions refer to that male face images account for 10\%, 20\%, 30\%, 40\%, and 50\% in a dataset, respectively.
To briefly express its meaning, we mark a dataset as CelebA-size-proportion, such as CelebA-1k-10\%. All datasets are resized to $64 \times 64$, considering the factors of computation efficiency.
DDPM and VPSDE are trained on 1k samples with five different proportions, while SMLD and VESDE are trained on 50k samples with five different proportions.
In total, 20 diffusion models are trained on training sets.
All target models with their best utility performance (FID score) are chosen to be attacked, and the FID scores range from 5 to 56 and are summarized in Table~\ref{tab:att_perf} in Appendix.

\begin{figure}[]
	\centering
	\includegraphics[width=0.60\linewidth]{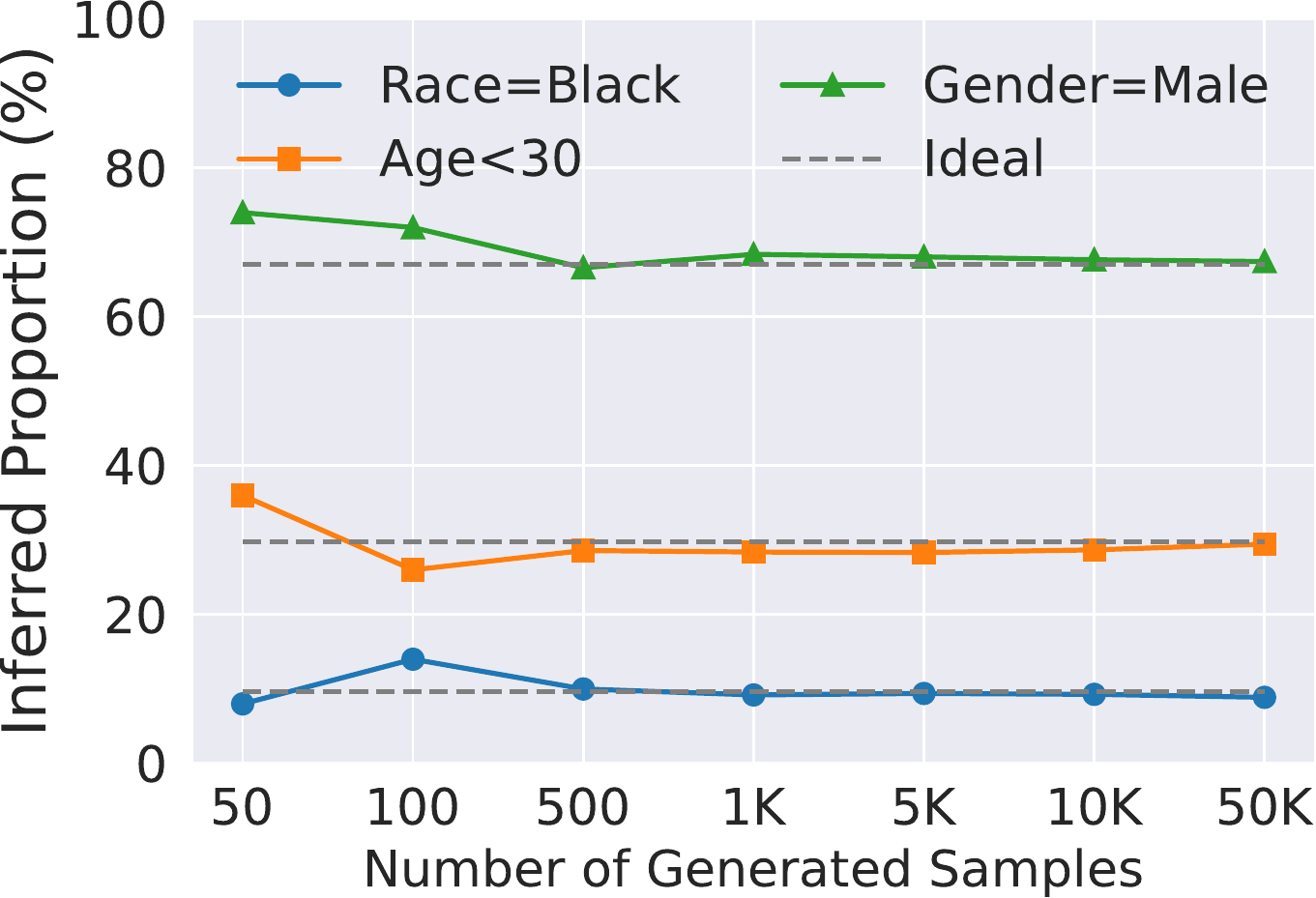}
	\caption{Attack performance to the number of generated samples. The target model is TabDDPM trained on Adult.}
	\label{fig:num_gen_samples_tabddpm}
\end{figure}  

\smallskip\noindent
\textbf{Attack performance on different samplers.} 
Figure~\ref{fig:att_perf} shows attack performance with regard to different samplers over four types of diffusion models.
Each type of diffusion model is trained on datasets with different proportions of the sensitive property Gender=Male.
Here, the real proportion of the  property Gender=Male is set as 10\%, 20\%, 30\%, 40\%, and 50\%, respectively.
An ideal attack means that the inferred proportion is equal to the real proportion. We take it as a reference and it is shown as the grey diagonal line in Figure~\ref{fig:att_perf}.
Overall, all types of samplers cannot defend against the property inference attack. 
Our inferred proportions are consistently close to the real proportions with the increase in the real proportion of the property male.
We do not show the attack performance on the ODE sampler and DPM sampler for SMLD and VESDE models, because both samplers do not support these models.

Table~\ref{tab:att_perf_sum} reports attack performance on different samplers.
We can see that our attacks show the best performance in the PC sampler and slightly inferior performance on the ODE sampler.
In a nutshell, our attacks can have at most a 2.78\% absolute difference among the three types of samplers.
   
\begin{figure*}[!t]
	\centering
	
	\subfigure[DDPM.]{
		\includegraphics[width=0.48\columnwidth]{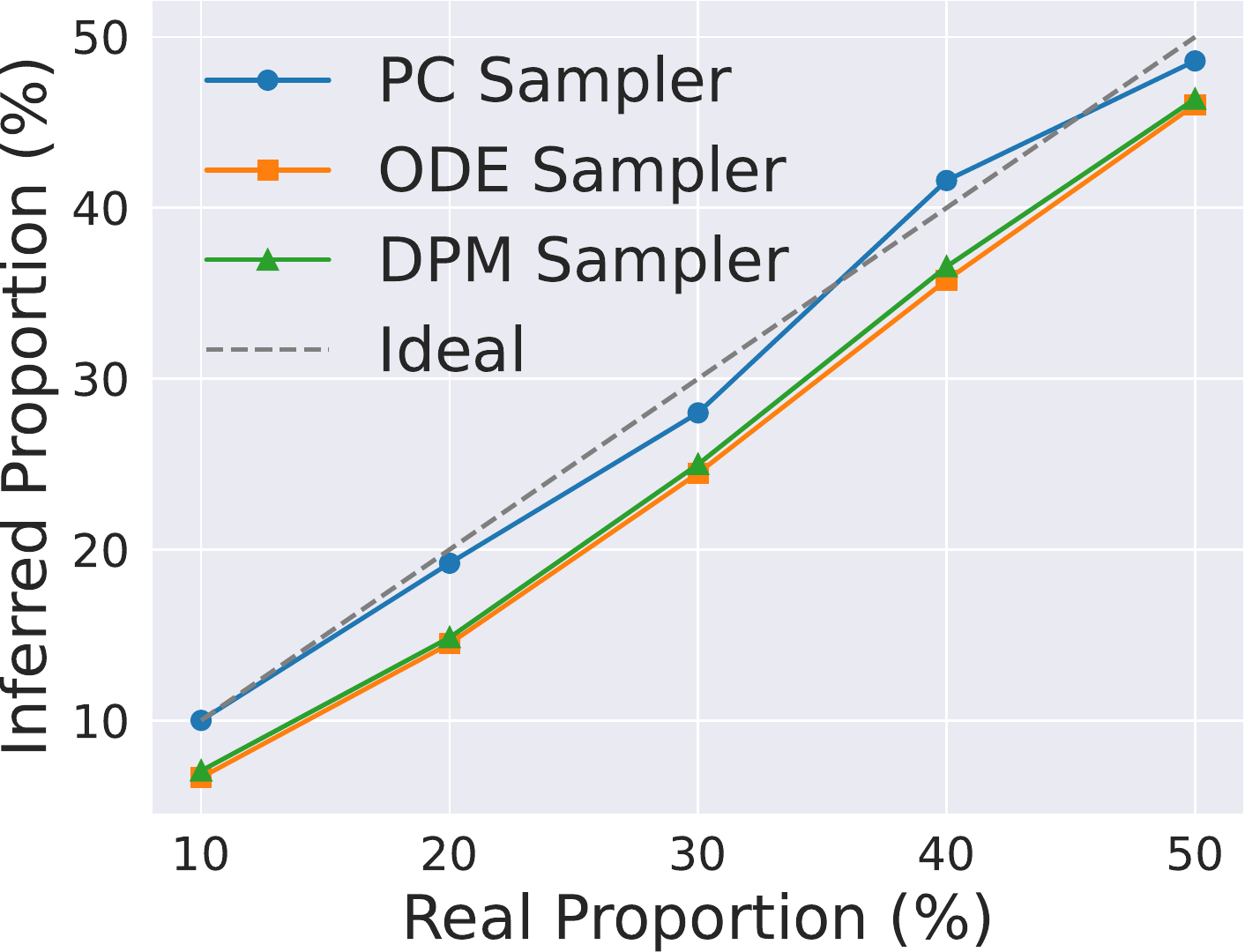}
		\label{fig:ddpm_celeba1k_male}
	}
	\subfigure[SMLD.]{
		\includegraphics[width=0.48\columnwidth]{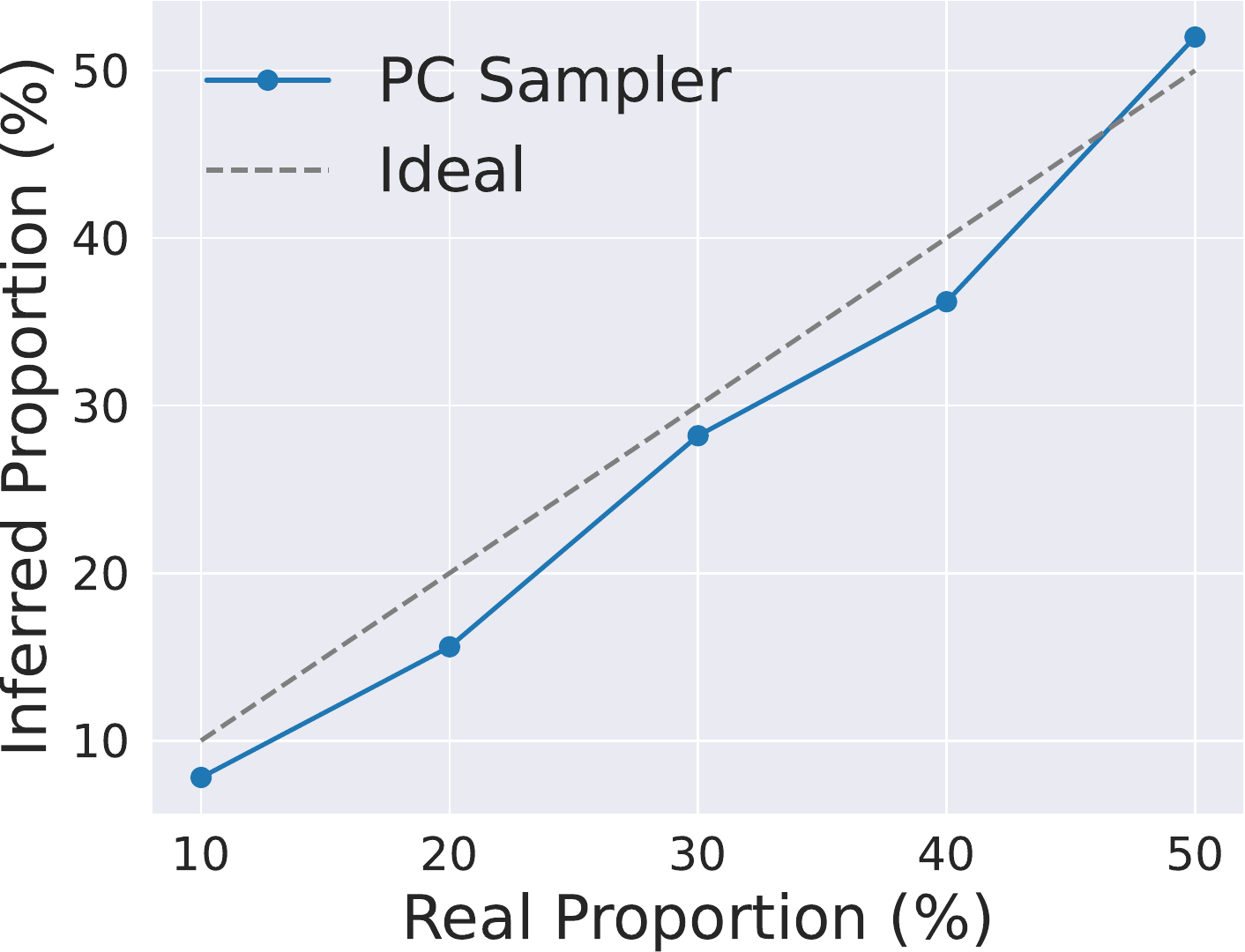}
		\label{fig:smld_celeba50k_male}
	}
	\subfigure[VPSDE.]{
		\includegraphics[width=0.48\columnwidth]{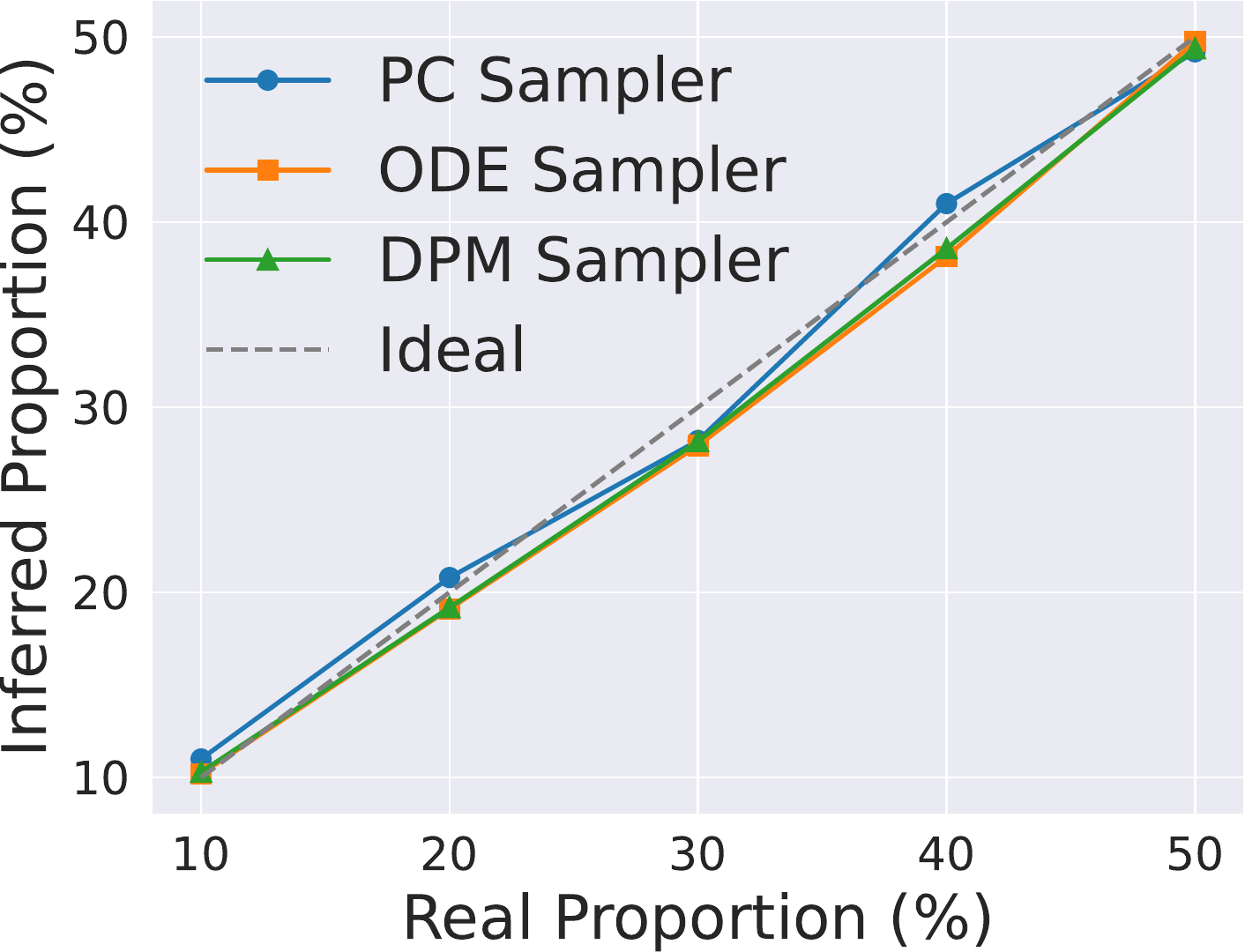}
		\label{fig:vpsde_celeba1k_male}
	}	
	\subfigure[VESDE.]{
		\includegraphics[width=0.48\columnwidth]{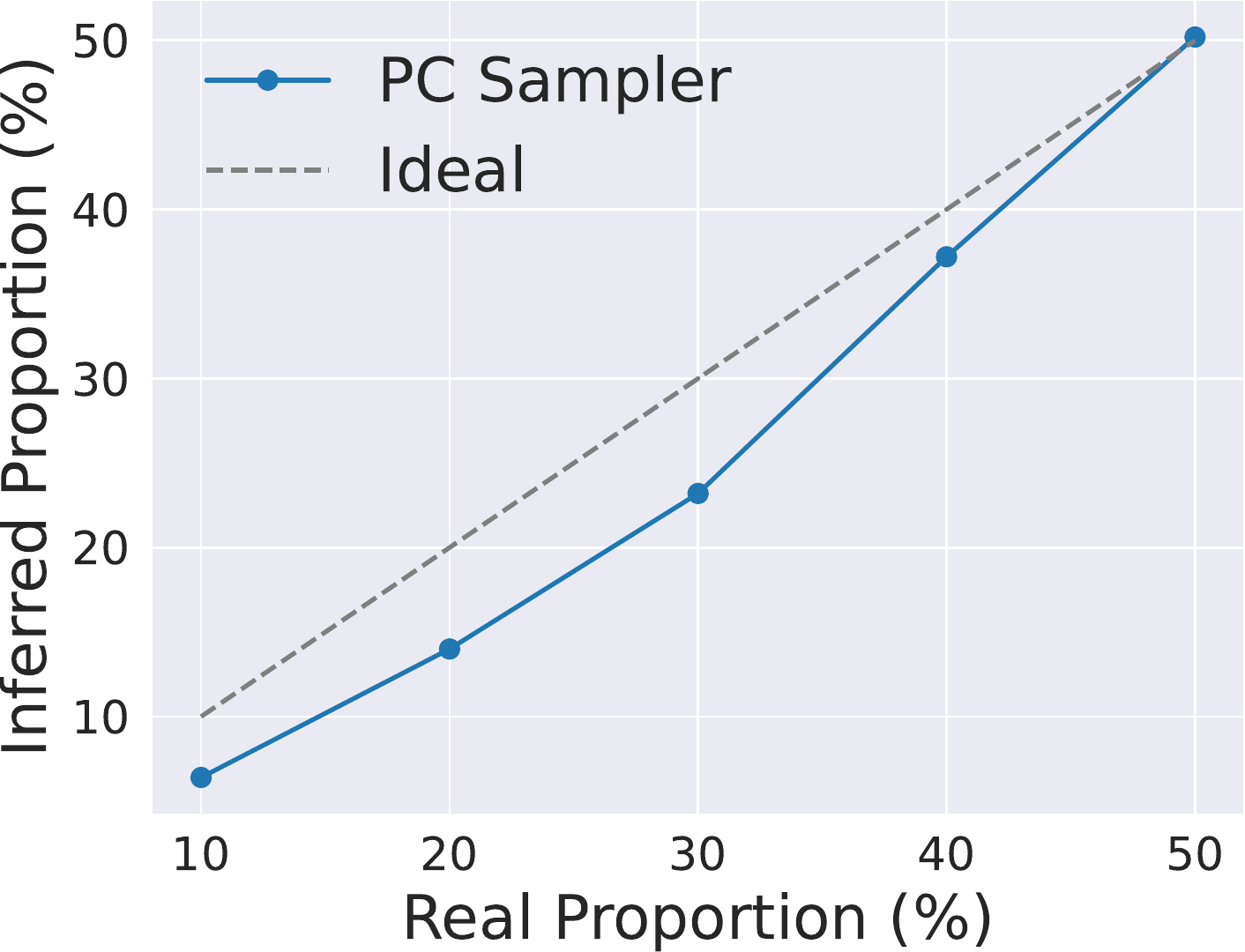}
		\label{fig:vesde_celeba50k_male}
	}
	\caption{Attack performance on different diffusion models, different samplers, and different proportions of the sensitive property. Here, the sensitive property is Gender=Male. Quantitative attack results are shown in Table~\ref{tab:att_perf} in Appendix.}
	\label{fig:att_perf}
\end{figure*}

\begin{figure*}[!t]
	\centering
	
	\subfigure[DDPM.]{
		\includegraphics[width=0.48\columnwidth]{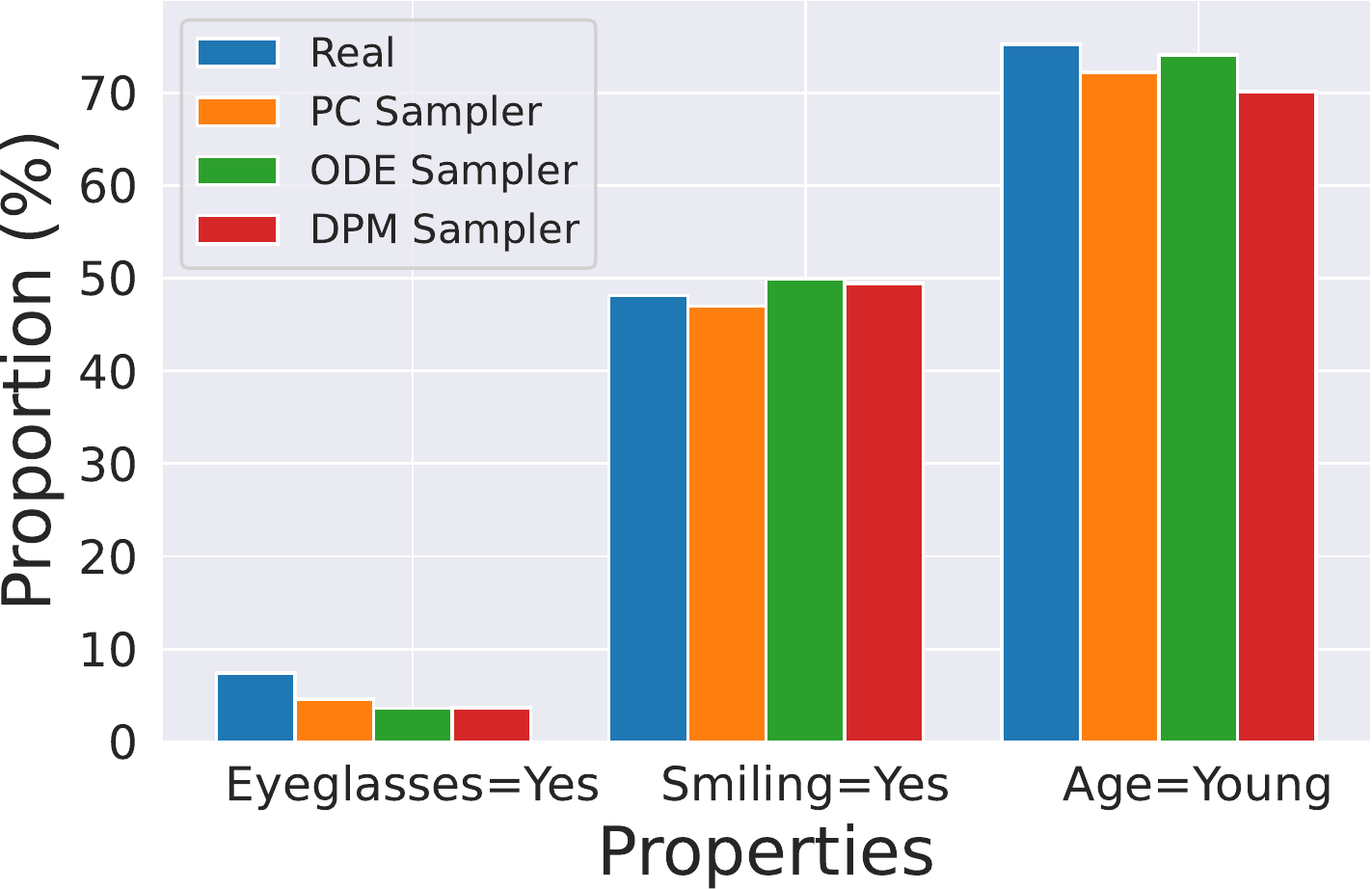}
		\label{fig:other_properties_ddpm_celeba1k_50}
	}
	\subfigure[SMLD.]{
		\includegraphics[width=0.48\columnwidth]{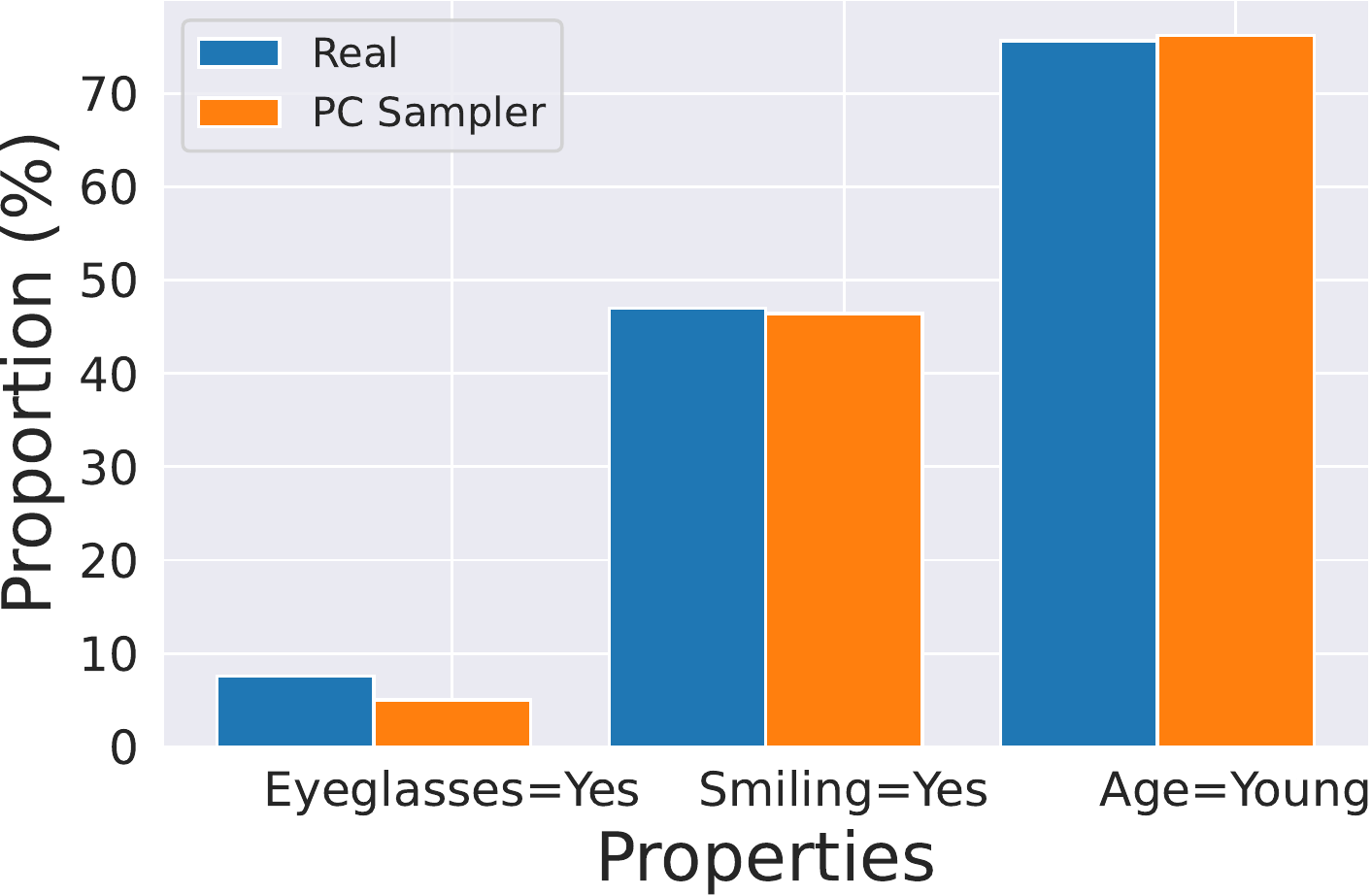}
		\label{fig:other_properties_smld_celeba50k_50}
	}
	\subfigure[VPSDE.]{
		\includegraphics[width=0.48\columnwidth]{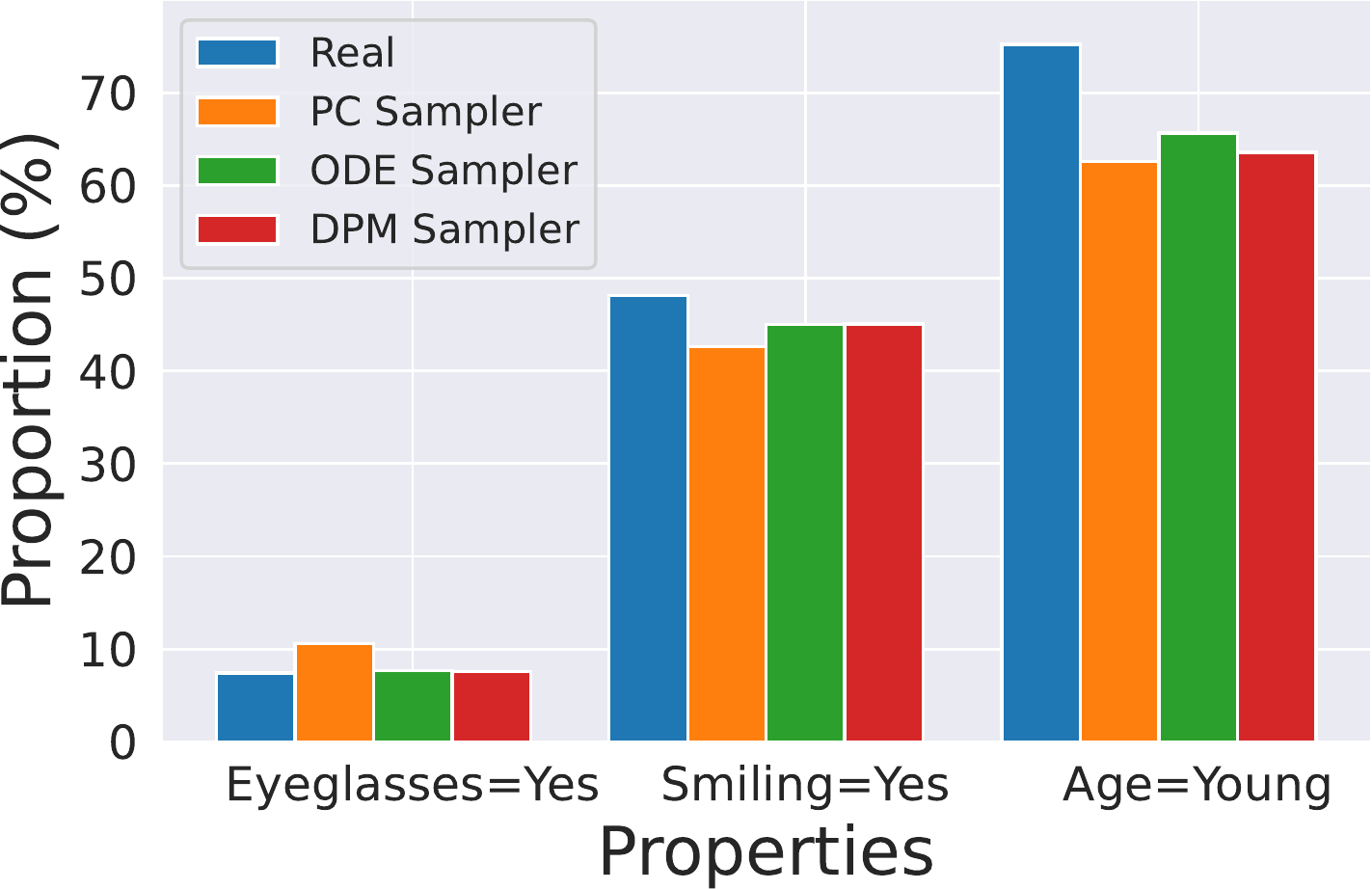}
		\label{fig:other_properties_vpsde_celeba1k_50}
	}	
	\subfigure[VESDE.]{
		\includegraphics[width=0.48\columnwidth]{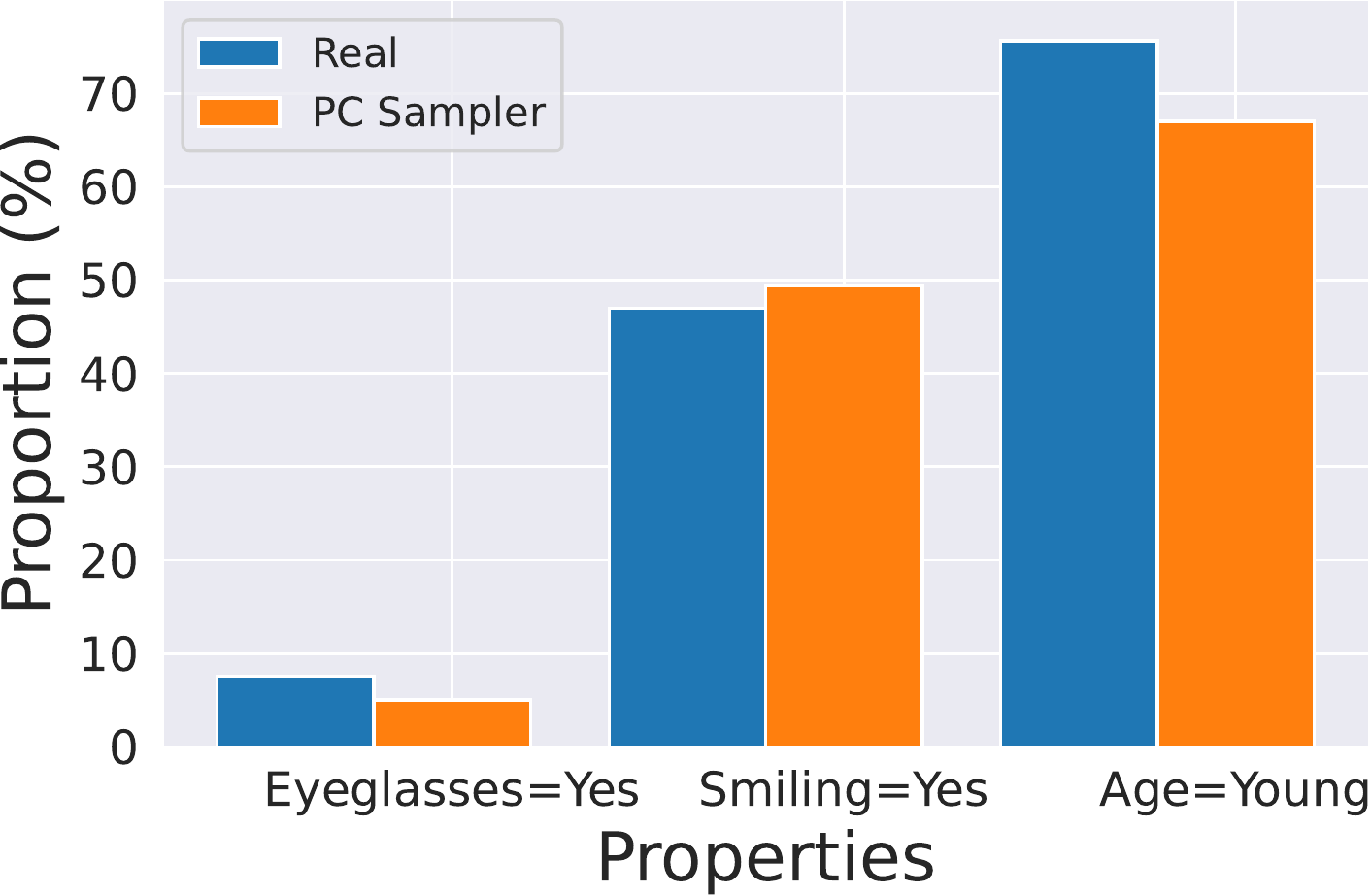}
		\label{fig:other_properties_vesde_celeba50k_50}
	}
	\caption{Attack performance on different properties across different diffusion models and samplers in image generation.}
	\label{fig:att_perf_other}
\end{figure*} 

\smallskip\noindent
\textbf{Attack performance on different diffusion models.} 
As shown in Figure~\ref{fig:att_perf}, we present the attack performance on twenty diffusion models encompassing four different types.
Each subfigure presents the attack performance of each type of diffusion model.
Overall, all diffusion models are vulnerable to property inference attacks.
Although each trained diffusion model can utilize different samplers, we can see that adversaries can still efficiently extract these sensitive information of a training set, regardless of the used samplers.
In particular, our attack can achieve almost perfect inference on VPSDE models for all samplers.
The reason why the property inference attack is effective on these diffusion models is that all existing samplers mainly focus on improving the quality of generated samples or sampling speed.
The other equally important issue, i.e. privacy, is not considered in their design.
In Section~\ref{sec:Defenses}, we will take the first step to provide privacy protection by developing a property aware sampling method for diffusion models.

Table~\ref{tab:att_perf_sum} reports attack results on different types of diffusion models.
We can see our attacks show the best performance on the VPSDE  with an average absolute difference of 1.04\%, and show marginally worse performance on the VESDE  where the average absolute difference is 3.88\%. 
\begin{table}[!t]
			\centering
	\caption{Summary of attack performances with different types of samplers and diffusion models. Here, we report the average absolute difference~(with standard deviation in parentheses) and the best and worst absolute difference.}
	\label{tab:att_perf_sum}
	\renewcommand{\arraystretch}{1.0}
	\scalebox{0.80}{
	\begin{tabular}{ll|ccc}
		\toprule
		&       & Average~(\%)     & Best~(\%) & Worst~(\%) \\
		\hline
		Sampler & PC    & 2.24 (1.84) & 0.00 & 6.80  \\
		& ODE   & 2.78 (2.02) & 0.19 & 5.53  \\
		& DPM   & 2.52 (1.78) & 0.29 & 5.15  \\
		\hline
		Model   & DDPM  & 3.24 (1.75) & 0.00 & 5.53  \\
		& VPSDE & 1.04 (0.62) & 0.19 & 2.07  \\
		& SMLD  & 2.84 (1.18) & 1.80 & 4.40  \\
		& VESDE & 3.88 (2.64) & 0.20 & 6.80 \\
		\bottomrule
	\end{tabular}

}
\end{table}

\smallskip\noindent
\textbf{Attack performance on different properties.} 
In addition to the property Gender=Male, we also choose three more properties based on their different proportions in CelebA.
The three properties are Eyeglasses=Yes, Smiling=Yes, Age=Young and their real proportions are roughly below 10\%, close to 50\%, and above 70\%, respectively.
The specific real proportions of these properties are plotted by blue bars in Figure~\ref{fig:att_perf_other}.
Here, we choose each type of the model with 50\% male as the target model.
Again, we can observe that our attack still remains effective on inferring the proportions of these properties on all diffusion models and samplers.
No matter what the smaller proportion of the property, such as eyeglasses, or the larger proportion of the property, such as age, the inferred proportions are very close to the real proportions.

\smallskip\noindent
\textbf{Attack performance on different numbers of generated samples.}
Figure~\ref{fig:num_gen_samples_ddpm_celeba1k_male_30} shows the attack performance on different numbers of generated samples.
Here, we choose the DDPM model trained on a dataset that contains 30\% male training samples, as the target model.
We can again observe that the attack performance will gradually become stable after 500 generated samples.

\begin{figure}[!t]
	\centering
	
	\subfigure[Attack performance on the number of generated samples. The target model is DDPM trained on CelebA-1k-30\%.]{
		\includegraphics[width=0.43\columnwidth]{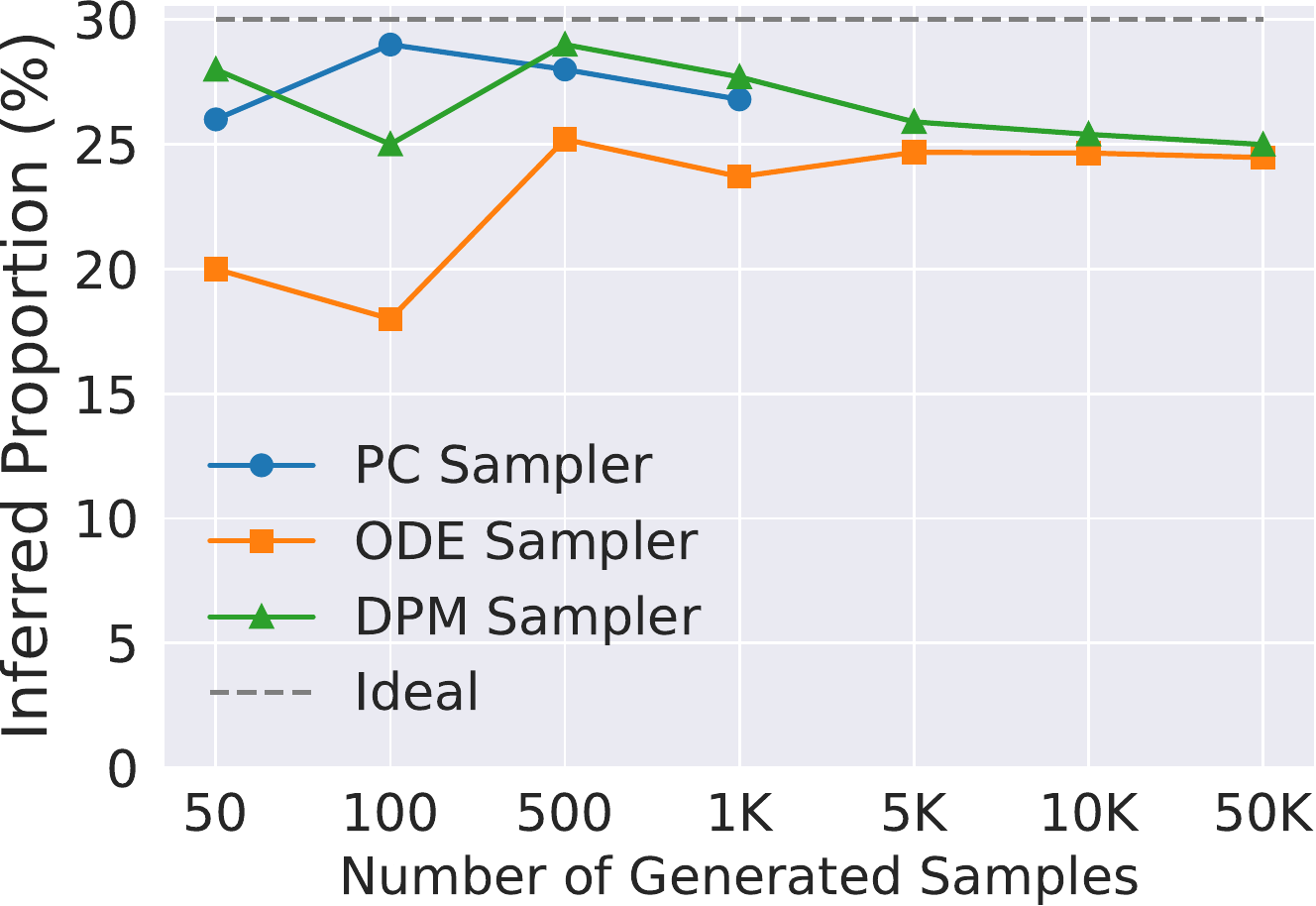}
		\label{fig:num_gen_samples_ddpm_celeba1k_male_30}
	}
	\subfigure[Attack performance on different FID values. The target model is VPSDE trained on CelebA-1k-30\%.]{
		\includegraphics[width=0.47\columnwidth]{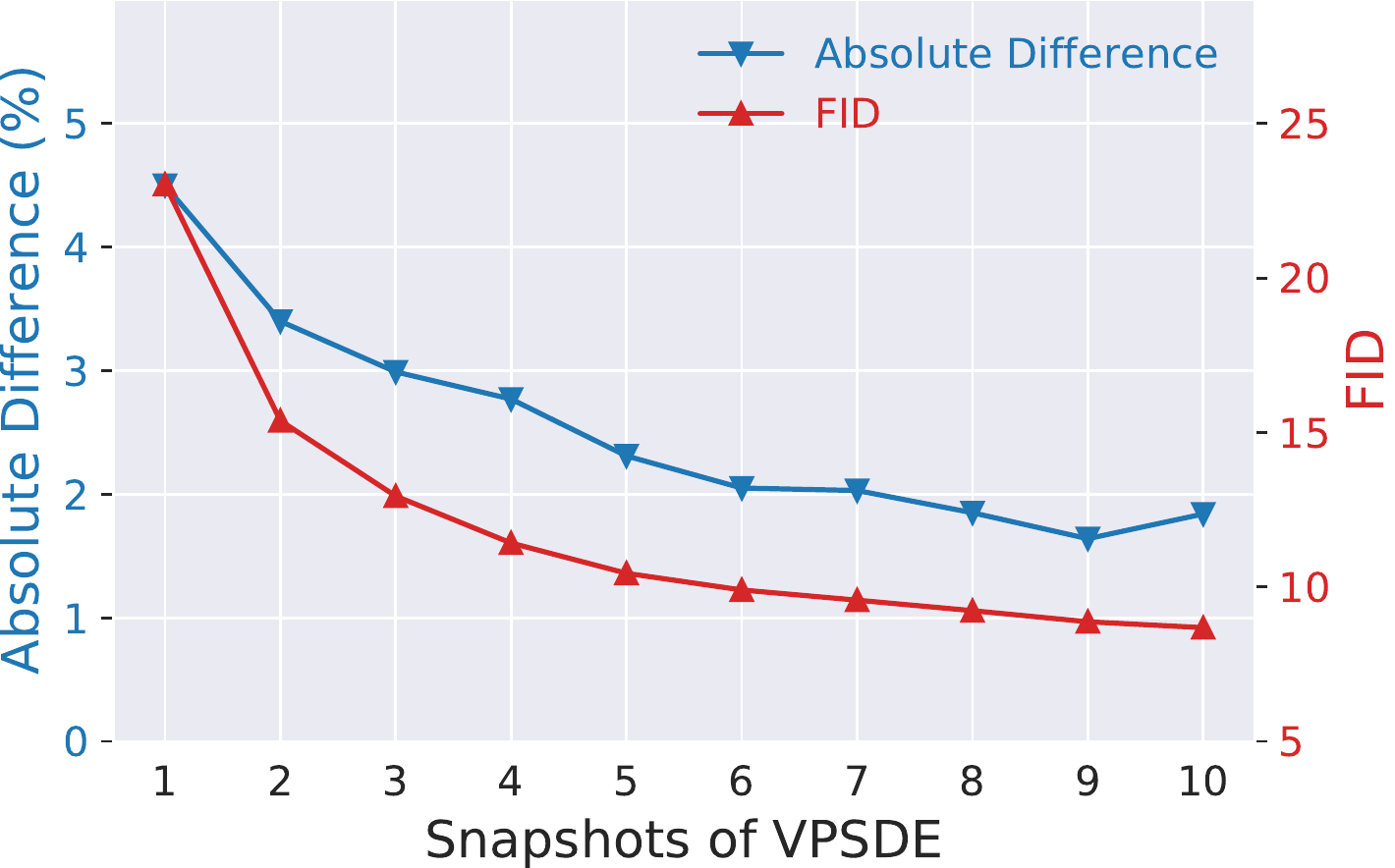}
		\label{fig:att_perf_to_fid_vpsde1k_30}
	}
	
	\caption{Attack performance with regard to the number of generated samples and FID scores.}
	\label{fig:att_img_ana}
\end{figure}

\smallskip\noindent
\textbf{Attack performance on different FID scores.} 
Figure~\ref{fig:att_perf_to_fid_vpsde1k_30} shows attack performance in terms of different FID scores of target models.
Here, we choose the DPM sampler and the VPSDE model trained on CelebA-1k-30\%.
Furthermore, we choose ten snapshots of VPSDE during the training process.
The left axis presents the absolute difference of a target model marked as the blue line, while the right axis shows the FID scores of a target model which is marked as the red line.
Overall, our attack becomes more accurate with the increase in the performance of target models. 
Note that a smaller FID means a better utility performance of a target model.
This also indicates that pursuing the good utility performance of a diffusion model can lead to more severe privacy risks.
Both model utility and privacy risks should be considered when diffusion models involve sensitive data.

\smallskip\noindent
\textbf{Takeaways.}
In summary, (1)  property inference attacks perform well on both tabular and image data.
(2) Both stochastic sampling and deterministic sampling are susceptible to property inference attacks.
(3) Different types of diffusion models cannot defend against this attack.
(4) No matter how large or small the proportion of certain properties is, our attack can precisely infer them.
(5) Inference performance becomes gradually stable after releasing 500 generated samples.
(6) The better the utility performance of a diffusion model is, the better the attack performance of property inference.

%------------------------------------------
\section{Defenses}
\label{sec:Defenses}
%------------------------------------------
In this section, we shift our focus to mitigating property inference attacks.
We first discuss several potential defenses.
Then we will introduce a property aware sampling method and present the defense results.
Finally, we discuss the defense of diffusion models trained with differential privacy.

%------------------------------------------
\subsection{Key Idea of Defenses}
%------------------------------------------
Property inference attacks leverage generated samples from a diffusion model to estimate the proportions of the properties.
To defend against this type of attack, model owners could manipulate the output of a diffusion model to disguise the real proportion of the property~$p_{s_i}$.

Therefore, the goal of our designed defenses is to make adversaries infer the proportion of a property~$\tilde p_{s_i}$ as close as one predefined value~$\gamma$ that the model owners wish.
In this work, we set the predefined value as the average of the number of values of one property: $\gamma = \frac{1}{k}$ where $k$ is the number of values, such as 0.5 for the binary property gender due to  Gender = \{Male, Female\} or 0.25 for one property containing four values.
There are at least two reasons for this. 
Firstly, it can disguise the real proportions of sensitive properties.
Secondly, because some properties, such as gender and race, are usually related to fairness, this choice can ensure the fairness of a diffusion model and achieve responsible synthetic data release. 

\begin{figure*}[!t]
	\centering
	\includegraphics[width=0.89\linewidth]{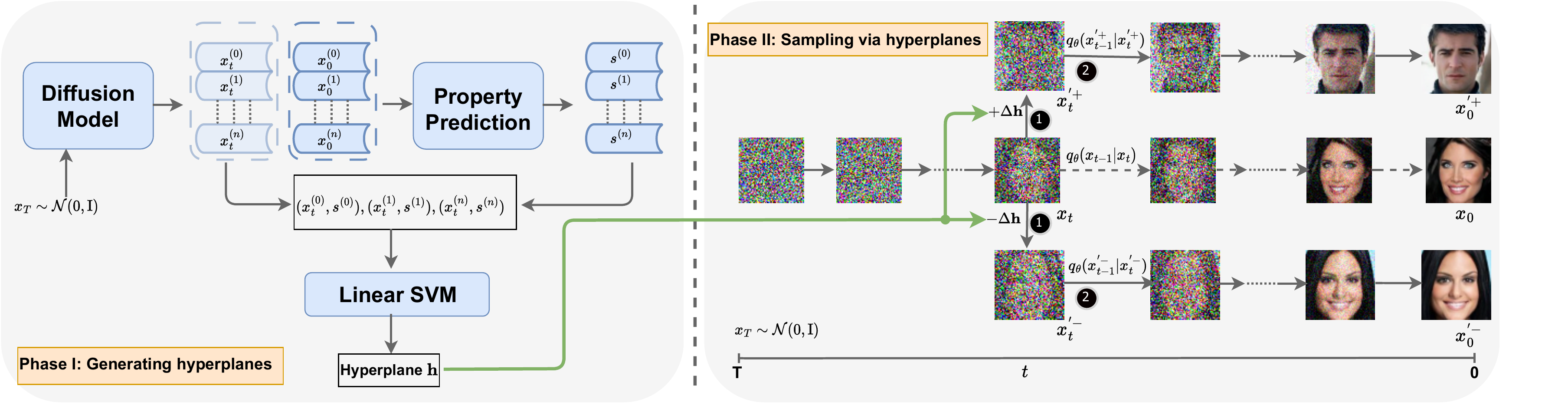}
	\caption{The process of the defense method {\sf PriSampler}. Phase I learns hyperplanes of sensitive properties. Phase II synthesizes samples via the learned hyperplanes. In \Circled{\small\textbf{1}}, at the $t$ diffusion step, {\sf PriSampler} changes the intermediate samples to the space of the specific sensitive property via the learned hyperplane. In \Circled{\small\textbf{2}}, {\sf PriSampler} continues to denoise samples step by step. Eventually, the desired samples~($x_0^{'+}$, $x_0^{'-}$) are obtained in the final step.}
	\label{fig:defense_flow}
\end{figure*} 

%------------------------------------------
\subsection{Potential Defenses}
%------------------------------------------
Based on the key idea that the designed defenses make adversaries infer the proportion of a property as the predefined value~$\gamma$, we discuss the following potential defenses.

\smallskip\noindent
\textbf{Dropping some samples of a larger proportion of the property.}
Take a binary property as an example, if the real proportion of a property is not equal to 0.5, model owners can choose to drop some generated samples of a larger proportion of the property to achieve a balance.
Specifically, model owners first collect all generated samples before releasing these samples.
Then, they calculate the real-time statistics and some generated samples that have a high proportion of the property will be dropped.

We assume there are $n$ binary and independent sensitive properties.
Furthermore, the proportion of the property~$s_i$ satisfies $p_{s_i} + \bar p_{s_i} = 1$ and  $p_{s_i} \in (0,0.5)$, where $p_{s_i}$ is the proportion containing this property and $\bar p_{s_i}$ is the proportion not containing this property.
In the worst case, the proportion of dropping samples is, at most  $1-2^{n}\prod_{i=1}^{n}p_{s_i}$. 
For instance, considering that there is one sensitive property and its real proportion is 10\%, i.e. $p_{s_1}=10\%$ and $n=1$, then
80\% samples among all generated samples will be discarded to achieve the balance, which is quite not economical. 
In particular, the sampling of diffusion models is time-consuming.
For the number of sensitive properties more than one, the number of dropped samples is exponentially increasing.
Thus, this method is simple but not scalable.

\smallskip\noindent
\textbf{Using a balanced dataset for sensitive properties.} 
This method requires model owners to prepare a balanced training dataset for sensitive properties, such as using a dataset containing 50\% male samples for the property Gender=Male.
Our experiments in Section~\ref{ssec:Results_image} show that this method indeed has a positive effect to some degree.
However, in some cases even for balanced properties, we observe that the learned diffusion models still produce imbalanced generated samples. 
For instance, the DPM sampler on the DDPM model trained on a 50\% male dataset produces male samples which are about 46.37\% of all generated samples.
Additionally, it is complicated and even impossible to collect sufficient training samples if we need to consider balancing more sensitive properties, such as collecting the same number of samples for the properties Martial-status=Married, Martial-status= Never-married and Martial-status=Divorced.

\smallskip
\noindent
\textbf{Property aware sampling.}
In addition to above discussed methods, we can design a new type of sampling mechanism that can automatically balance the proportion of sensitive properties. 
In this way, we can avoid the waste of generated samples and fastidious dataset selection.
We illustrate this method in the next subsection.

\subsection{Defense Method --- {\sf PriSampler}}
\label{ssec:PriSampler}

The main purpose of the property aware sampling method is to balance the proportion of sensitive properties in the sampling process of diffusion models.
As a result, the inferred proportion always remains at the predefined value~$\gamma$.
Our method is inspired by the semantic latent space of generative adversarial network~(GANs) for image generation in this work~\cite{shen2020interpreting}.
Due to the essential difference in the sampling process between diffusion models and GANs, we adapt the method to be well suitable for diffusion models
and propose {\sf PriSampler} as a general-purpose defense for diffusion models on both tabular and image data scenarios.

The key idea of {\sf PriSampler} is to guide one sampler to generate novel samples in the latent space of sensitive properties.
For instance, for the property Gender=Male, given the corresponding property hyperplane, samplers generate male samples on the side of this hyperplane and female samples on the opposite of this hyperplane.
In order to find such a hyperplane, we use a linear support vector machine~(SVM) to learn a decision boundary for each sensitive property.
This is due to the simplicity and efficiency of the linear SVM, and the hyperplane of the trained linear SVM can easily be utilized as well.
Then, given a base sampler, we can use the hyperplane to guide the sampler to synthesize new samples.
Here, the base sampler can be any sampler that is used for sampling in prior diffusion models.

\smallskip\noindent
\textbf{The process of~{\sf PriSampler}.} We will first illustrate our defense process on image data. The defense for tabular data follows the almost same process but is slightly modified to better suit the characteristics of tabular data.

\smallskip\noindent
\textbf{$\bullet$~Image data.}
Figure~\ref{fig:defense_flow} shows the process of our defense {\sf PriSampler} on the image data scenario.
It consists of two phases: generating hyperplanes and sampling via hyperplanes.
In phase I, our method aims to find a latent space in terms of a sensitive property from the diffusion model.
To achieve this goal, we leverage a linear SVM to learn the hyperplane of the sensitive property.
In detail, given a diffusion model, we can get many different types of generated samples from different diffusion steps.
As shown in the left part of Figure~\ref{fig:defense_flow}, starting from Gaussian noise sample~$x_T \sim \mathcal{N}(0,\mathrm{I})$, the diffusion model can produce the sample~$x_t$ at intermediate diffusion step $t$ and the final sample~$x_0$ at the $t=0$ step.
The final sample~$x_0$ is also the sample that we finally use.
Then, the samples $X_0 = \{x_0^{(0)}, x_0^{(1)},...,x_0^{(n)}\}$ are inputted to a property prediction classifier and the corresponding prediction scores can be obtained, i.e.  $S = \{s^{(0)}, s^{(1)},...,s^{(n)}\}$.
Instead of using $X_0$, we pair $X_t$ and $S$. 
The data samples $(X_t, S)$ will be used for training a linear SVM.
A hyperplane corresponding to this property can be obtained from the well-trained SVM.

In phase II, our method aims to sample via the learned hyperplane.
As shown in the right part of Figure~\ref{fig:defense_flow}, our method manipulates samples at the $t$ diffusion step.
To be specific, given Gaussian noise sample~$x_T \sim \mathcal{N}(0,\mathrm{I})$, we can get the sample~$x_t$.
At the $t$ diffusion step, we change the sampling direction via the learned hyperplane, and the samples with and without the corresponding sensitive property can be obtained.
In the remaining diffusion step, the samples will continue to synthesize in the specific sensitive property latent space.
Finally, samples with the balanced sensitive property can be generated.

\smallskip\noindent
\textbf{$\bullet$~Tabular data.}
The defense process on tabular data is similar to that on image data. However, there are two differences.
In phase I, tabular generated samples do not need a property prediction because their properties are explicitly shown and we can directly check the properties of these generated samples.
The second difference is that we choose the last step (i.e. $t=0$) to conduct property aware samplings for tabular data.
This is because each property is each column in tabular data and our extensive empirical experiments in Section~\ref{ssec:Defenses_res_tabular} will show that it still performs well on both utility and defense performance.

\begin{figure*}[!t]
	\centering
	
	\subfigure[Adult.]{
		\includegraphics[width=0.55\columnwidth]{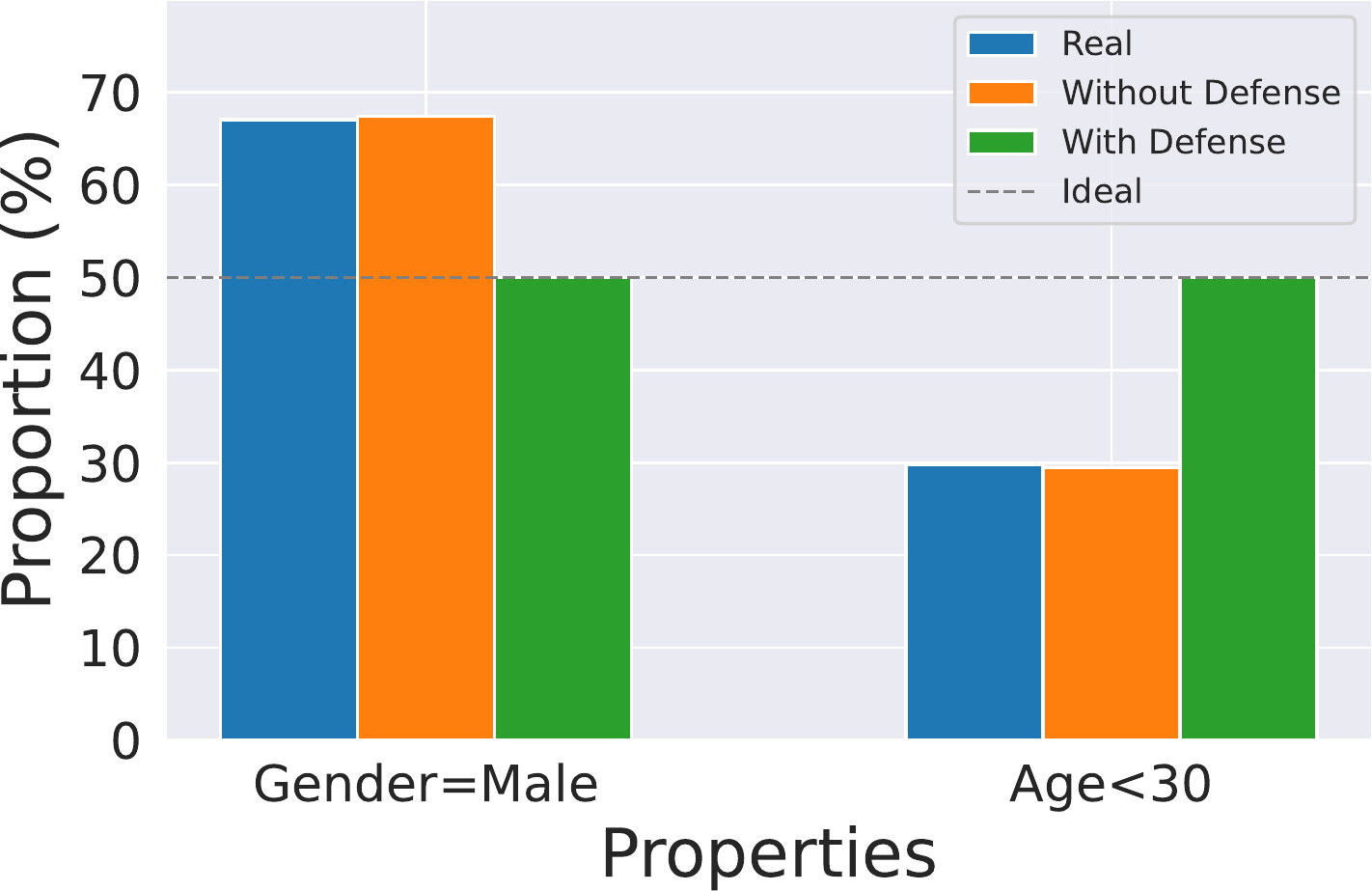}
		\label{fig:defense_tabddpm_adult_single}
	}
	\subfigure[Churn.]{
		\includegraphics[width=0.55\columnwidth]{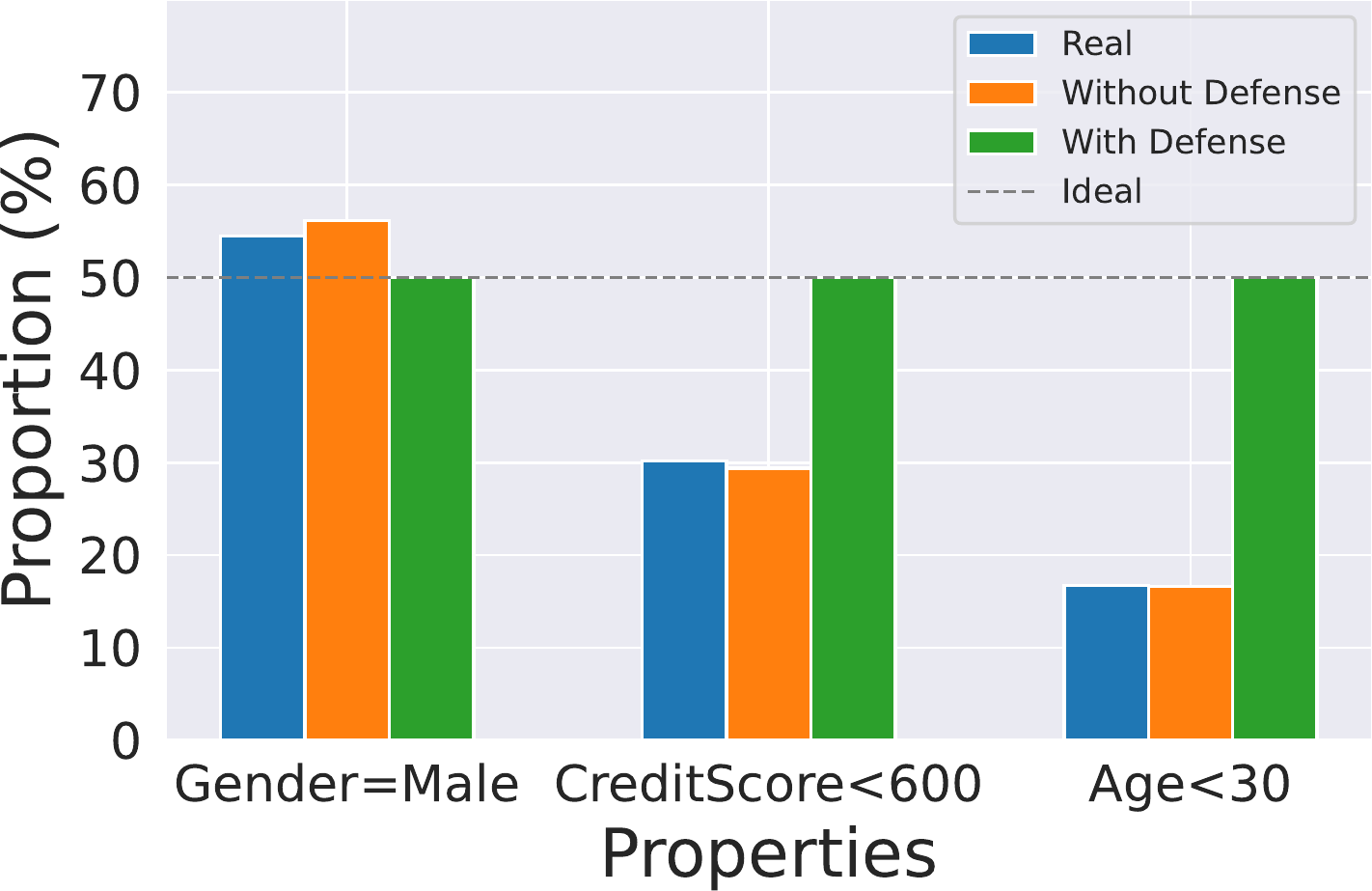}
		\label{fig:defense_tabddpm_churn_single}
	}
	\subfigure[Cardio.]{
		\includegraphics[width=0.55\columnwidth]{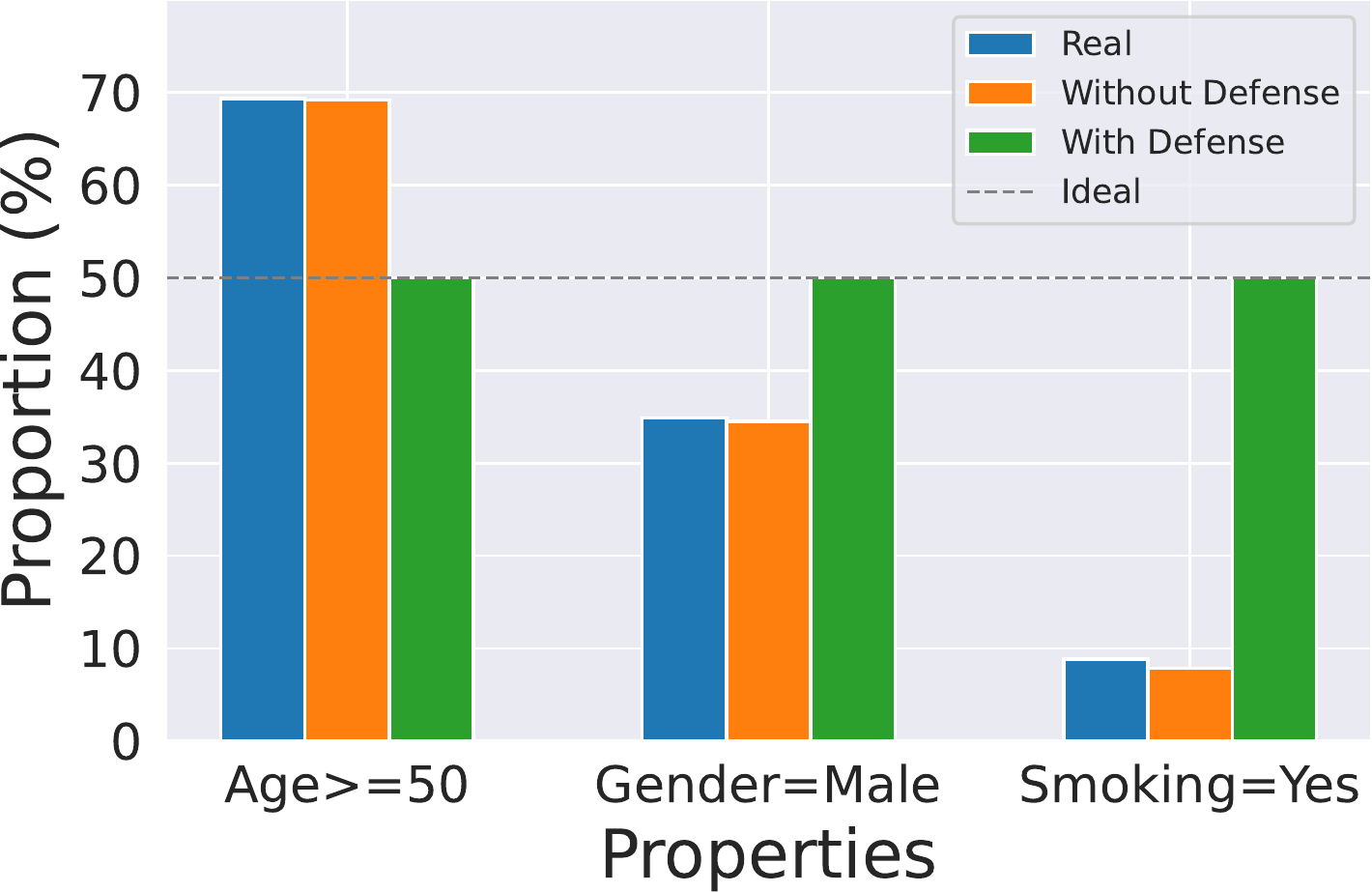}
		\label{fig:defense_tabddpm_cardio_single}
	}	
	\caption{Defense performance for the binary property on TabDDPM across different datasets.}
	\label{fig:def_perf_tab_single}
\end{figure*} 

\smallskip\noindent
\textbf{Details for different properties.}
For different numbers of properties, the generated samples are calculated as follows.

\smallskip\noindent
\textbf{$\bullet$~Single property.}
Given a hyperplane~$\mathbf h$ obtained in phase I, and a sample~$x_t$ at step $t$, we can get $x_t'$:
\begin{equation}
x_t' = x_t + \alpha \mathbf h.
	\label{eq:single}
\end{equation} 
%$x_t' = x_t + \alpha \mathbf h$. 
$\alpha$ is a hyperparameter, and a value $\alpha >0$ means that a positive sample~$x_{t}^{'+}$ is obtained and it has this property.
A value $\alpha <0$ means a negative sample~$x_{t}^{'-}$ is obtained and does not have this property.

For one binary property, such as Gender = \{Male, Female\}, we can synthesize the male sample~$x_{t}^{'+}$ and the female sample~$x_{t}^{'-}$ by choosing a value $\alpha>0$ and a value $\alpha<0$, respectively.
In this work, depending on different base samplers and diffusion models, we choose different $\alpha$.
We provide the details in Appendix~\ref{ssec:Details_Algorithm}.
For other types of properties, we can transform them into the case of binary properties and we detail them as follows.

For one multi-categorical property that has more than two values, such as Martial-status = \{Married, Divorced, Never-married\}, we transform this case into the case of one binary property via the one-vs-rest strategy.
Specifically, given a multi-categorical property that has $k$ values, we can still obtain two types of samples: $x_{t}^{'+}$ that contains this property and $x_{t}^{'-}$ that does not contain this property, which is the same with the case of the binary property.
However, we only choose the sample~$x_{t}^{'+}$ containing this property.
This is because samples~$x_{t}^{'-}$ that do not contain this property have many cases.
For instance, for the property Martial-status=Divorced,  $x_{t}^{'-}$ includes samples that have the properties Martial-status=Divorced and Martial-status=Never-married.
However, we can ensure that the sample~$x_{t}^{'+}$ only has this property Martial-status=Divorced.
Thus, a multi-categorical property that has $k$ values requires $k$ hyperplanes and for each hyperplane, we only choose samples~$x_{t}^{'+}$.

For one numeric property, such as age $\in (0,100)$ and CreditScore $\in (0,1000)$, we can also transform it into one categorical property by splitting its values into $k$ parts.
For example, the property age $\in (0,100)$ can be transformed into one binary property, i.e. age$<$30 and age$>=$30.
Then, the subsequent process is the same as the case in binary or multi-categorical property.

\begin{table}
	\centering
	\caption{Utility performance on TabDDPM.}
	\label{tab:def_tabddpm_utility}
	\renewcommand{\arraystretch}{1.0}
\scalebox{0.65}{	
	
\begin{tabular}{l|l|l|lc|lc} 
	\toprule
	\multirow{2}{*}{Dataset} & \multirow{2}{*}{Property} &                                                                                      & \multicolumn{2}{c|}{Without Defense}                                                                                                             & \multicolumn{2}{c}{With Defense}                                                                                                              \\ 

	&                           & \begin{tabular}[c]{@{}l@{}}Real\\Prop. (\%)\end{tabular}                             & \begin{tabular}[c]{@{}l@{}}Inferred\\Prop. (\%)\end{tabular}                         & \begin{tabular}[c]{@{}c@{}}Utility \\F1 (\%) $\uparrow$\end{tabular} & \begin{tabular}[c]{@{}l@{}}Inferred\\Prop. (\%)\end{tabular}                      & \begin{tabular}[c]{@{}c@{}}Utility \\F1 (\%) $\uparrow$\end{tabular}  \\ 
	\hline
	\multirow{6}{*}{Adult}   & Gender=Male               & 67.05                                                                                & 67.43                                                                                & 79.58                                                     & 50.00                                                                             & 79.18                                                      \\ 
	
	& Age$<$30                  & 29.81                                                                                & 29.46                                                                                & 79.58                                                     & 50.00                                                                             & 79.16                                                      \\ 
	\cline{2-7}
	& Race=\{1,2,3,4,5\}        & \begin{tabular}[c]{@{}l@{}}85.34, \\9.64, \\3.21, \\0.98, \\0.83\end{tabular}        & \begin{tabular}[c]{@{}l@{}}88.15,\\8.88,\\1.90,\\0.62, \\0.45\end{tabular}           & 79.58                                                     & \begin{tabular}[c]{@{}l@{}}20.00,\\20.00,\\20.00,\\20.00,\\20.00\end{tabular}     & 78.59                                                      \\ 
	\hline\hline
	\multirow{7}{*}{Churn}   & Gender=Male               & 54.47                                                                                & 56.16                                                                                & 75.75                                                     & 50.00                                                                             & 74.19                                                      \\ 
	
	& CreditScore$<$600         & 30.19                                                                                & 29.42                                                                                & 75.75                                                     & 50.00                                                                             & 75.12                                                      \\ 
	
	& Age$<$30                  & 16.75                                                                                & 16.64                                                                                & 75.75                                                     & 50.00                                                                             & 69.33                                                      \\ 
	\cline{2-7}
	& Gender+Geography          & \begin{tabular}[c]{@{}l@{}}(54.47, \\45.53)+\\(49.64, \\25.20, \\25.16)\end{tabular} & \begin{tabular}[c]{@{}l@{}}(56.16, \\43.84)+\\(51.30, \\25.71, \\22.99)\end{tabular} & 75.75                                                     & \begin{tabular}[c]{@{}l@{}}(50.00,\\50.00)+\\(33.34,\\33.33,\\33.33)\end{tabular} & 72.96                                                      \\ 
	\hline\hline
	\multirow{3}{*}{Cardio}  & Gender=Male               & 34.87                                                                                & 34.50                                                                                & 73.54                                                     & 50.00                                                                             & 73.67                                                      \\ 
	
	& Age$\geq$50               & 69.34                                                                                & 69.18                                                                                & 73.54                                                     & 50.00                                                                             & 72.78                                                      \\ 

	& Smoking=Yes               & 8.84                                                                                 & 7.84                                                                                 & 73.54                                                     & 50.00                                                                             & 73.44                                                      \\
	\bottomrule
\end{tabular}

}
\end{table}

\smallskip\noindent
\textbf{$\bullet$~Multiple properties.} 
When there are multiple sensitive properties, the key idea is that we manipulate one property while keeping others unchanged. 
That is, we need to find a new hyperplane that is orthogonal to other hyperplanes.
Take two properties as an example, we first manipulate the first one and manipulate the second condition on the first one.
In this way, we can get generated samples with balanced properties.
To be more specific, given two hyperplanes~$\mathbf{h_1}$ and $\mathbf{h_2}$ obtained in phase I and a sample~$x_t$.
We first get a new hyperplane:
\begin{equation}
	\scalemath{0.83}{
	\mathbf{h_2'} = \mathbf{h_2-(h_2^Th_1)h_1},
}
	\label{eq:new_hyper}
\end{equation} 
where $\mathbf{(h_2^Th_1)h_1}$ is the projection of $\mathbf h_2$ onto $\mathbf h_1$.
The new hyperplane~$\mathbf{h_2'}$ equals the vector difference between $\mathbf h_2$ and the projection of $\mathbf h_2$ onto $\mathbf h_1$. 
Therefore, $\mathbf{h_2'}$ is orthogonal to $\mathbf{h_1}$.
In other words, $\mathbf{h_2'}$
can achieve that the second property is changed without impacting the first property.
Then, based on Equation~\ref{eq:single}, we can get $x_t'$ through $x_t$ and $\mathbf{h_1}$.
Given $x_t'$ and $\mathbf{h_2'}$, we can obtain $x_t''$.
Here, we require that the hyperplanes of multiple properties are independent, i.e. they are not in the same space. Otherwise, it is hard to find a hyperplane to guarantee that manipulating samples in this hyperplane does not affect the others.

%------------------------------------------
\subsection{Experimental Setup}
%------------------------------------------
We apply our defense on the stochastic sampling method for TabDDPM in tabular data generation.
We also apply our defense on two types of base samplers for image generation: one PC sampler for stochastic sampling and one DPM sampler for deterministic sampling.
We directly use trained diffusion models that are attacked in Section~\ref{sec:method} as the protected target models.
In our defense, we use the library Sklearn to implement Linear SVM.
For image data, we also utilize trained property classifiers used in attacking to predict scores~$S$. 
We choose different diffusion steps for different samplers of diffusion models to manipulate. 
We summarize them in Appendix~\ref{ssec:Details_Algorithm}.
For tabular data generation, 50k generated samples are used for computing F1 scores and defense performance.
For image generation, 500 generated samples for stochastic sampling and 50k generated samples for deterministic sampling are used for computing FID scores and defense performance.
In terms of defense performance, we report the predicted values and the absolute difference. However, in the context of defense, absolute difference refers to the absolute difference between the predicted value and the predefined value.

%------------------------------------------
\subsection{Defense Results on Tabular Data}
\label{ssec:Defenses_res_tabular}
%------------------------------------------

\begin{figure}[!t]
	\centering
	\includegraphics[width=0.60\linewidth]{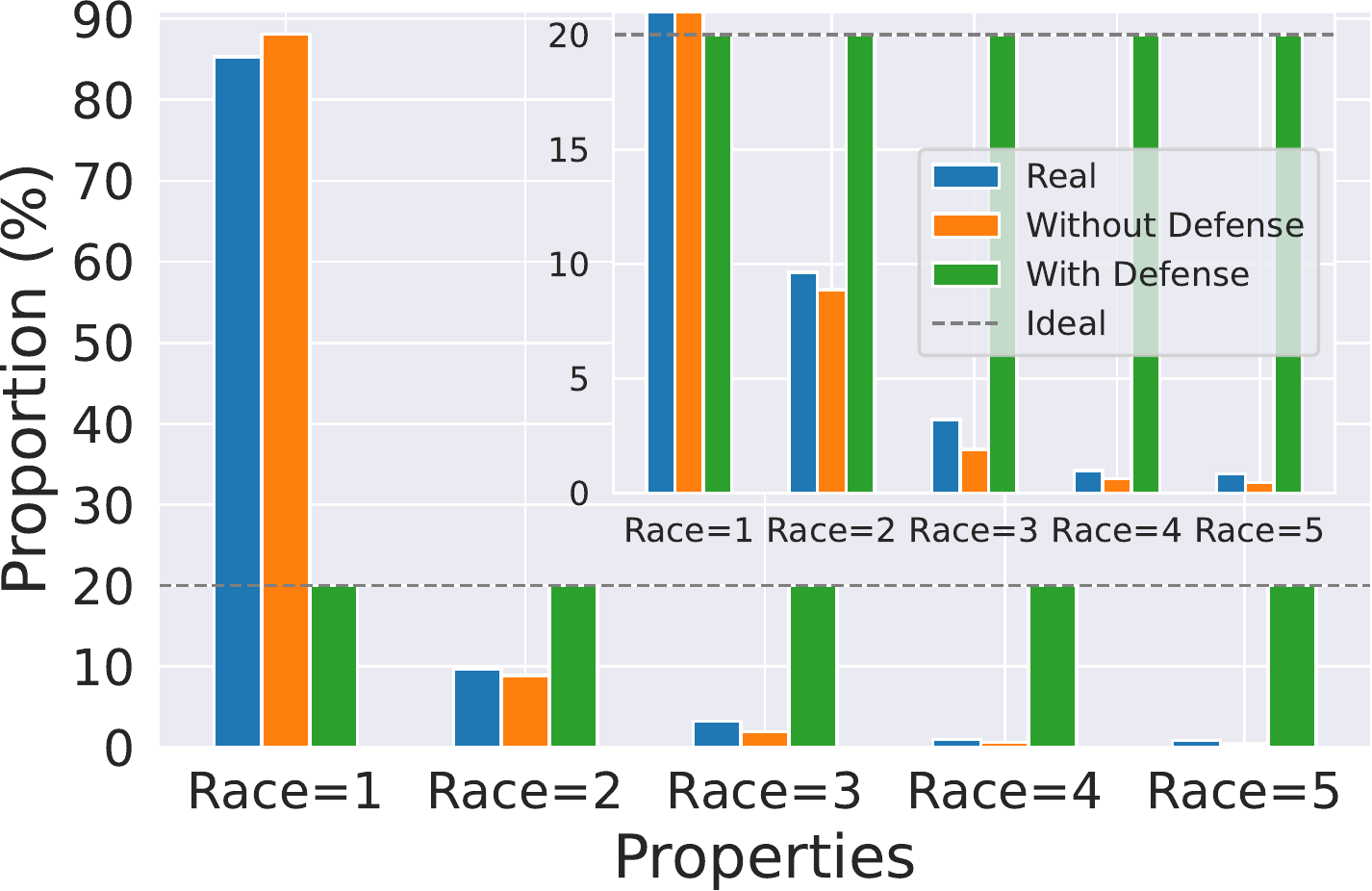}
	\caption{Defense performance for the multi-categorical property on TabDDPM on Adult. Here, Race = \{1, 2, 3, 4, 5\} refers to Race = \{White, Black, Asian-Pac-Islander, Amer-Indian-Eskimo, Others\}.}
	\label{fig:defense_tabddpm_adult_multi_cat_race}
\end{figure} 

\noindent
\textbf{Defense performance on different properties. }  
Figure~\ref{fig:def_perf_tab_single} shows the defense performance for different binary properties on TabDDPM across different datasets.
The blue bar is the real proportion of each property in a dataset.
The orange bar is the inferred proportion for TabDDPM without defense, while the green bar is the inferred proportion for TabDDPM with defense, i.e. {\sf PriSampler} on TabDDPM.
The grey dashed line is the ideal proportion, i.e. the predefined value~$\gamma$.
Overall, we can observe that our defense can achieve perfect performance.
Our {\sf PriSampler} can make the inferred proportions of all properties across three datasets remain at a predefined value of 50\%.

Figure~\ref{fig:defense_tabddpm_adult_multi_cat_race} shows the defense performance for multi-categorical properties.
Here, the protected property is race which has five values with different proportions. Thus, the predefined value~$\gamma$ is set as 20\%, which ensures each property is equal.
Again, we can see that our {\sf PriSampler}  can still perform perfectly on all proportions.
For the property Race=White whose real proportion is about 85\%, our method can still generate the samples with a desired proportion of 20\%.
Even for the extremely small proportion of properties, such as Race=Amer-Indian-Eskimo with the real proportion of 0.98\%, our defense still makes its inferred proportions remain 20\%.
This also indicates that our method can make synthetic data release fair in terms of the property race.
In addition, as shown in Table~\ref{tab:def_tabddpm_utility}, its F1 score with defense is 78.59\%, while the original F1 score without defense is 79.58\%. It means that our defense does not significantly compromise the model utility.
Additional results on the properties martial-status and geography are shown in Figure~\ref{fig:def_tabddpm_mul} in Appendix.

\begin{figure}[!t]
	\centering
	\includegraphics[width=0.60\linewidth]{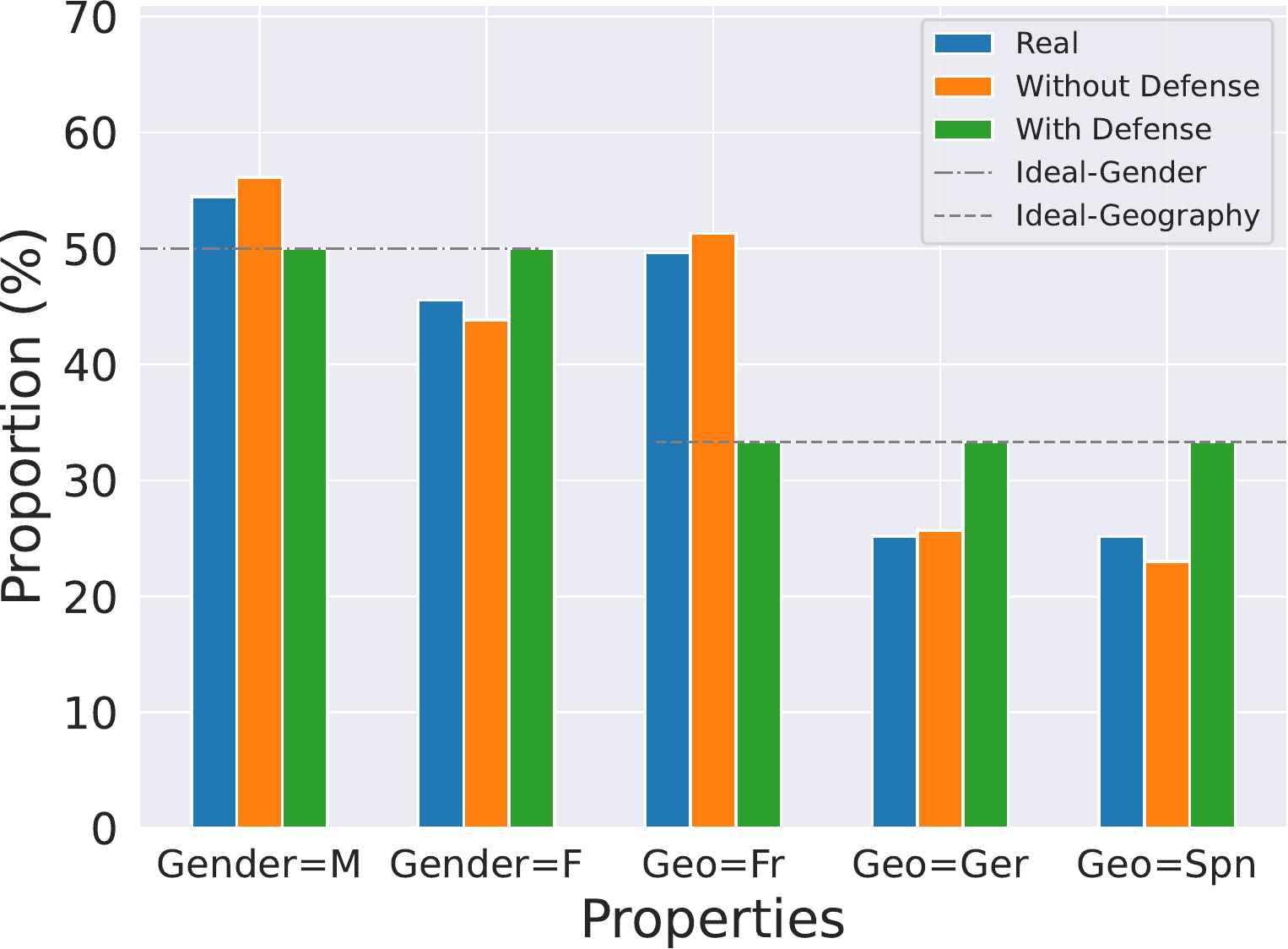}
	\caption{Defense performance for multiple properties on TabDDPM on Churn. Gender and Geography are protected properties. Here, Gender = \{M, F\} refers to Gender = \{Male, Female\}, Geo = \{Fr, Ger, Spn\} refers to Geography = \{France, Germany, Spain\}.}
	\label{fig:defense_tabddpm_churn_geographygender}
\end{figure}

\begin{figure*}[!t]
	\centering
	
	\subfigure[DDPM.]{
		\includegraphics[width=0.48\columnwidth]{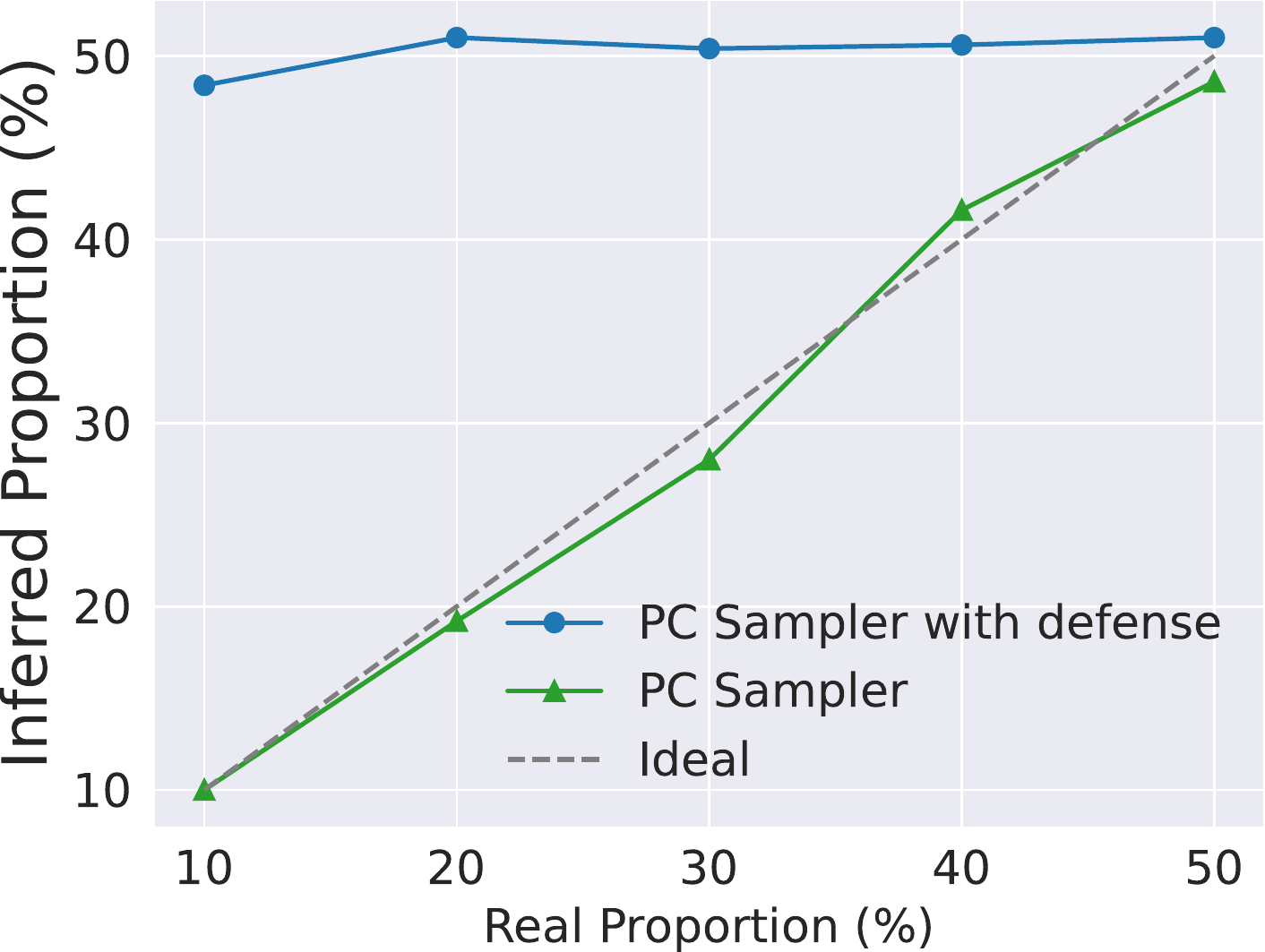}
		\label{fig:defense_sde_ddpm_celeba1k_male}
	}
	\subfigure[SMLD.]{
		\includegraphics[width=0.48\columnwidth]{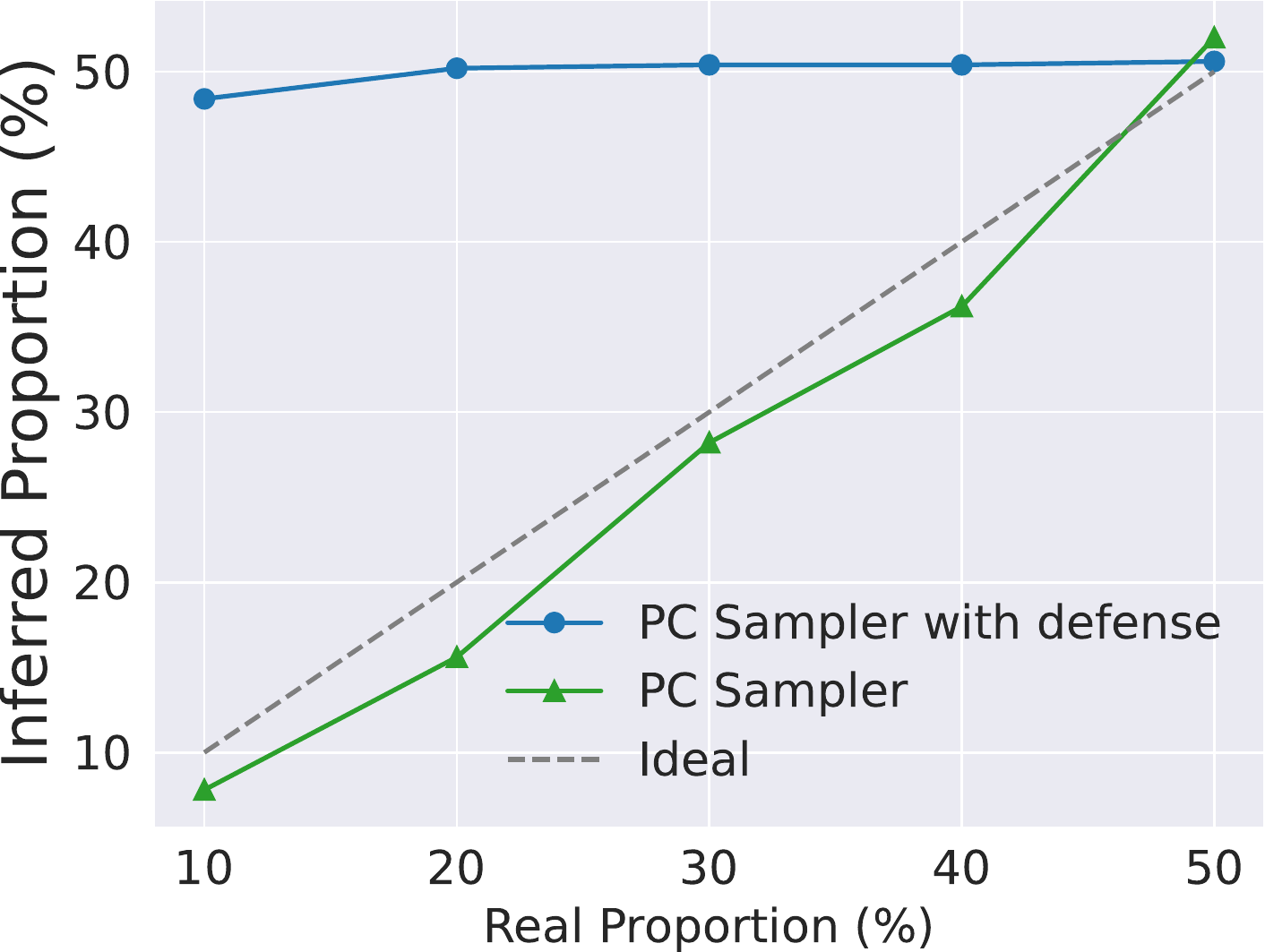}
		\label{fig:defense_sde_smld_celeba50k_male}
	}
	\subfigure[VPSDE.]{
		\includegraphics[width=0.48\columnwidth]{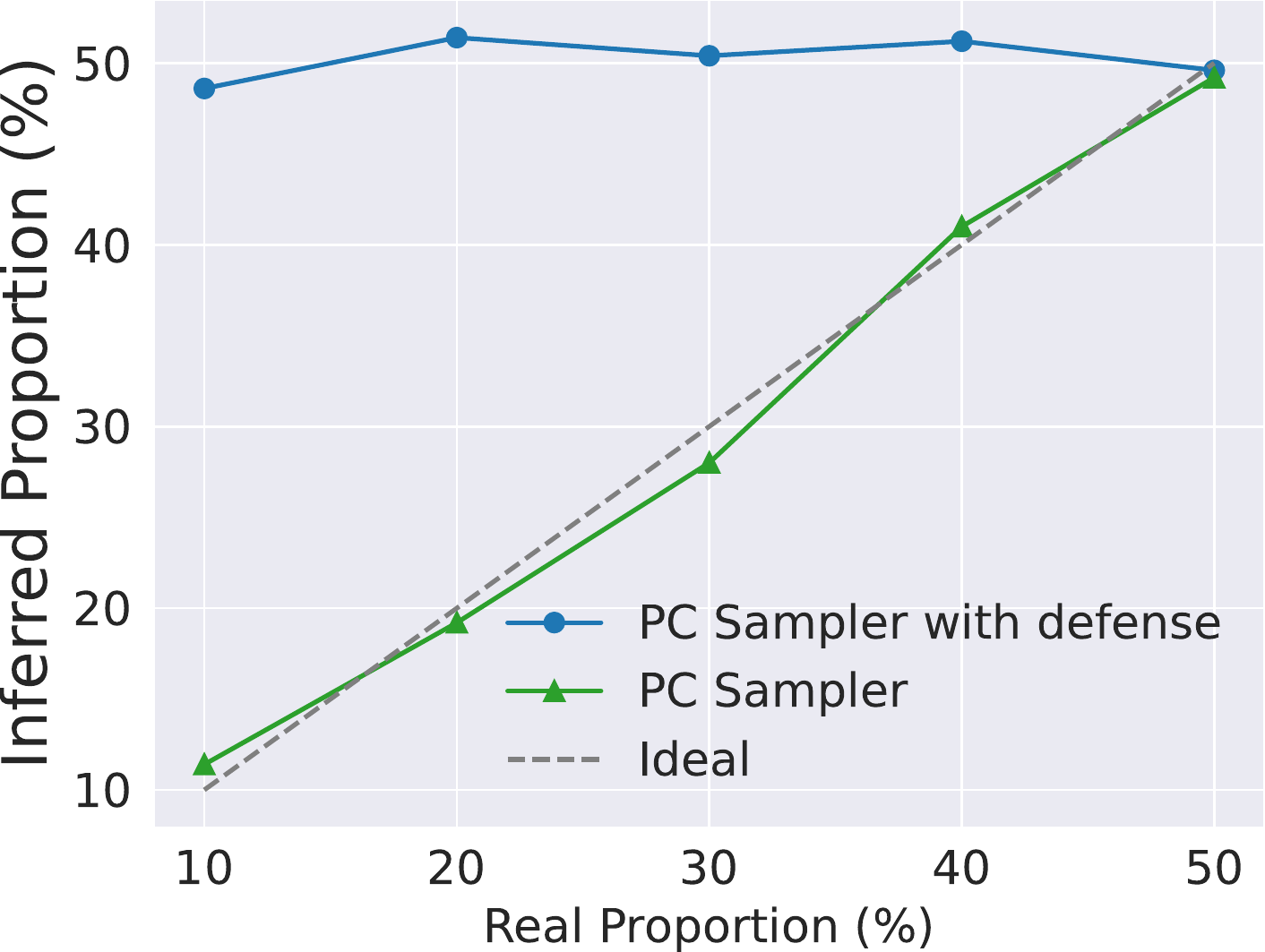}
		\label{fig:defense_sde_vpsde_celeba1k_male}
	}	
	\subfigure[VESDE.]{
		\includegraphics[width=0.48\columnwidth]{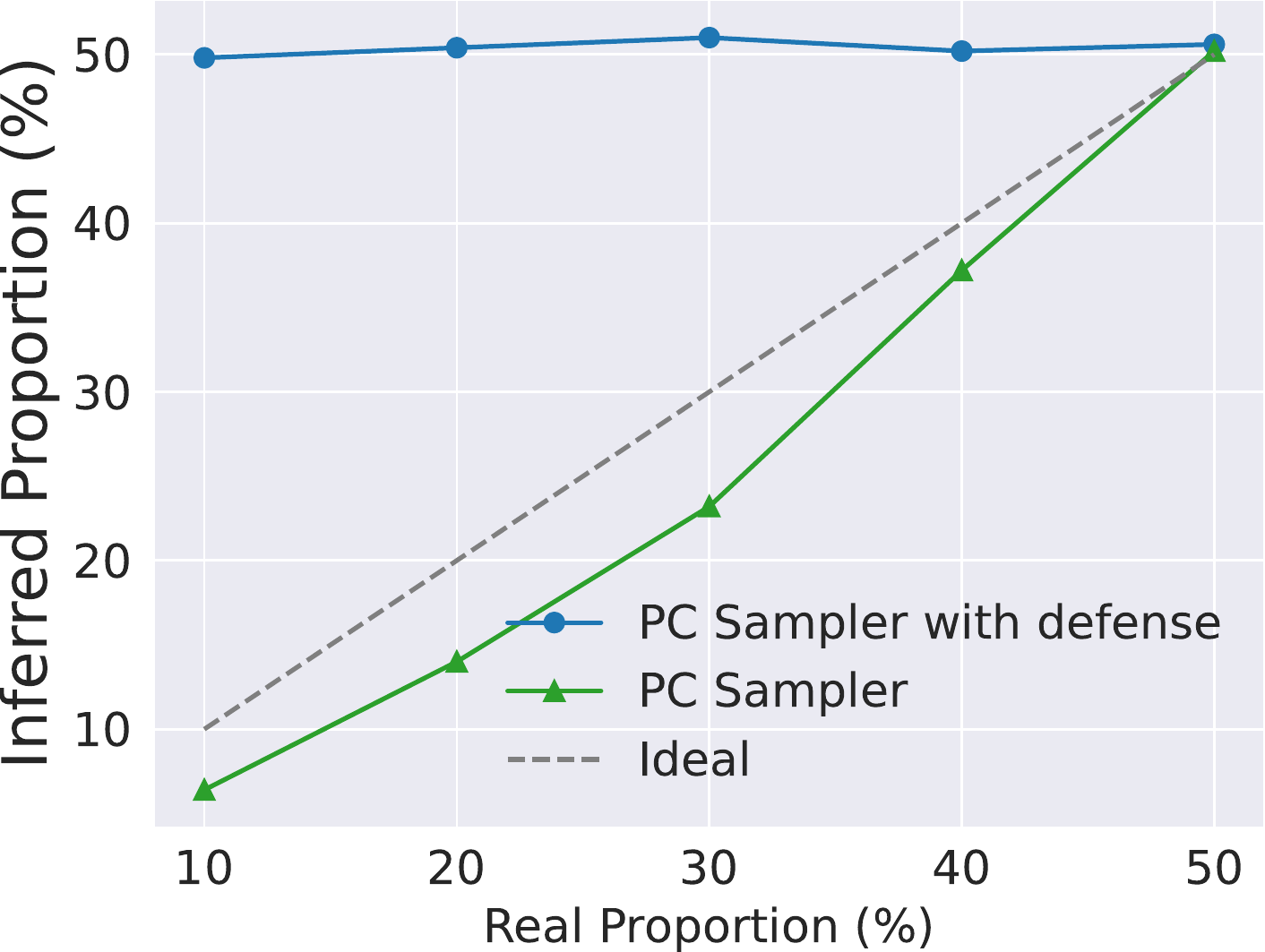}
		\label{fig:defense_sde_vesde_celeba50k_male}
	}
	\caption{Defense performance on the PC sampler. Quantitative results are reported in Table~\ref{tab:def_perf} in Appendix.}
	\label{fig:def_perf_pc}
\end{figure*}

\smallskip\noindent
\textbf{Defense performance on more than one property. }  
Figure~\ref{fig:defense_tabddpm_churn_geographygender} shows the defense performance for multiple properties on the TabDDPM on Churn.
The protected properties are Gender and Geography which have two and three values respectively. Thus, the corresponding ideal proportions are 50\%, and 33.33\%.
Again, we can see the inferred proportions of all properties can remain predefined values under our defense method.
In contrast, the model without defense is vulnerable to property inference attacks and adversaries can precisely infer the proportions of all properties.
We also present the results on the properties Gender and Race in Figure~\ref{fig:defense_tabddpm_adult_racegender} in Appendix.

\smallskip\noindent
\textbf{Utility performance.}
Table~\ref{tab:def_tabddpm_utility} describes the utility performance for TabDDPM across different properties and datasets. Overall, our defense cannot significantly impact the model utility and can provide excellent protection.
For instance, for the property Gender=Male on Adult, the utility performance with defense is 79.18\%, which only drops 0.4\% compared to that without defense. However, we can achieve a privacy-preserving and fair data generation, i.e. a proportion of 50\% for the protected property Gender=Male.

%------------------------------------------
\subsection{Defense Results on Image Data}
%------------------------------------------
\label{ssec:Defenses_res_image}

\noindent
\textbf{Defense performance on different samplers.} 
Figure~\ref{fig:def_perf_pc} and Figure~\ref{fig:def_perf_dpm} present our defense performance to protect the sensitive property Gender=Male for the PC sampler and the DPM sampler, respectively.
Overall, our defense can achieve excellent performance in image generation scenarios. 
Even if the real proportion is 10\%, our method can make adversaries get an inferred proportion of 50\%.
Table~\ref{tab:def_perf} in Appendix shows the corresponding quantitative results.

Table~\ref{tab:def_perf_sum} summarizes the results for each sampler among different diffusion models.
For the single property male, the average absolute difference is 0.75\% for the PC sampler and 0.64\% for the DPM sampler.
It indicates that our defense can control the error of an inferred proportion below 1.00\%.

\begin{figure}[!t]
	\centering
	
	\subfigure[DDPM.]{
		\includegraphics[width=0.45\columnwidth]{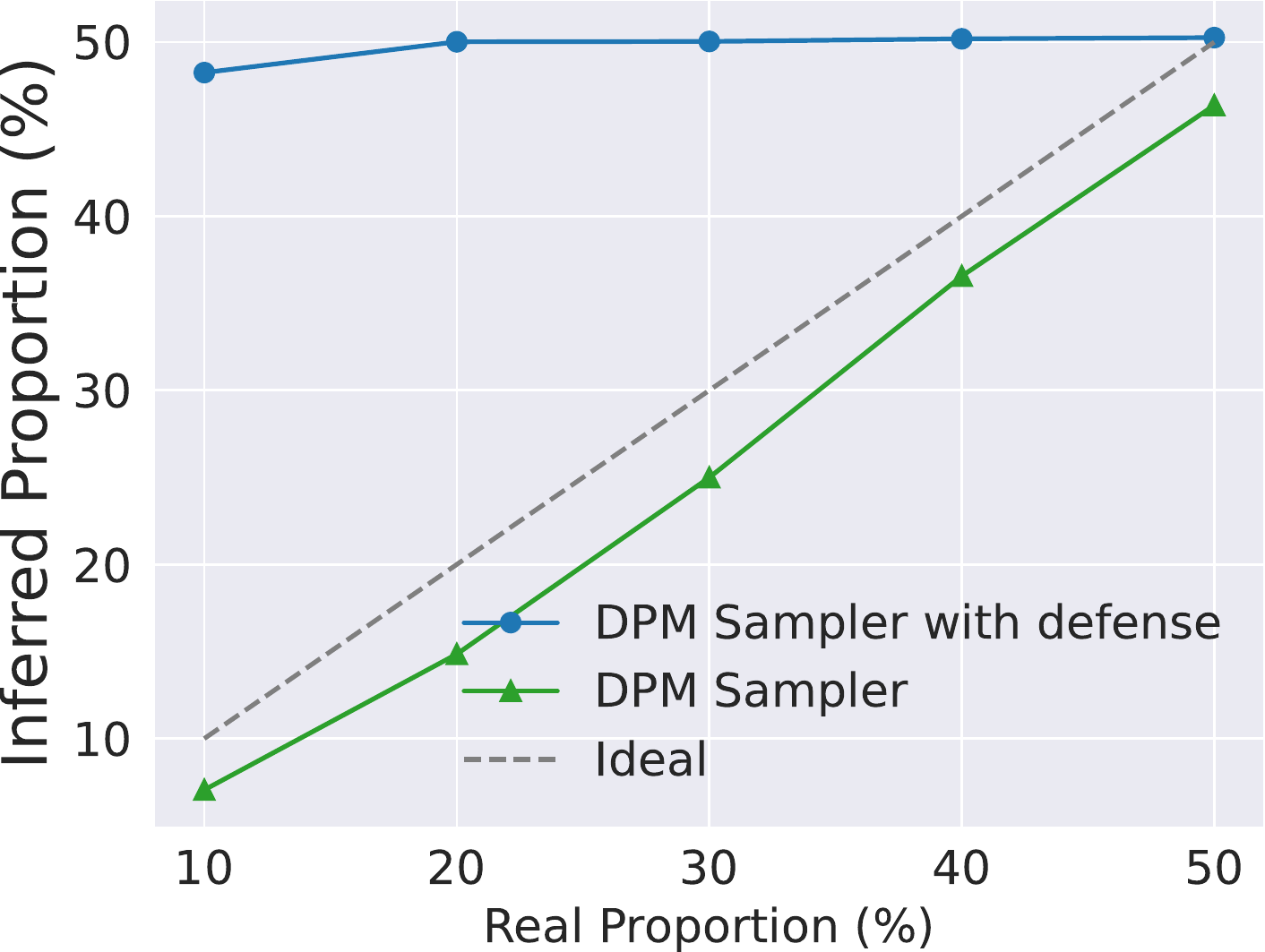}
		\label{fig:defense_dpm_ddpm_celeba1k_male}
	}
	\subfigure[VPSDE.]{
		\includegraphics[width=0.45\columnwidth]{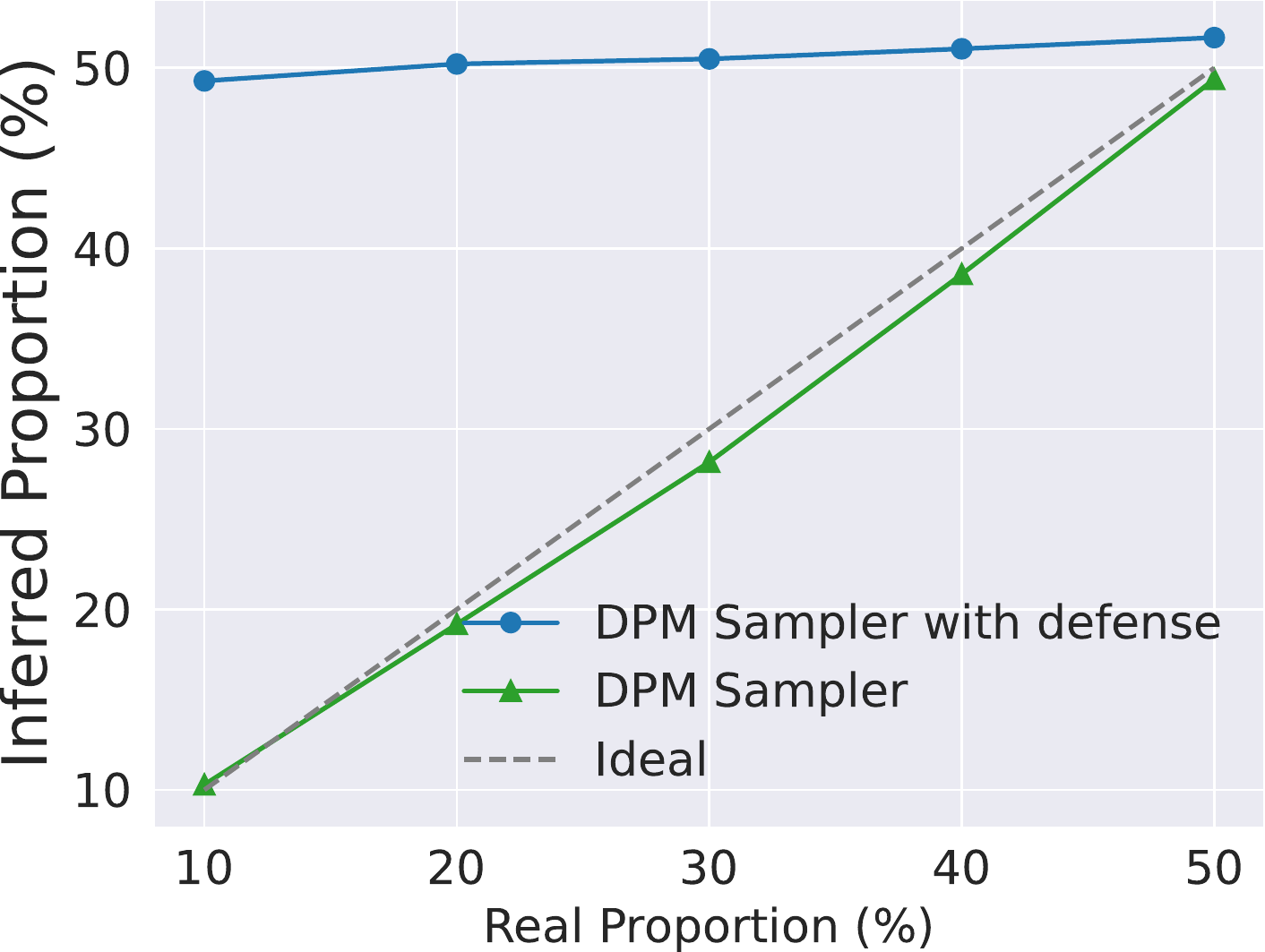}
		\label{fig:defense_dpm_vpsde_celeba1k_male}
	}
	
	\caption{Defense performance on the DPM sampler. }
	\label{fig:def_perf_dpm}
\end{figure} 

\smallskip\noindent
\textbf{Defense performance on different diffusion models.} 
Figure~\ref{fig:def_perf_pc} and Figure~\ref{fig:def_perf_dpm} present our defense performance on four types of diffusion models.
Similarly, we can observe the inferred proportions almost remain 50\% for diffusion models trained on different datasets.
This means our method can be effectively applied to different diffusion models.

\begin{figure}[!t]
	\centering
	
	\subfigure[PC sampler.]{
		\includegraphics[width=0.45\columnwidth]{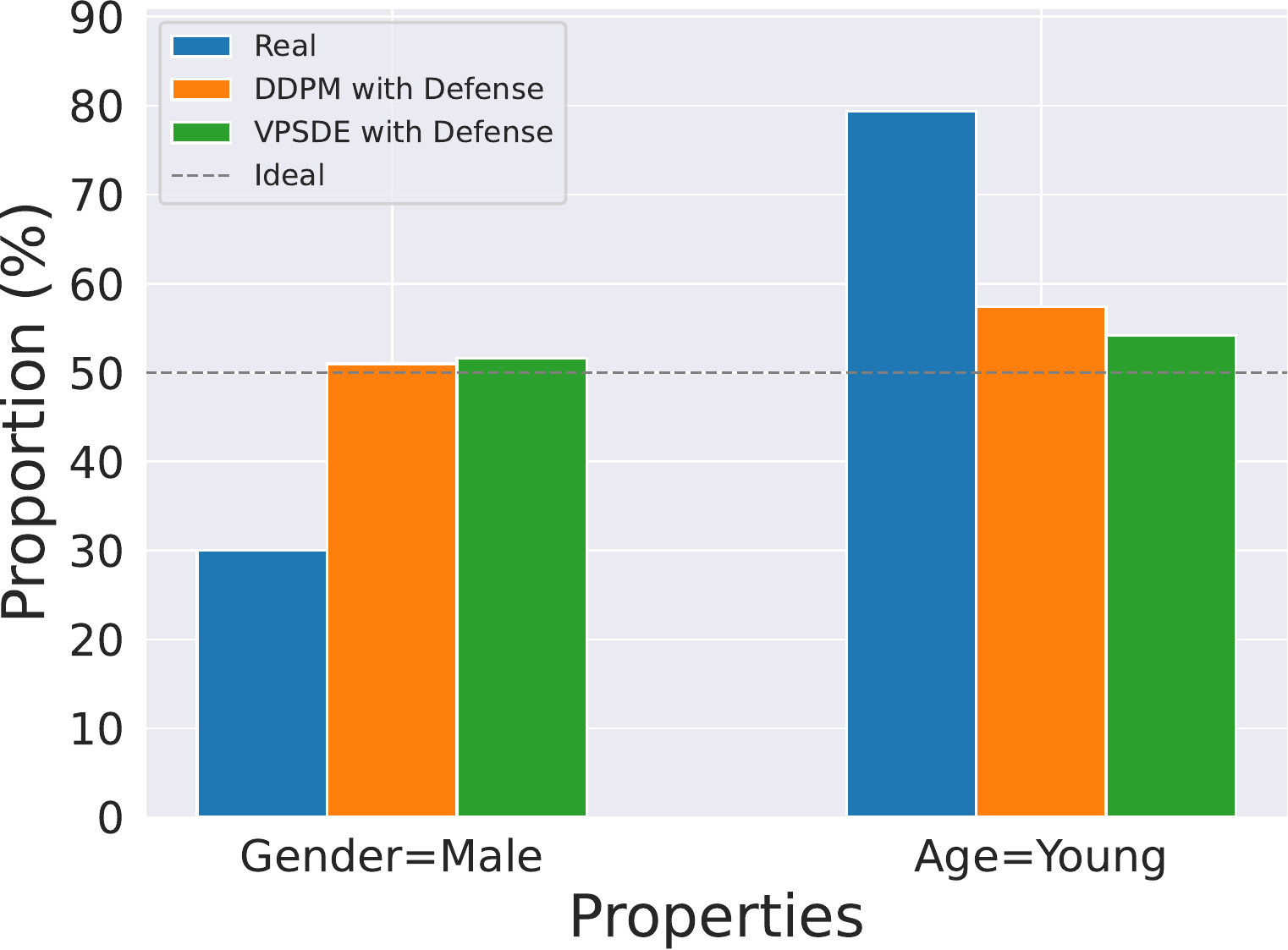}
		\label{fig:defense_two_sde}
	}
	\subfigure[DPM sampler.]{
		\includegraphics[width=0.45\columnwidth]{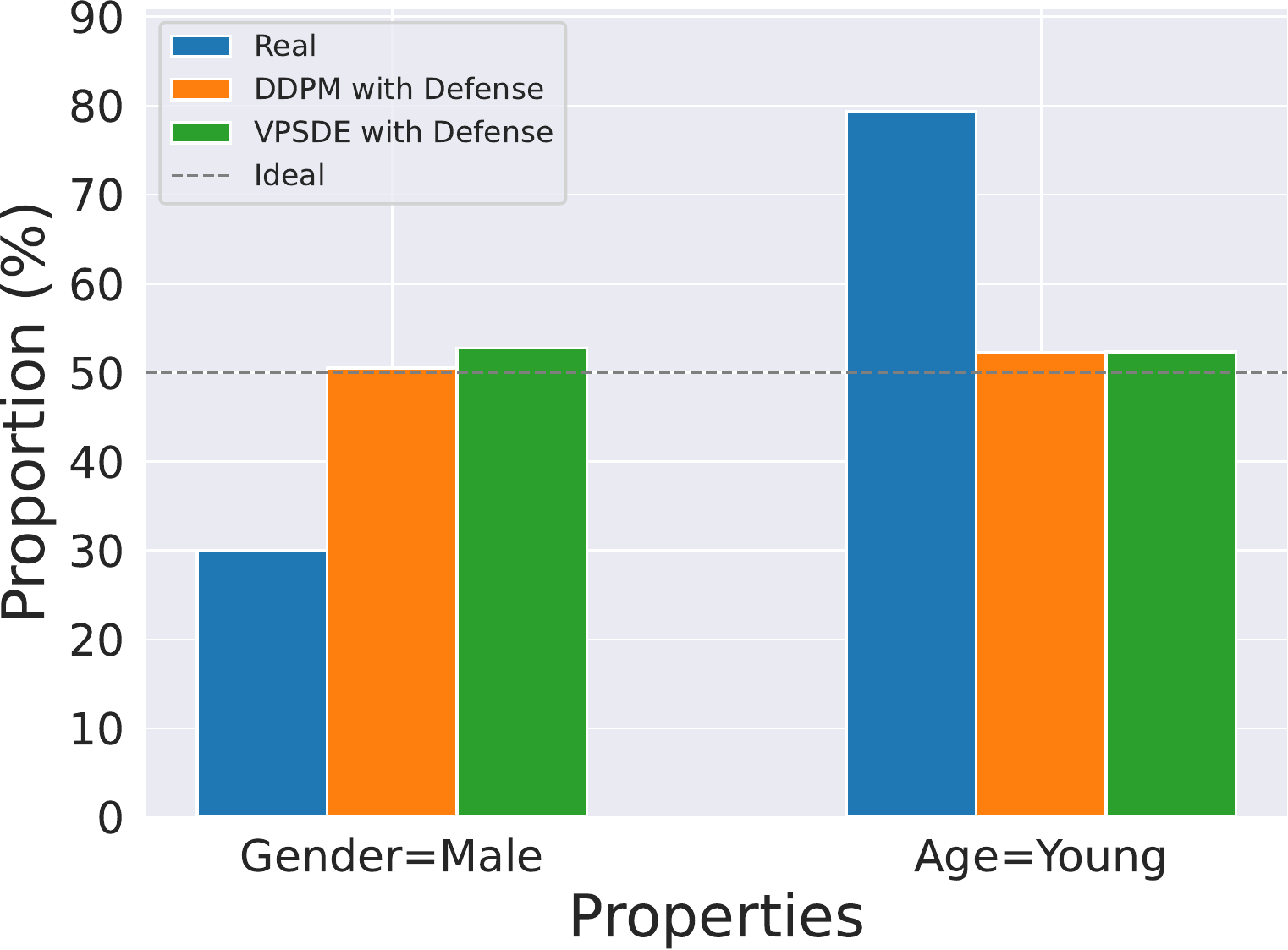}
		\label{fig:defense_two_dpm}
	}
	
	\caption{Defense performance for multiple properties. }
	\label{fig:def_perf_two}
\end{figure}

\smallskip\noindent
\textbf{Defense performance on more than one property.} 
Figure~\ref{fig:def_perf_two}
shows the defense performance to protect two sensitive properties Gender and Age.
Here, we choose the models trained on CelebA-1k-30\% as target models.
That is, the real proportion of the property Gender=Male is 30\%. The corresponding real proportion of the sensitive property Age=Young is 79.4\%.
Overall, we can observe the inferred proportions of both properties are about 50\%, no matter the larger proportion or the smaller proportion.
Table~\ref{tab:def_perf_two} in Appendix shows the corresponding quantitative results.

Table~\ref{tab:def_perf_sum} summarizes the result for Gender+Age properties on both samplers.
We can see that the average absolute difference is 3.55\% for the PC sampler and 1.97\% for the DPM sampler.
We also note that the absolute difference on two properties is larger than that on the single property.
One of the reasons is that these properties are entangled together in image generation.
As a result, it might lead the hyperplane not to completely separate these properties.
We will take this as our future work, i.e. designing disentangled generative algorithms for diffusion models from the perspective of the training process to improve defense performance on multiple properties in image generation.

\smallskip\noindent
\textbf{Utility performance.}
Table~\ref{tab:def_img_utility} presents the utility performance of DDPM trained on CelebA. 
The utility of all models in image generation is also shown along with the qualitative results in Appendix~\ref{ssec:add_res_defense}.
Overall, we can see that the utility performance drops on all properties for PC and DPM samplers. 
This is because our defense restricts samplers to synthesizing images in the sensitive property space, which reduces the diversity of images to some degree.
Thus, our defense induces higher FID scores where FID scores consider the quality and diversity of generated images.
However, the quality of generated images under our defense is still realistic, such as Figure~\ref{fig:def_perf_cmp_partial}(b) for DDPM on CelebA-1k-30\%.
In  appendix~\ref{ssec:add_res_defense}, we present additional results about defense performance about the number of generated samples and different diffusion steps.

\begin{table}[]
	\centering
	\caption{Summary of defense performances in image generation. Here, we report the average absolute difference~(with standard deviation in parentheses) and the best and worst absolute difference.}
	\label{tab:def_perf_sum}
	\renewcommand{\arraystretch}{1.1}
	\scalebox{0.75}{
		
		\begin{tabular}{ll|ccc} 
			\toprule
			Property                    & Sampler & Average~(\%) & Best~(\%) & Worst~(\%)  \\ 
			\hline
			\multirow{2}{*}{Gende=Male} & PC      & 0.75 (0.48)  & 0.20      & 1.60        \\
			& DPM     & 0.64 (0.65)  & 0.02      & 1.75        \\ 
			\hline
			\multirow{2}{*}{Gender+Age} & PC      & 3.55 (2.92)  & 1.00      & 7.40        \\
			& DPM     & 1.97 (0.98)  & 0.53      & 2.73        \\
			\bottomrule
		\end{tabular}
		
	}
\end{table}

\subsection{Comparison with Differential Privacy}
\label{ssec:cmp_dp}
Differential privacy~\cite{abadi2016deep,dwork2008differential} is a common measure for defending against privacy attacks.
In this subsection, we explore the feasibility of differential privacy to defend against property inference attacks.
Furthermore, we make a comparison with our method {\sf PriSampler}.

We use differentially private diffusion models~(DPDMs) proposed by Dockhorn et. al~\cite{dockhorn2022differentially}, because they are the first to apply differentially private stochastic gradient descent~(DPSGD)~\cite{abadi2016deep} to diffusion models and can generate meaningful images.
We do not explore differentially private diffusion models for tabular data due to the lack of such works while DPDMs are specifically designed for image generation.
We adopt their suggested hyperparameters to train DPDMs.
We set the number of epochs and batch sizes as 100, and 128 respectively. 
The image size is fixed at 64, and we choose different sizes of training sets, i.e. CelebA-1k-30\% and CelebA-50k-30\%, and different privacy budgets~$\epsilon$, i.e. $\epsilon =10$ and $\epsilon=50$. We fix $\delta$ as $10^{-6}$ for all models.
Here, we synthesize samples by stochastic sampling because DPDMs~\cite{dockhorn2022differentially} analyze that it can obtain better FID scores under differential privacy conditions.

Table~\ref{tab:def_perf_cmp} presents the comparison between our method and DPDM.
Figure~\ref{fig:def_perf_cmp_partial} visually shows synthetic samples.
For the CelebA-1k-30\% dataset, DPDM almost cannot generate meaningful images, which leads to an FID score of 446.35.
In contrast, our method can achieve a 40.00 FID value and the inferred proportion is 50.50\%. 
Figure~\ref{fig:def_perf_cmp_partial}(b) also shows the good quality of synthetic samples for our method {\sf PriSampler}.
For the CelebA-50k-30\% dataset, we can clearly see that the generated samples from DPDM only have a vague shape of the human face.
Even if we increase the privacy budget ~$\epsilon$ from 10 to 50, the synthetic human face samples are still distorted, although we can see that FID decreases from 121.56 to 103.64.
Here, note that $\epsilon = 10$ is usually considered as low amounts of privacy.
We also observe that the inferred proportion for DPDM under $\epsilon = 10$ is 44.00\%, while that for DPDM under $\epsilon = 50$ is 24.20\%.
It indicates that DPDM under smaller privacy budgets can disguise the real proportion of certain properties to some extent.
However, the quality of the generated samples is too vague. 
In contrast, our method {\sf PriSampler} can still synthesize meaningful samples with a balanced proportion.

\begin{table}
	\centering
	\caption{Utility of DDPM trained on CelebA.}
	\label{tab:def_img_utility}
	\renewcommand{\arraystretch}{1.1}
\scalebox{0.75}{

\begin{tabular}{cl|c|cr|cr} 
	\toprule
	&                              &                                                          & \multicolumn{2}{r}{Without Defense}                                                                                                       & \multicolumn{2}{|c}{With\ Defense}                                                                                                               \\
	Sampler & \multicolumn{1}{c|}{Property} & \begin{tabular}[c]{@{}c@{}}Real\\Prop. (\%)\end{tabular} & \begin{tabular}[c]{@{}c@{}}Inferred\\Prop. (\%)\end{tabular} & \multicolumn{1}{c|}{\begin{tabular}[c]{@{}c@{}}Utility \\FID~$\downarrow$\end{tabular}} & \begin{tabular}[c]{@{}c@{}}Inferred\\Prop. (\%)\end{tabular} & \multicolumn{1}{c}{\begin{tabular}[c]{@{}c@{}}Utility \\FID~$\downarrow$\end{tabular}}  \\
	\hline
	PC      & Gender=Male                  & 20.00                                                    & 19.20                                                        & 24.85                                                                      & 51.00                                                        & 46.80                                                                      \\
	& Gender=Male                  & 30.00                                                    & 28.00                                                        & 25.55                                                                      & 50.40                                                        & 40.00                                                                      \\
	& Gender+Age                   & 30.00, 79.40                                             & 28.00, 79.40                                                 & 25.55                                                                      & 51.00, 57.40                                                 & 48.74                                                                      \\
	\hline
	DPM     & Gender=Male                  & 20.00                                                    & 14.85                                                        & 23.47                                                                      & 50.02                                                        & 47.87                                                                      \\
	& Gender=Male                  & 30.00                                                    & 24.99                                                        & 23.34                                                                      & 50.04                                                        & 44.41                                                                      \\
	& Gender+Age                   & 30.00, 79.40                                             & 24.99, 77.80                                                 & 23.34                                                                      & 50.53, 52.28                                                 & 40.25                                                                      \\
	\bottomrule
\end{tabular}

}
\end{table}

\begin{table}[]
	
			\centering
\caption{Comparison between {\sf PriSampler} and DPDM. DDPM* means {\sf PriSampler} is applied to the DDPM model. SMLD* means {\sf PriSampler} is applied to the SMLD model.}
\label{tab:def_perf_cmp}
\renewcommand{\arraystretch}{1.1}
\scalebox{0.75}{
	
	\begin{tabular}{l|c|c||cc|l}
		\toprule
		& \multicolumn{2}{c||}{CelebA-1k-30\%} & \multicolumn{3}{c}{CelebA-50k-30\%}   \\
		& DPDM ($\epsilon=10$)        & DDPM*        & DPDM ($\epsilon=10$) & DPDM ($\epsilon=50$) & SMLD*  \\
		\hline
		FID                                                      & 446.35              & 40.00       & 121.56        & 103.64        & 45.52 \\
		\hline
		\begin{tabular}[c]{@{}l@{}}Inferred\\ Prop.\end{tabular} & 100.00               & 50.40       & 44.00         & 24.20         & 50.40\\
		\bottomrule
	\end{tabular}

}
\end{table}

\begin{figure}[]
	\centering
	\includegraphics[width=0.70\linewidth]{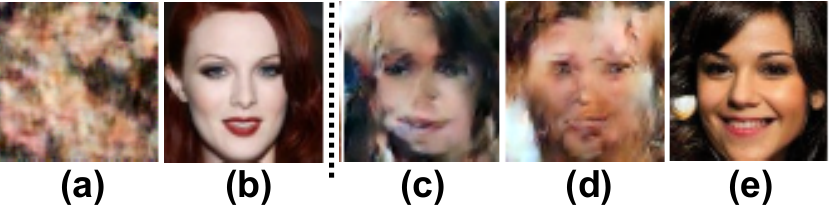}
	\caption{Visualization of synthetic samples under the defense DPDM and our method {\sf PriSampler}. (a) $\epsilon = 10$, DPDM trained on CelebA-1k-30\%.
(b) {\sf PriSampler} for DDPM trained on CelebA-1k-30\%.
(c) $\epsilon = 10$, DPDM trained on CelebA-50k-30\%.
(d) $\epsilon = 50$, DPDM trained on CelebA-50k-30\%.
(e) {\sf PriSampler} for SMLD trained on CelebA-50k-30\%. More visualizations are presented in Figure~\ref{fig:def_perf_cmp_whole} in Appendix.}
	\label{fig:def_perf_cmp_partial}
\end{figure}

%------------------------------------------
\section{Discussion}
\label{sec:Discussion}
%------------------------------------------
Our method {\sf PriSampler} aims to navigate
a sampler in the property space and is operated in the sampling process.
Thus, it is a training-free method.
Furthermore, it is a model-agnostic method and can be used as a plug-in for a wide range of diffusion models. 
In this section, we discuss limitations and future work.

\smallskip\noindent
\textbf{Model utility.} 
Although our method can guarantee the defense performance, i.e. achieving the predefined proportions, it will sacrifice model utility to some extent. 
Nevertheless, we take the first step to protect diffusion models from property inference attacks.
Furthermore, our defense method is still promising and competitive, compared to diffusion models trained with differential privacy.

\smallskip\noindent
\textbf{Entangled properties.}
When our defense is applied to protect multiple sensitive properties, the defense performance on image data is worse than that on tabular data. 
More precisely, 
the absolute difference between predicted value and predefined value~$\gamma$ on image data is larger than that on tabular data.
This might be because these properties in image generation are entangled together.
As a result, it is difficult to find an ideal hyperplane to completely differentiate them.
In contrast, properties in tabular data are explicitly represented in different columns.
This can be considered that these properties are disentangled and operating one property does not affect others.
Thus, our defense on tabular data can provide perfect protection. 
In the future, for image generation, we intend to design diffusion models with disentangled properties, which aim to separate entangled properties as large as possible in the training process.
In that way, we expect that the defense performance on image generation can be further improved.

\smallskip\noindent
\textbf{Membership inference.} Membership inference and property inference are two main types of privacy attacks, but their attack goals are different. 
Membership inference involves the privacy of individual training samples of a training set while property inference involves the privacy of global properties of a training set. 
In future, we plan to study the relationship between two types of privacy attacks and provide a holistic defense measure.

\smallskip\noindent
\textbf{Adaptive Attacks}.
Our defense inherently limits the feasibility for adversaries to infer the real proportion of sensitive properties accurately, even if they are aware of our defense mechanisms. This is because {\sf PriSampler} significantly obfuscates sensitive properties that could lead to successful inference attacks.

\smallskip\noindent
\textbf{Attacks via weights.}
Our attack method
only utilizes the synthetic data from a diffusion model to mount property inference attacks.
Prior work on classification models~\cite{ganju2018property} has proposed to infer the sensitive properties by the weights of a full-connected neural network. 
It is interesting to investigate whether this is also feasible for diffusion models.

%------------------------------------------
\section{Related Work}
\label{sec:Related work}
%------------------------------------------

\noindent
\textbf{Diffusion models.} 
Diffusion models~\cite{sohl2015deep,ho2020denoising} have recently drawn immense attention to academia and industry due to their high success in synthesizing realistic images.
Subsequently, various methods~\cite{song2019generative,song2020score,songdenoising,karras2022elucidating,lu2022dpm,lu2022dpm} are proposed to further improve the performance of diffusion models from the perspective of the training process, sampling mechanisms.
Beyond image synthesis, diffusion models have been investigated for tabular data generation~\cite{kotelnikov2023tabddpm,sattarov2023findiff,kim2023stasy}. 
However, these works focus on improving the generative performance of diffusion models. 
In this work, we study diffusion models from the viewpoint of privacy.

\smallskip\noindent
\textbf{Property inference attacks.} 
Property inference attacks allow adversaries to infer global sensitive information of the training set from a machine learning model~\cite{hu2023sok,liu2022ml}.
They are firstly studied by Ateniese et al.~\cite{ateniese2015hacking} on simple machine learning models, such as SVM and Hidden Markov Models.
Since then, there are a more increasing number of works focusing on property inference in neural network models, such as fully-connected neural networks~\cite{ganju2018property}, convolution neural networks~\cite{suri2022formalizing,chaudhari2022snap,mahloujifar2022property}, generative adversarial networks\cite{zhou2021property}, graph neural networks~\cite{zhang2022inference}, and federated learning models~\cite{melis2019exploiting}.
However, these works mainly focus on attacks, and their attack methods heavily rely on shadow models which require a large amount of computation. 
Property inference attacks on emerging diffusion models have not yet been extensively studied.
In this work, we take the first step to explore property inference attacks against various types of diffusion models.
Our work extends beyond existing works by focusing on diffusion models under more realistic attack scenarios and affordable attack costs, and designing new effective defenses to safeguard diffusion models.

There are several works studying privacy attacks against diffusion models through the lens of membership inference attacks~\cite{wu2022membership,carlini2023extracting,hu2023membership,matsumoto2023membership,duan2023diffusion,zhu2023data,pang2023black}.
However, membership inference attacks aim to infer whether a sample was used for training a machine learning model and focus more on the privacy of the individual training sample of the training set.
In contrast, this work endeavors to study property inference attacks which aim to infer the globally sensitive information of the training set of diffusion models.

%------------------------------------------
\section{Conclusion}
\label{sec:Conclusion}
%------------------------------------------

In this work, we present the inaugural study on property inference of diffusion models.
Under the property inference attack which only utilizes synthetic samples, we investigate the property inference risks for both tabular and image data generation.
Our extensive empirical analysis shows that various diffusion models and their samplers are vulnerable to property inference attacks.
For instance, as few as 500 generated samples can precisely infer the real proportion of a property.
More severely, we observe that better performance of diffusion models can lead to a more accurate estimation of property inference, highlighting a critical privacy trade-off in the pursuit of improved generative capabilities.
To address these vulnerabilities, we introduce a model-agnostic plug-in defense method {\sf PriSampler}.
Our evaluations demonstrate that {\sf PriSampler} not only effectively mitigates the risks of property inference across different types of samplers and diffusion models, but also surpasses diffusion models trained with differential privacy in terms of model utility and defense performance. This study underscores the importance of integrating privacy considerations in the development of advanced diffusion models from the perspective of property inference.

We have also identified several directions for future work, including developing attack methods via weights, designing diffusion models with disentangled properties, and constructing a holistic defense by exploring the relationship between property inference and membership inference.

\section*{Acknowledgments}
This research was funded in whole by the Luxembourg National Research Fund (FNR), grant reference 13550291.
%-------------------------------------------------------------------------
%\clearpage
\bibliographystyle{ACM-Reference-Format}
\bibliography{egbib}

\clearpage
\appendix
\section{Appendix}

\subsection{Implementation Details of {\sf PriSampler}}
\label{ssec:Details_Algorithm}

Our method {\sf PriSampler} has two hyperparameters: $\alpha$ and $t$.
$\alpha$ controls the distance of the desired samples~$x^{'}_{t}$ from an intermediate sample~$x_t$.
$t$ is the diffusion step in which we manipulate an intermediate sample in the $t$ step. 
For tubular data, we fix $t$ as 0 for all properties.
In terms of $\alpha$, we summarize them in Table~\ref{tab:Hyperparameters_prisampler_tabular}.
For the base sampler, we use the stochastic sampling method provided by TabDDPM.
The number of sampling steps for Adult and Churn is 100 while that for Cardio is 1,000.

For image data, Table~\ref{tab:Hyperparameters_prisampler_image} shows the hyperparameters $\alpha$ and $t$ used for different samplers and diffusion models.
For base samplers, the total number of sampling steps is 40 for the DPM sampler and 1,000 for the PC sampler. 
In terms of the base sampler --- DPM sampler, we set its hyperparameter `{dpm\_solver\_method}' as `singlestep', and `{dpm\_solver\_order}' as `3'.
Therefore, for {\sf PriSampler} applied to the DPM sampler,
we set $\alpha$ and $t$ as 50 and 6. Note, here, $t=6$ is the index of diffusion steps rather than actual diffusion steps, because the step size of the DPM sampler is equal to 3, i.e. `{dpm\_solver\_order}' = `3'.
In terms of the base sampler --- PC sampler, we set its hyperparameter `{predictor}' as `ReverseDiffusionPredictor', and `corrector' as `LangevinCorrector'. 
For {\sf PriSampler} applied to the PC sampler, we choose different $\alpha$ and $t$ for different diffusion models, as shown in Table~\ref{tab:Hyperparameters_prisampler_image}. 
This is because the PC sampler is stochastic sampling where fresh noise will be added in the sampling process, which may affect the generated samples in the protected property space.
Therefore, we adjust $\alpha$ and $t$ to achieve the desired proportion.

\begin{table}[!h]
	\centering
	\caption{Hyperparameters of {\tt PriSampler} for different properties on TabDDPM on tabular data.}
	\label{tab:Hyperparameters_prisampler_tabular}
	\renewcommand{\arraystretch}{1.3}
	\scalebox{1.0}{
		
	\begin{tabular}{ll|l} 
		\toprule
		Dataset & Property           & $\alpha$                  \\
		\hline
		\multirow{3}{*}{Adult}     & Gender=Male        & 50                 \\
		& Age$<$30              & 1                  \\
		& Race=\{1,2,3,4,5\} & 5, 50, 50, 50, 50  \\
		\hline
		\multirow{4}{*}{Churn}   & Gender=Male        & 50                 \\
		& CreditScore$<$600     & 1                  \\
		& Age$<$30              & 1                  \\
		& Gender+Geography   & 50+(5, 100, 100)   \\
		\hline
		\multirow{3}{*}{Cardio}  & Gender=Male        & 50                 \\
		& Age$\geq$50              & 1                  \\
		& Smoking=Yes        & 50                 \\
		\bottomrule
	\end{tabular}

}
\end{table}

\begin{table*}
\centering
\caption{Hyperparameters ($\alpha,t$) of {\tt PriSampler} for different samplers and diffusion models on image data.}
\label{tab:Hyperparameters_prisampler_image}
\renewcommand{\arraystretch}{1.3}
\scalebox{0.95}{

\begin{tabular}{l|l|r|r|r|r|r|r|r|r|r|r} 
\hline
Sampler & Model & \multicolumn{2}{r|}{\textcolor[rgb]{0.2,0.2,0.2}{CelebA-1k-10}}  & \multicolumn{2}{r|}{\textcolor[rgb]{0.2,0.2,0.2}{CelebA-1k-20}}  & \multicolumn{2}{r|}{\textcolor[rgb]{0.2,0.2,0.2}{CelebA-1k-30}}  & \multicolumn{2}{r|}{\textcolor[rgb]{0.2,0.2,0.2}{CelebA-1k-40}}  & \multicolumn{2}{r}{\textcolor[rgb]{0.2,0.2,0.2}{CelebA-1k-50}}   \\ 
\hline
        &       & $\alpha$   & $t$                                                          & $\alpha$   & $t$                                                          & $\alpha$   & $t$                                                          & $\alpha$   & $t$                                                          & $\alpha$   & $t$                                                           \\ 
\hline
\multirow{2}{*}{PC}      & DDPM  & 150 & 699                                                        & 150 & 699                                                        & 150 & 699                                                        & 140 & 699                                                        & 140 & 699                                                         \\ 
\cline{2-12}
        & VPSDE & 220 & 699                                                        & 150 & 699                                                        & 170 & 699                                                        & 150 & 699                                                        & 140 & 699                                                         \\ 
\hline
\multirow{2}{*}{DPM}    & DDPM  & 50  & 6                                                          & 50  & 6                                                          & 50  & 6                                                          & 50  & 6                                                          & 50  & 6                                                           \\ 
\cline{2-12}
        & VPSDE & 50  & 6                                                          & 50  & 6                                                          & 50  & 6                                                          & 50  & 6                                                          & 50  & 6                                                           \\ 
\hline\hline
Sampler & Model & \multicolumn{2}{r|}{\textcolor[rgb]{0.2,0.2,0.2}{CelebA-50k-10}} & \multicolumn{2}{r|}{\textcolor[rgb]{0.2,0.2,0.2}{CelebA-50k-20}} & \multicolumn{2}{r|}{\textcolor[rgb]{0.2,0.2,0.2}{CelebA-50k-30}} & \multicolumn{2}{r|}{\textcolor[rgb]{0.2,0.2,0.2}{CelebA-50k-40}} & \multicolumn{2}{r}{\textcolor[rgb]{0.2,0.2,0.2}{CelebA-50k-50}}  \\ 
\hline
        &       & $\alpha$   & $t$                                                          & $\alpha$   & $t$                                                          & $\alpha$   & $t$                                                          & $\alpha$   & $t$                                                          & $\alpha$   & $t$                                                           \\ 
\hline
\multirow{2}{*}{PC}      & SMLD  & 40  & 500                                                        & 40  & 500                                                        & 40  & 500                                                        & 40  & 500                                                        & 40  & 500                                                         \\ 
\cline{2-12}
        & VESDE & 25  & 549                                                        & 25  & 549                                                        & 25  & 549                                                        & 15  & 549                                                        & 10  & 549                                                         \\
\hline
\end{tabular}

}
\end{table*}

\subsection{Additional Results on Attacks}
\label{ssec:add_res}
\noindent
\textbf{Attack performance on different sizes of training sets.} 
Figure~\ref{fig:size_of_sets_ddpm_celeba_male_30} plots attack performance in terms of sizes of training sets.
Here, the target models are the DDPM models trained on a dataset containing 30\% male training samples.
Therefore, the real proportion of the property male is 30\%.
We can see that the inference performance slightly decreases with the increase in the size of training sets. 
For example, for the DPM sampler, when the size of training sets increases from 10k to 50k, the inferred proportions decrease from about 29\% to around 26\%.
Overall, the inferred proportions for all samplers fluctuate between 25\% and 30\%.

\begin{figure}[!t]
	\centering
	\includegraphics[width=0.85\linewidth]{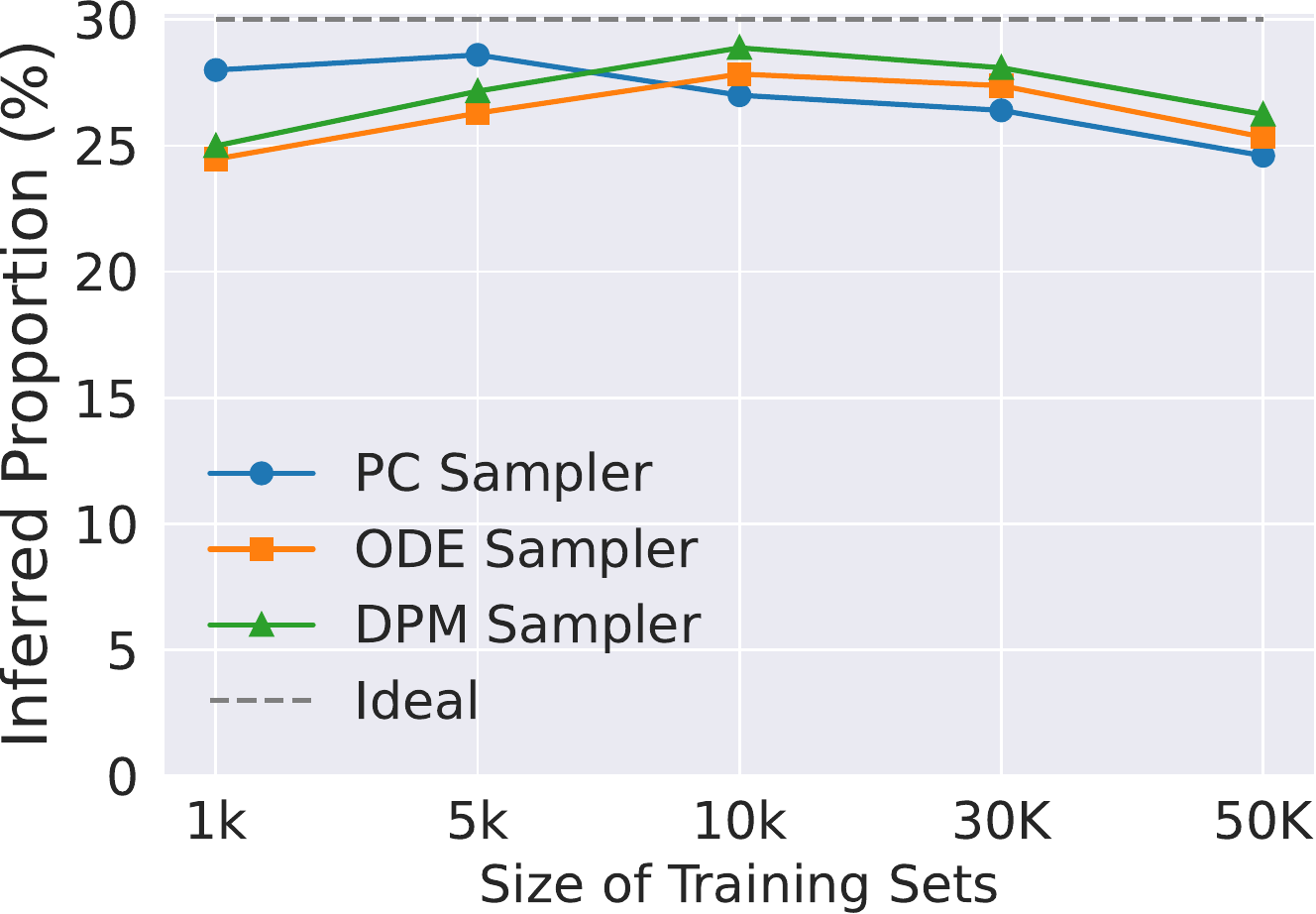}
	\caption{Attack performance with respect to different sizes of training sets. The target models are DDPM models trained on CelebA with the property Gender=Male of 30\%.}
	\label{fig:size_of_sets_ddpm_celeba_male_30}
\end{figure}

\begin{figure}[!t]
	\centering
	
	\subfigure[VPEDM.]{
		\includegraphics[width=0.85\columnwidth]{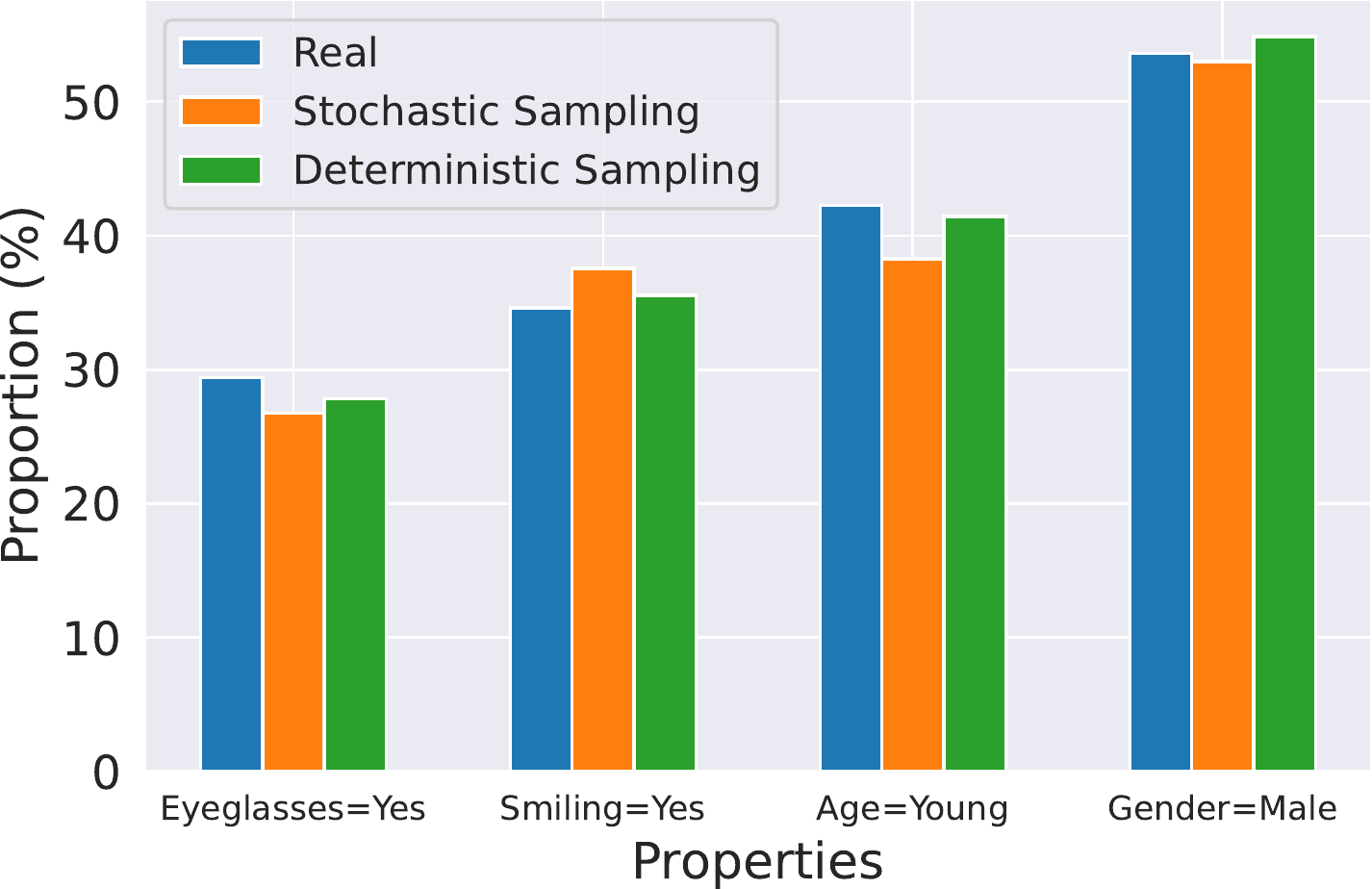}
		\label{fig:case_edm_vpsde}
	}
	\subfigure[VEEDM.]{
		\includegraphics[width=0.85\columnwidth]{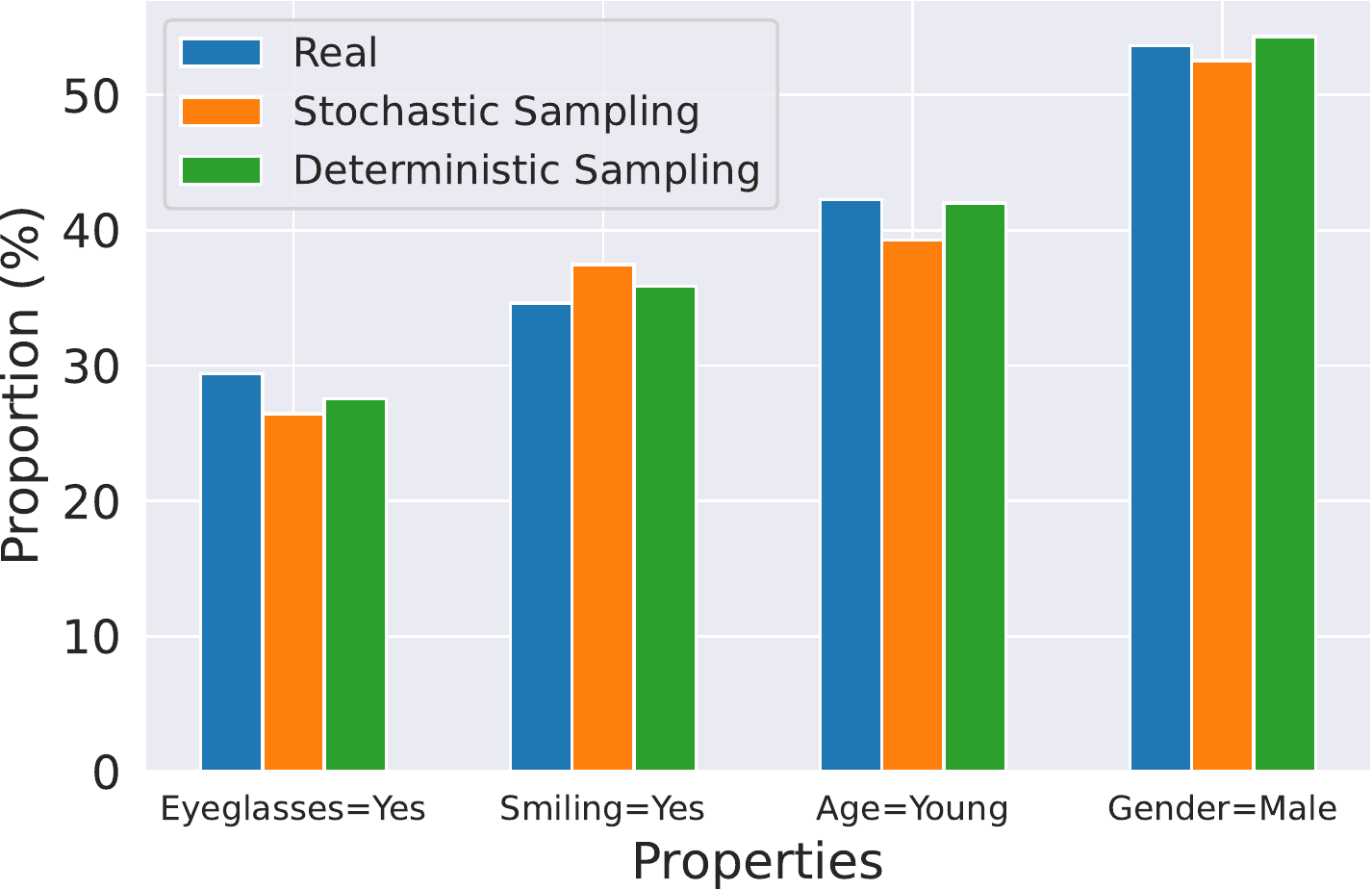}
		\label{fig:case_edm_vesde}
	}
	
	\caption{Attack performance on the EDM models.}
	\label{fig:case_edm}
\end{figure}

%------------------------------------------
\subsection{Case Study: Attacks in Practice}
\label{ssec:case}
%------------------------------------------
In this section, we demonstrate the property inference risks in practice through one case study in which we perform property inference attacks against publicly available well-trained diffusion models.

We choose EDM models proposed by Karras et. al ~\cite{karras2022elucidating} as target models.
EDM models achieve competitive performance in image synthesis by a design space to decouple complex components. 
Similar to SSDE, EDM models include VP and VE formulations, and in this work, we call them VPEDM and VEEDM, respectively.
For each model type, they also have two types of sampling methods to synthesize samples: stochastic sampling and deterministic sampling. 
In our experiments, we conduct property inference attacks on VPEDM and VEEDM.
They are both trained by their original authors on the Flickr-Faces-HQ~(FFHQ) dataset which contains 70,000 human face images~\cite{StyleGAN12019style}.
All samples of the FFHQ dataset used for training have $64 \times 64$ resolution.

Similarly, we assume that only generated samples can be obtained by adversaries.
Because the FFHQ dataset does not annotate the properties of each image, here we use the proportion of the property in the training set inferred by our property inference classifier as the real proportion.
Although this might bring some errors due to a lack of human annotation, we report the attack performance by both the inferred proportion and the absolute difference.
The absolute difference can eliminate this type of error because it shows attack performance by how close the real and inferred proportions are.
In this case study, we directly use the property classifiers used in Section~\ref{sec:method} to infer the proportion of different properties, which also aims to illustrate 
that the assumption about shadow datasets is not always required.
We can relax this by directly using a pre-trained classifier.
50,000 generated samples for all sampling methods and diffusion models are used to perform the property inference.
We choose four properties: Eyeglasses=Yes, Smiling=Yes, Age=Young and Gender=Male.

\noindent
\textbf{Results.} 
Figure~\ref{fig:case_edm} presents the performance of property inference attacks against EDM models over four properties.
Overall, our attack can achieve a rather precise estimation for the proportion of each property.
Although the real proportion of these four properties has wide ranges from 29\% to 53\%, we can observe that the inferred proportions of various private properties are all close to the real proportions.
In addition, different types of sampling methods show similar high privacy risks in all properties and diffusion models.

Table~\ref{tab:case_edm}  describes quantitative attack results on the EDM models. 
We can see that all samplers can achieve a good performance, obtaining an FID score between 2 and 3.
We also report the absolute difference.
The minimal absolute difference is 0.23\%, which can be seen in inferring the property Age=Young on VEEDM using deterministic sampling.
When inferring the property Age=Young in VPSDE under the stochastic sampling, our attack shows a little inferior performance with an absolute difference of 3.97\%.
To sum up, our attack on the EDM models can achieve at most a 4\% absolute difference.

\begin{table}[]
	\centering
	\caption{Quantitative attack results on the EDM models. Stoch.: Stochastic; Deter.: Deterministic.}
	\label{tab:case_edm}
	\renewcommand{\arraystretch}{1.3}
	\scalebox{0.85}{

\begin{tabular}{ccll|rrr} 
	\toprule
	Model                  & Sampler                 & \begin{tabular}[c]{@{}l@{}}Utility\\FID $\downarrow$\end{tabular} & Property       & \begin{tabular}[c]{@{}r@{}}~Real\\ Prop. \\ (\%)\end{tabular} & \begin{tabular}[c]{@{}r@{}}Inferred\\ Prop. \\ (\%)\end{tabular} & \begin{tabular}[c]{@{}r@{}}Abs.\\ Diff. \\ (\%) $\downarrow$\end{tabular}  \\ 
	\hline
	\multirow{8}{*}{VPEDM} & \multirow{4}{*}{Stoch.} & \multirow{4}{*}{2.87}                                & Eyeglasses=Yes & 29.39                                                         & 26.73                                                            & 2.66                                                          \\
	&                         &                                                      & Smiling=Yes    & 34.61                                                         & 37.54                                                            & 2.93                                                          \\
	&                         &                                                      & Age=Young      & 42.25                                                         & 38.28                                                            & 3.97                                                          \\
	&                         &                                                      & Gender=Male    & 53.62                                                         & 53.01                                                            & 0.61                                                          \\ 
	\cmidrule{2-7}
	& \multirow{4}{*}{Deter.} & \multirow{4}{*}{2.47}                                & Eyeglasses=Yes & 29.39                                                         & 27.83                                                            & 1.56                                                          \\
	&                         &                                                      & Smiling=Yes    & 34.61                                                         & 35.54                                                            & 0.93                                                          \\
	&                         &                                                      & Age=Young      & 42.25                                                         & 41.44                                                            & 0.81                                                          \\
	&                         &                                                      & Gender=Male    & 53.62                                                         & 54.85                                                            & 1.23                                                          \\ 
	\hline\hline
	\multirow{8}{*}{VEEDM} & \multirow{4}{*}{Stoch.} & \multirow{4}{*}{2.85}                                & Eyeglasses=Yes & 29.39                                                         & 26.44                                                            & 2.95                                                          \\
	&                         &                                                      & Smiling=Yes    & 34.61                                                         & 37.47                                                            & 2.86                                                          \\
	&                         &                                                      & Age=Young      & 42.25                                                         & 39.27                                                            & 2.98                                                          \\
	&                         &                                                      & Gender=Male    & 53.62                                                         & 52.55                                                            & 1.07                                                          \\ 
	\cmidrule{2-7}
	& \multirow{4}{*}{Deter.} & \multirow{4}{*}{2.57}                                & Eyeglasses=Yes & 29.39                                                         & 27.54                                                            & 1.85                                                          \\
	&                         &                                                      & Smiling=Yes    & 34.61                                                         & 35.85                                                            & 1.24                                                          \\
	&                         &                                                      & Age=Young      & 42.25                                                         & 42.02                                                            & 0.23                                                          \\
	&                         &                                                      & Gender=Male    & 53.62                                                         & 54.3                                                             & 0.68                                                          \\
	\bottomrule
\end{tabular}
		
	}
\end{table}

\begin{table*}[]
	\centering
	\caption{The qualitative attack results for the sensitive property Gender=Male. Prop.: proportion. Abs. Diff. : absolute difference. $\downarrow$ means smaller is better.}
	\label{tab:att_perf}
	\renewcommand{\arraystretch}{1.3}
	\scalebox{0.75}{	

\begin{tabular}{llllc|llllc||llllc|llllc} 
	\toprule
	Model                  & \begin{tabular}[c]{@{}l@{}}Real\\ Prop. \\ (\%)\end{tabular} & \begin{tabular}[c]{@{}l@{}}Inferred\\ Prop. \\ (\%)\end{tabular} & \begin{tabular}[c]{@{}l@{}}Abs.\\ Diff. \\(\%) $\downarrow$\end{tabular} & \begin{tabular}[c]{@{}c@{}}Utility\\FID $\downarrow$\end{tabular} & Model                  & \begin{tabular}[c]{@{}l@{}}Real\\ Prop.\\ (\%)\end{tabular} & \begin{tabular}[c]{@{}l@{}}Inferred\\ Prop.\\ (\%)\end{tabular} & \begin{tabular}[c]{@{}l@{}}Abs.\\ Diff.\\ (\%) $\downarrow$\end{tabular} & \begin{tabular}[c]{@{}c@{}}Utility\\FID $\downarrow$\end{tabular} & Model                 & \begin{tabular}[c]{@{}l@{}}Real\\ Prop.\\ (\%)\end{tabular} & \begin{tabular}[c]{@{}l@{}}Inferred\\ Prop.\\ (\%)\end{tabular} & \begin{tabular}[c]{@{}l@{}}Abs.\\ Diff.\\ (\%) $\downarrow$\end{tabular} & \begin{tabular}[c]{@{}c@{}}Utility\\FID $\downarrow$\end{tabular} & Model                  & \begin{tabular}[c]{@{}l@{}}Real\\ Prop.\\ (\%)\end{tabular} & \begin{tabular}[c]{@{}l@{}}Inferred\\ Prop.\\ (\%)\end{tabular} & \begin{tabular}[c]{@{}l@{}}Abs.\\ Diff.\\ (\%) $\downarrow$\end{tabular} & \begin{tabular}[c]{@{}c@{}}Utility\\FID $\downarrow$\end{tabular}  \\ 
	\hline
	\multicolumn{10}{c||}{PC Sampler}                                                                                                                                                                                                                                                                                                                                                                                                                                                                                                                                   & \multicolumn{10}{c}{ODE Sampler}                                                                                                                                                                                                                                                                                                                                                                                                                                                                                                                                  \\ 
	\hline
	\multirow{5}{*}{DDPM}  & 10                                                           & 10.00                                                            & 0.00                                                          & 24.45                                                  & \multirow{5}{*}{SMLD}  & 10                                                          & 7.80                                                            & 2.20                                                          & 23.24                                                  & \multirow{5}{*}{DDPM} & 10                                                          & 6.66                                                            & 3.34                                                          & 16.80                                                  & \multirow{5}{*}{VPSDE} & 10                                                          & 10.19                                                           & 0.19                                                          & 5.59                                                    \\
	& 20                                                           & 19.20                                                            & 0.80                                                          & 24.85                                                  &                        & 20                                                          & 15.60                                                           & 4.40                                                          & 24.64                                                  &                       & 20                                                          & 14.49                                                           & 5.51                                                          & 17.12                                                  &                        & 20                                                          & 19.11                                                           & 0.89                                                          & 5.64                                                    \\
	& 30                                                           & 28.00                                                            & 2.00                                                          & 25.55                                                  &                        & 30                                                          & 28.20                                                           & 1.80                                                          & 24.72                                                  &                       & 30                                                          & 24.47                                                           & 5.53                                                          & 17.41                                                  &                        & 30                                                          & 27.93                                                           & 2.07                                                          & 5.97                                                    \\
	& 40                                                           & 41.60                                                            & 1.60                                                          & 26.57                                                  &                        & 40                                                          & 36.20                                                           & 3.80                                                          & 25.08                                                  &                       & 40                                                          & 35.77                                                           & 4.23                                                          & 17.61                                                  &                        & 40                                                          & 38.14                                                           & 1.86                                                          & 5.95                                                    \\
	& 50                                                           & 48.60                                                            & 1.40                                                          & 28.96                                                  &                        & 50                                                          & 52.00                                                           & 2.00                                                          & 25.48                                                  &                       & 50                                                          & 46.03                                                           & 3.97                                                          & 18.46                                                  &                        & 50                                                          & 49.75                                                           & 0.25                                                          & 6.22                                                    \\ 
	\hline
	\multicolumn{10}{c||}{PC Sampler}                                                                                                                                                                                                                                                                                                                                                                                                                                                                                                                                   & \multicolumn{10}{c}{DPM Sampler}                                                                                                                                                                                                                                                                                                                                                                                                                                                                                                                                  \\ 
	\hline
	\multirow{5}{*}{VPSDE} & 10                                                           & 11.00                                                            & 1.00                                                          & 19.22                                                  & \multirow{5}{*}{VESDE} & 10                                                          & 6.40                                                            & 3.60                                                          & 37.75                                                  & \multirow{5}{*}{DDPM} & 10                                                          & 7.05                                                            & 2.95                                                          & 22.60                                                  & \multirow{5}{*}{VPSDE} & 10                                                          & 10.29                                                           & 0.29                                                          & 8.26                                                    \\
	& 20                                                           & 20.80                                                            & 0.80                                                          & 20.97                                                  &                        & 20                                                          & 14.00                                                           & 6.00                                                          & 42.08                                                  &                       & 20                                                          & 14.85                                                           & 5.15                                                          & 23.47                                                  &                        & 20                                                          & 19.18                                                           & 0.82                                                          & 8.54                                                    \\
	& 30                                                           & 28.20                                                            & 1.80                                                          & 20.22                                                  &                        & 30                                                          & 23.20                                                           & 6.80                                                          & 35.94                                                  &                       & 30                                                          & 24.99                                                           & 5.01                                                          & 23.34                                                  &                        & 30                                                          & 28.16                                                           & 1.84                                                          & 8.68                                                    \\
	& 40                                                           & 41.00                                                            & 1.00                                                          & 20.62                                                  &                        & 40                                                          & 37.20                                                           & 2.80                                                          & 55.89                                                  &                       & 40                                                          & 36.56                                                           & 3.44                                                          & 23.43                                                  &                        & 40                                                          & 38.57                                                           & 1.43                                                          & 8.75                                                    \\
	& 50                                                           & 49.20                                                            & 0.80                                                          & 21.65                                                  &                        & 50                                                          & 50.20                                                           & 0.20                                                          & 39.31                                                  &                       & 50                                                          & 46.37                                                           & 3.63                                                          & 24.47                                                  &                        & 50                                                          & 49.37                                                           & 0.63                                                          & 8.96                                                    \\
	\bottomrule
\end{tabular}
		
	}
\end{table*}

\subsection{Additional Results on Defenses}
\label{ssec:add_res_defense}

\smallskip\noindent
\textbf{Additional defense results on tabular data.} 
Figure~\ref{fig:def_tabddpm_mul} presents defense performance for the multi-categorical property on TabDDPM.
Specifically, Figure~\ref{fig:defense_tabddpm_adult_multi_cat_martial} shows defense performance on the sensitive property Martial-status on the Adult dataset.
The property Martial-status has seven values, and its ideal proportion is about 14.28, i.e. 1/7.
Figure~\ref{fig:defense_tabddpm_adult_multi_cat_martial} shows defense performance on the sensitive property Geography on the Churn dataset. The ideal proportion is about 33.33, i.e. 1/3.
Again, we can see that our defense can achieve perfect performance.
Figure~\ref{fig:defense_tabddpm_adult_racegender} shows defense performance for multiple properties on the TabDDPM on Adult. The protected properties are Gender and Race.
We can observe that the desired proportion can be achieved under our defense method.

\smallskip\noindent
\textbf{Defense performance on different numbers of generated samples.} 
Figure~\ref{fig:defense_num_gen_samples_ddpm_celeba1k_male_30} shows the defense performance on different numbers of generated samples.
Here, we choose the DDPM model trained on CelebA-1k-30\% as the target model.
The sensitive property is male.
We can clearly see that both types of samplers can provide good protection for the property male even if model owners only release as few as 50 samples.
Although the PC sampler shows a slight fluctuation in the phase of releasing a few samples, it is gradually stable after 500 samples.
The DPM sampler extremely stabilizes no matter how many samples are released. 
One of the reasons might be that random noise added during the sampling process for the PC sampler has some effects on generated samples because the PC sampler belongs to stochastic sampling while the DPM sampler belongs to deterministic sampling.

\begin{figure}[!t]
	\centering
	
	\subfigure[The property Martial-status on Adult. Noth that the real proportion of Mar=7 is 0.06\% and its inferred proportion without defense is 0.01\%. Mar = \{1, 2, 3, 4, 5, 6, 7\} refers to Martial-status = \{Married-civ-spouse, Never-married, Divorced, Separated, Widowed, Married-spouse-absent, Married-AF-spouse\}.]{
	\includegraphics[width=0.80\columnwidth]{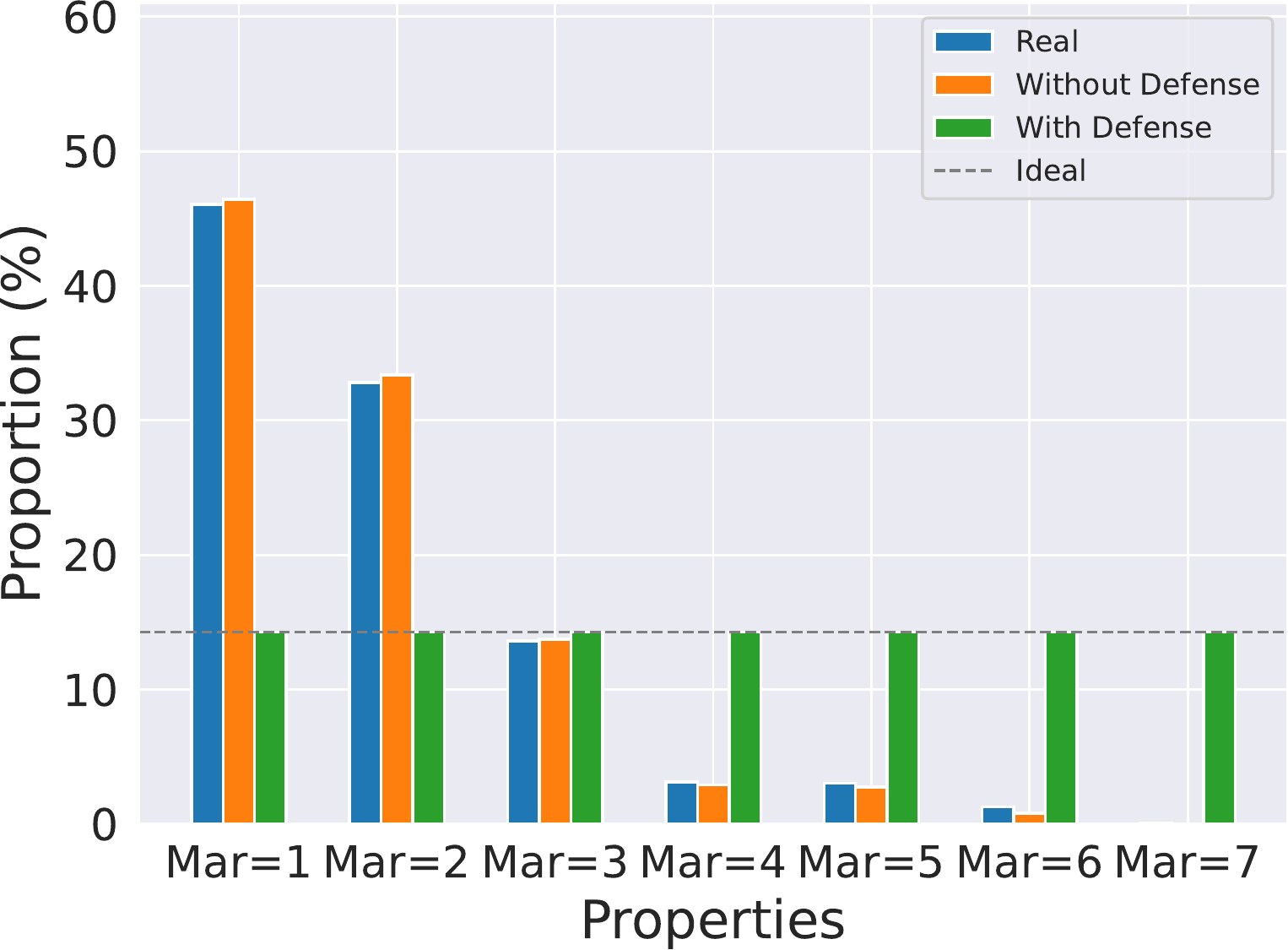}
	\label{fig:defense_tabddpm_adult_multi_cat_martial}
}	
	\subfigure[The property Geography on Churn. Geo = \{Fr, Ger, Spn\} refers to Geography = \{France, Germany, Spain\}.]{
		\includegraphics[width=0.80\columnwidth]{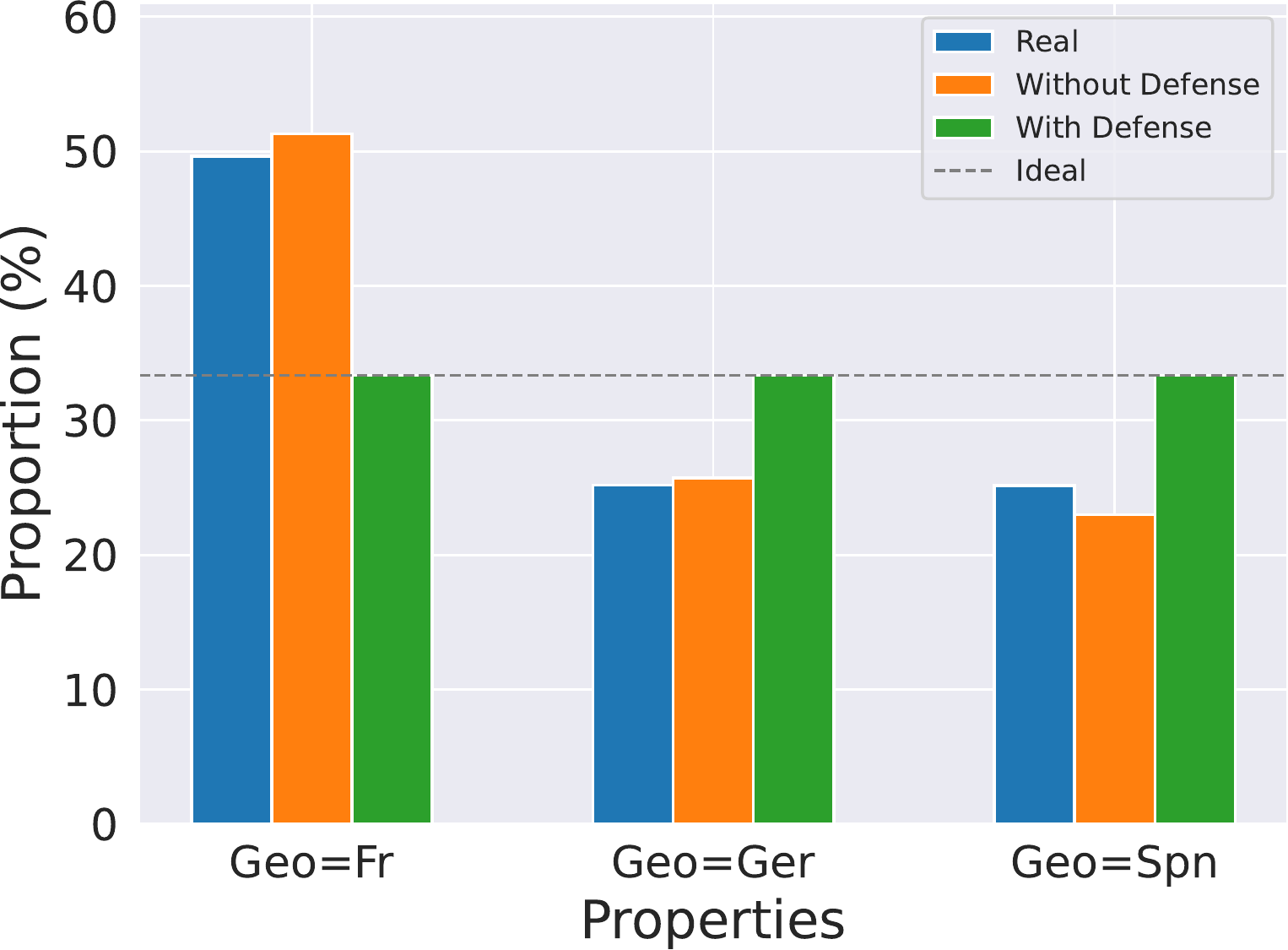}
		\label{fig:defense_tabddpm_churn_multi_cat_geograph}
	}

	\caption{Defense performance for the multi-categorical property on TabDDPM.}
	\label{fig:def_tabddpm_mul}
\end{figure}

\begin{figure}[!t]
	\centering
	\includegraphics[width=0.85\linewidth]{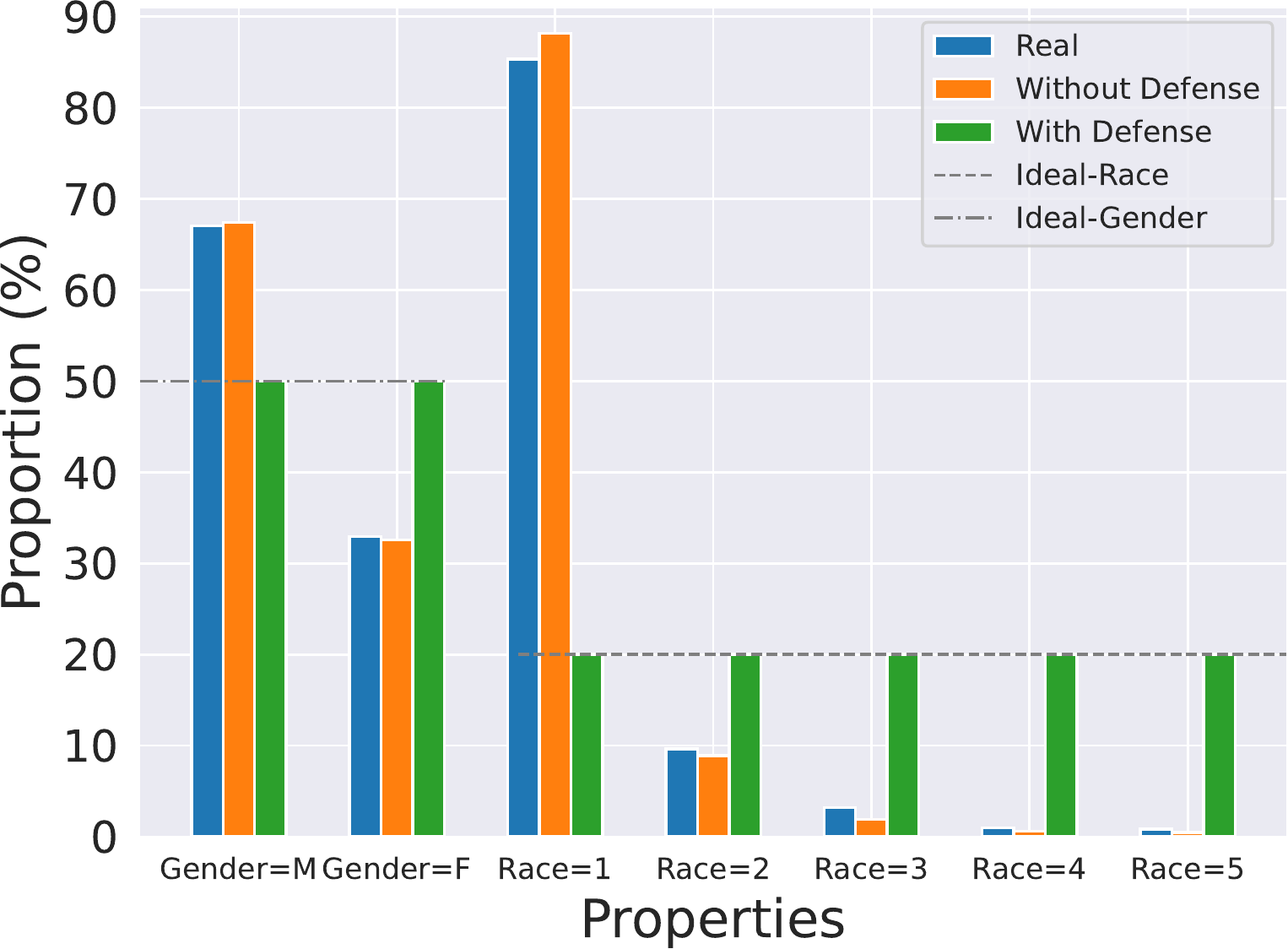}
	\caption{Defense performance for multiple properties on the TabDDPM on Adult. Gender and  Race are protected properties. Here,  Gender = \{M, F\} refers to Gender = \{Male, Female\} and Race = \{1, 2, 3, 4, 5\} refers to Race = \{White, Black, Asian-Pac-Islander, Amer-Indian-Eskimo, Others\}.}
	\label{fig:defense_tabddpm_adult_racegender}
\end{figure}

\begin{table}[]
	
	\centering
	\caption{Defense performances on a single property. Desi. Prop.: Desired Proportion.}
	\label{tab:def_perf}
	\renewcommand{\arraystretch}{1.2}
	\scalebox{0.65}{	
		
		\begin{tabular}{llllll||llllll}
			\toprule
			Model                                     & \begin{tabular}[c]{@{}l@{}}Real\\ Prop.\\ (\%)\end{tabular} & \begin{tabular}[c]{@{}l@{}}Desi.\\ Prop.\\ (\%)\end{tabular} & \begin{tabular}[c]{@{}l@{}}Inferred\\ Prop.\\ (\%)\end{tabular} & \begin{tabular}[c]{@{}l@{}}Abs.\\ Diff.\\ (\%)\end{tabular} & FID   & Model                  & \begin{tabular}[c]{@{}l@{}}Real\\ Prop.\\ (\%)\end{tabular} & \begin{tabular}[c]{@{}l@{}}Desi.\\ Prop.\\ (\%)\end{tabular} & \begin{tabular}[c]{@{}l@{}}Inferred\\ Prop.\\ (\%)\end{tabular} & \begin{tabular}[c]{@{}l@{}}Abs.\\ Diff.\\ (\%)\end{tabular} & FID   \\
			\hline
			\multicolumn{12}{c}{PC Sampler}                                                                                                                                                                                                                                                                                                                                                                                                                                                                                                                                                                              \\
			\hline
			\multicolumn{1}{c}{\multirow{5}{*}{DDPM}} & 10                                                          & 50                                                           & 48.40                                                           & 1.60                                                        & 56.35 & \multirow{5}{*}{VPSDE} & 10                                                          & 50                                                           & 48.60                                                           & 1.40                                                        & 58.72 \\
			\multicolumn{1}{c}{}                      & 20                                                          & 50                                                           & 51.00                                                           & 1.00                                                        & 46.80 &                        & 20                                                          & 50                                                           & 51.40                                                           & 1.40                                                        & 41.26 \\
			\multicolumn{1}{c}{}                      & 30                                                          & 50                                                           & 50.40                                                           & 0.40                                                        & 40.00 &                        & 30                                                          & 50                                                           & 50.40                                                           & 0.40                                                        & 38.89 \\
			\multicolumn{1}{c}{}                      & 40                                                          & 50                                                           & 50.60                                                           & 0.60                                                        & 48.04 &                        & 40                                                          & 50                                                           & 51.20                                                           & 1.20                                                        & 51.61 \\
			\multicolumn{1}{c}{}                      & 50                                                          & 50                                                           & 51.00                                                           & 1.00                                                        & 49.58 &                        & 50                                                          & 50                                                           & 49.60                                                           & 0.40                                                        & 42.82 \\
			\hline
			\multirow{5}{*}{SMLD}                     & 10                                                          & 50                                                           & 48.40                                                           & 1.60                                                        & 38.40 & \multirow{5}{*}{VESDE} & 10                                                          & 50                                                           & 49.80                                                           & 0.20                                                        & 61.32 \\
			& 20                                                          & 50                                                           & 50.20                                                           & 0.20                                                        & 44.68 &                        & 20                                                          & 50                                                           & 50.40                                                           & 0.40                                                        & 68.95 \\
			& 30                                                          & 50                                                           & 50.40                                                           & 0.40                                                        & 45.52 &                        & 30                                                          & 50                                                           & 51.00                                                           & 1.00                                                        & 57.47 \\
			& 40                                                          & 50                                                           & 50.40                                                           & 0.40                                                        & 45.17 &                        & 40                                                          & 50                                                           & 50.20                                                           & 0.20                                                        & 77.57 \\
			& 50                                                          & 50                                                           & 50.60                                                           & 0.60                                                        & 46.95 &                        & 50                                                          & 50                                                           & 50.60                                                           & 0.60                                                        & 49.75 \\
			\hline
			\multicolumn{12}{c}{DPM Sampler}                                                                                                                                                                                                                                                                                                                                                                                                                                                                                                                                                                             \\
			\hline
			\multirow{5}{*}{DDPM}                     & 10                                                          & 50                                                           & 48.25                                                           & 1.75                                                        & 44.70 & \multirow{5}{*}{VPSDE} & 10                                                          & 50                                                           & 49.28                                                           & 0.72                                                        & 52.44 \\
			& 20                                                          & 50                                                           & 50.02                                                           & 0.02                                                        & 47.87 &                        & 20                                                          & 50                                                           & 50.22                                                           & 0.22                                                        & 52.78 \\
			& 30                                                          & 50                                                           & 50.04                                                           & 0.04                                                        & 44.41 &                        & 30                                                          & 50                                                           & 50.50                                                           & 0.50                                                        & 50.66 \\
			& 40                                                          & 50                                                           & 50.19                                                           & 0.19                                                        & 45.90 &                        & 40                                                          & 50                                                           & 51.06                                                           & 1.06                                                        & 53.00 \\
			& 50                                                          & 50                                                           & 50.26                                                           & 0.26                                                        & 47.68 &                        & 50                                                          & 50                                                           & 51.68                                                           & 1.68                                                        & 45.33\\
			\bottomrule
		\end{tabular}
		
	}
\end{table}

\begin{figure}[!t]
	\centering
	\includegraphics[width=0.85\linewidth]{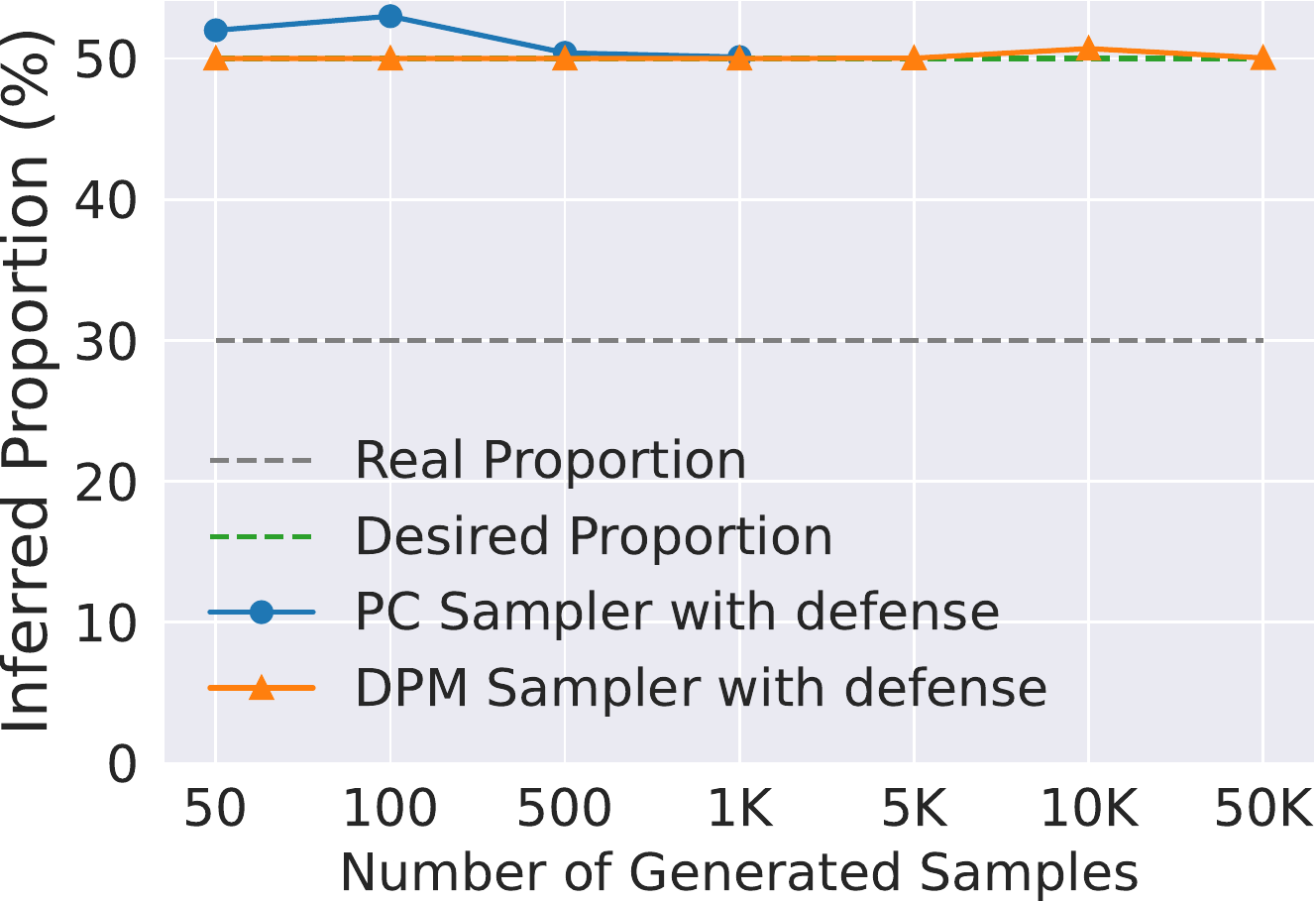}
	\caption{Defense Performance on the different numbers of generated samples. }
	\label{fig:defense_num_gen_samples_ddpm_celeba1k_male_30}
\end{figure}  

\begin{table}[]
	
	\centering
	\caption{Defense performances on multiple properties.}
	\label{tab:def_perf_two}
	\renewcommand{\arraystretch}{1.3}
	\scalebox{0.55}{
		
		\begin{tabular}{llllll|llllll} 
			\toprule
			Model &             & \begin{tabular}[c]{@{}l@{}}Real\\ Prop.\\ (\%)\end{tabular} & \begin{tabular}[c]{@{}l@{}}Desi.\\ Prop.\\ (\%)\end{tabular} & \begin{tabular}[c]{@{}l@{}}Inferred\\ Prop.\\ (\%)\end{tabular} & \begin{tabular}[c]{@{}l@{}}Abs.\\ Diff.\\ (\%)\end{tabular} &           & \begin{tabular}[c]{@{}l@{}}Real\\ Prop.\\ (\%)\end{tabular} & \begin{tabular}[c]{@{}l@{}}Desi.\\ Prop.\\ (\%)\end{tabular} & \begin{tabular}[c]{@{}l@{}}Inferred\\ Prop.\\ (\%)\end{tabular} & \begin{tabular}[c]{@{}l@{}}Abs.\\ Diff.\\ (\%)\end{tabular} & FID    \\ 
			\hline
			\multicolumn{12}{c}{PC sampler}                                                                                                                                                                                                                                                                                                                                                                                                                                                                                                                                    \\ 
			\hline
			DDPM  & Gender=Male & 30                                                          & 50                                                           & 51.00                                                           & 1.00                                                        & Age=Young & 79.40                                                       & 50                                                           & 57.40                                                           & 7.40                                                        & 48.74  \\
			VPSDE & Gender=Male & 30                                                          & 50                                                           & 51.60                                                           & 1.60                                                        & Age=Young & 79.40                                                       & 50                                                           & 54.20                                                           & 4.20                                                        & 45.95  \\ 
			\hline
			\multicolumn{12}{c}{DPM sampler}                                                                                                                                                                                                                                                                                                                                                                                                                                                                                                                                   \\ 
			\hline
			DDPM  & Gender=Male & 30                                                          & 50                                                           & 50.53                                                           & 0.53                                                        & Age=Young & 79.40                                                       & 50                                                           & 52.28                                                           & 2.28                                                        & 40.25  \\
			VPSDE & Gender=Male & 30                                                          & 50                                                           & 52.73                                                           & 2.73                                                        & Age=Young & 79.40                                                       & 50                                                           & 52.33                                                           & 2.33                                                        & 42.29  \\
			\bottomrule
		\end{tabular}

	}
\end{table}

\smallskip\noindent
\textbf{Defense performance on different diffusion steps.}
Figure~\ref{fig:defense_diffSteps_ddpm_celeba1k_male_30} shows defense performance on different diffusion steps. Here, the target model is DDPM trained on CelebA-1k-30\% and we use the PC sampler and the total number of sampling steps is 1,000, and the sensitive property is male.
The blue line and the left axis show the inferred proportion while the red line and the right axis present the corresponding FID values.
Generated samples in the 0 diffusion step are pure Gaussian noise while generated samples in the 999 step are realistic samples.
Overall, we can see that defense performance and model utility in the latter stage of diffusion steps are better than that of the former stage.
Generally, choosing late middle diffusion steps can obtain a good balance in defense performance, FID scores, and the meaningfulness of generated images.

\begin{figure}[!t]
	\centering
	\includegraphics[width=0.85\linewidth]{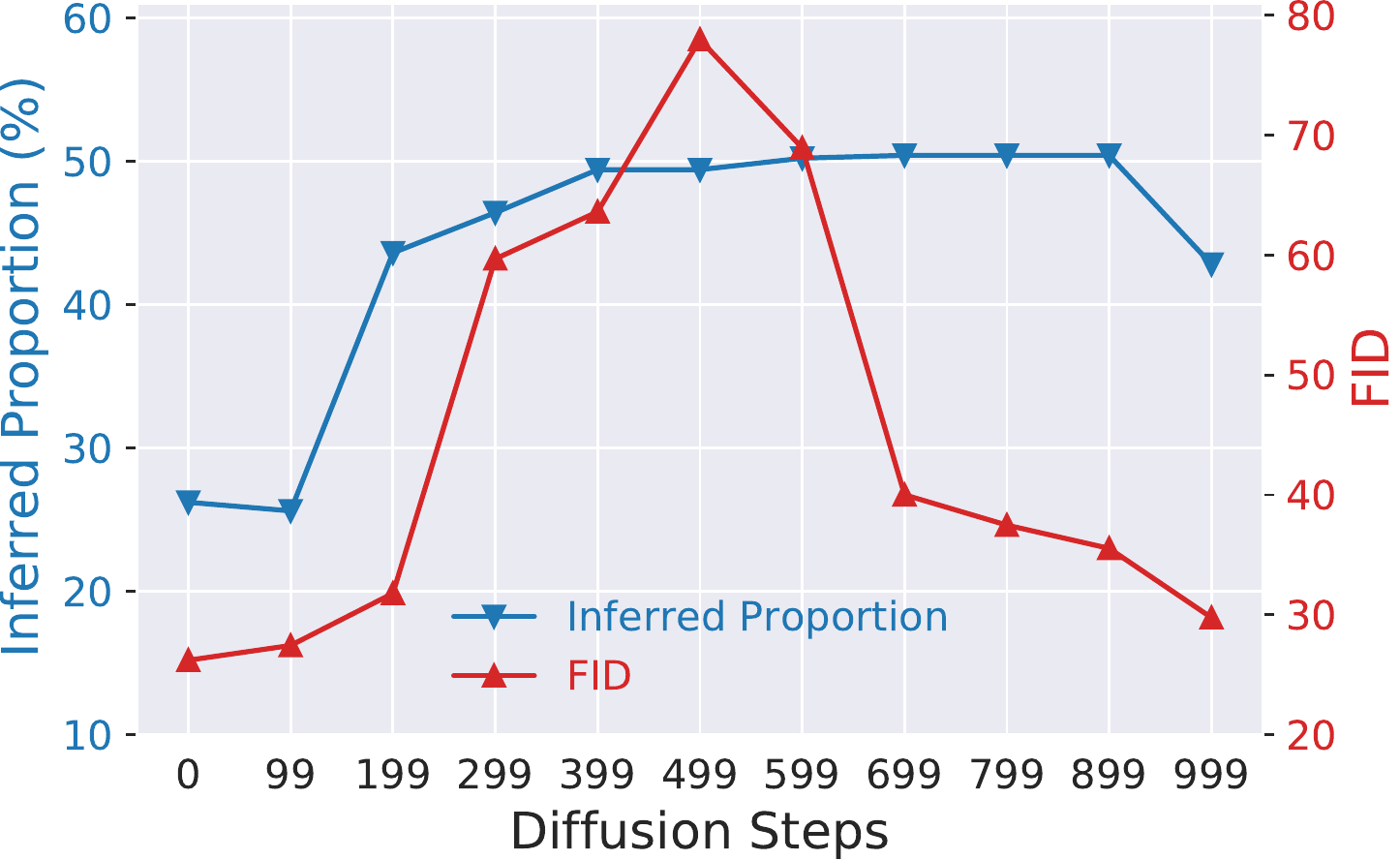}
	\caption{Defense performance on different diffusion steps.}
	\label{fig:defense_diffSteps_ddpm_celeba1k_male_30}
\end{figure}

\begin{figure}[!t]
	\centering
	
	\subfigure[$\epsilon = 10$, DPDM trained on CelebA-1k-30\%.]{
		\includegraphics[width=0.72\columnwidth]{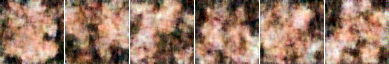}
		\label{fig:dpdm_celeba_1k_30_eps10}
	}
	\subfigure[{\sf PriSampler} for DDPM trained on CelebA-1k-30\%.]{
		\includegraphics[width=0.72\columnwidth]{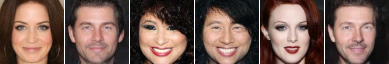}
		\label{fig:ddpm_celeba_1k_30}
	}
	\subfigure[$\epsilon = 10$, DPDM trained on CelebA-50k-30\%.]{
		\includegraphics[width=0.72\columnwidth]{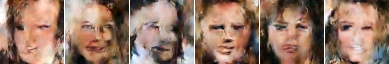}
		\label{fig:dpdm_celeba_50k_30_eps10}
	}	
	\subfigure[$\epsilon = 50$, DPDM trained on CelebA-50k-30\%.]{
		\includegraphics[width=0.72\columnwidth]{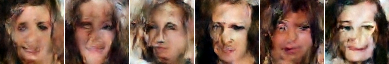}
		\label{fig:dpdm_celeba_50k_30_eps50}
	}
	\subfigure[{\sf PriSampler} for SMLD trained on CelebA-50k-30\%.]{
		\includegraphics[width=0.72\columnwidth]{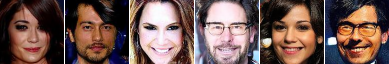}
		\label{fig:smld_celeba_50k_30}
	}
	\caption{More visualizations of synthetic samples under the defense DPDM and our defense method {\sf PriSampler}.}
	\label{fig:def_perf_cmp_whole}
\end{figure} 

\end{document}